\documentclass{aa}
\usepackage[utf8]{inputenc}  
\usepackage[dvipsnames]{xcolor}
\usepackage[T1]{fontenc}
\usepackage{lmodern}  
\usepackage{graphicx}
\usepackage{hyperref}
\usepackage{placeins}
\usepackage{float}
\usepackage{makecell}
\usepackage{subfigure}
\usepackage{stfloats} 
\usepackage[normalem]{ulem}
\usepackage{amsmath}
\definecolor{elizabra1}{RGB}{150, 80, 200}
\definecolor{elizabra}{RGB}{0, 200, 150}

\definecolor{elizabra2}{RGB}{0, 100, 150}
\hypersetup{
  colorlinks=true,    
  linkcolor=elizabra2,
  citecolor=elizabra1,     
  urlcolor=elizabra2       
}

\usepackage{txfonts}

\newcommand{\eliza}[1]{\textcolor{black}{#1}}

\newcommand{\eliz}[1]{\textcolor{black}{ #1}}

\numberwithin{subsection}{section}

\begin{document} 

  \titlerunning{Cosmic-ray impact on optical and mid-infrared emission-line diagnostics in NGC 5728}
  \authorrunning{E. Koutsoumpou et al.}
  
 \title{Cosmic-ray impact on optical and mid-infrared emission-line diagnostics in NGC 5728}

\author{E. Koutsoumpou\inst{1}\fnmsep\thanks{\email{evkoutso@phys.uoa.gr}}, J. A. Fernández-Ontiveros\inst{2}, K. M. Dasyra\inst{1}, and L. Spinoglio\inst{3}
          }

   \institute{Section of Astrophysics, Astronomy \& Mechanics, Department of Physics, National and Kapodistrian University of Athens, University Campus Zografos, GR 15784, Athens, Greece
           \and
            Centro de Estudios de F\'isica del Cosmos de Arag\'on (CEFCA), Plaza San Juan 1, E--44001, Teruel, Spain
        \and
            Istituto di Astrofisica e Planetologia Spaziali (INAF--IAPS), Via Fosso del Cavaliere 100, I--00133 Roma, Italy
            }

\date{Received 5 June 2025 / Accepted 5 September 2025}



\abstract
{Cosmic rays (CRs), produced by the jets of active galactic nuclei (AGN) and supernovae (SNe), serve as a significant feedback mechanism influencing emission lines in narrow line region (NLR) clouds. These highly energetic particles, propelled by shocks, not only heat the interstellar medium (ISM) but also modify its chemical composition. This study investigates the role of CRs, particularly in their ability to excite gas and align with observed line flux ratios across UV and optical diagnostic diagrams. We employed {\sc {CLOUDY}} simulations to explore the CR ionization rate, the ionization parameter, and the initial hydrogen density effects on optical and mid-infrared (MIR) emission. Our analysis includes high-quality optical data from the Multi Unit Spectroscopic Explorer (MUSE) on the Very Large Telescope (VLT) for NGC 5728, supplemented by infrared observations from the \textit{James Webb Space Telescope} (\textit{JWST}). 
This multiwavelength approach provides a deeper understanding of CR impact on the ISM. Our previous results indicate that CRs are instrumental in heating the inner regions of gas clouds, enhancing the emission of low-ionization optical lines such as [\ion{N}{ii}] and [\ion{S}{ii}]. Additionally, the integration of mid-infrared (MIR) data reveals that emission lines susch as [\ion{Ar}{ii}] and [\ion{Ne}{ii}] within the \textit{JWST} Mid-Infrared Instrument (MIRI) field of view are sensitive to CRs. \eliz{In contrast, high-ionization lines (e.g., [\ion{Ne}{v}], [\ion{O}{iv}]) serve as robust tracers of photoionization being insensitive to CR effects. Moreover, mixed optical–MIR diagnostic diagrams offer insight into the relative roles of CRs and shocks, which often produce similar signatures in emission lines. We find that while both mechanisms can elevate certain line ratios, their influence on MIR diagnostics diverges: shocks and CRs affect low-ionization lines differently, allowing for a better understanding when multiwavelength data are available. Our approach not only helps to resolve the degeneracy between metallicity and CR ionization but also enables the potential differentiation of shocks and CR-driven processes in AGN host galaxies.}
}
 
 \maketitle

\section{Introduction}\label{intro}

Emission-line diagnostics have been a very useful tool in investigating the different and complex mechanisms of gas excitation within galaxies. The most commonly used diagnostics for disentangling
gas ionized by radiation of active galactic nuclei (AGN) from stellar-driven ionization by O and B stars are based on optical lines such as $[\ion{O}{iii}]\lambda5007\rm \mathring{A}$, $\,[\ion{O}{i}]\lambda6300\rm \mathring{A}$, H$\alpha$, H$\beta$, $[\ion{N}{ii}]\lambda6584\rm \mathring{A}$, and $[\ion{S}{ii}]\lambda\lambda 6716,6731\rm \mathring{A}$; they are known as the Baldwin, Phillips, and Terlevich (BPT) diagrams \citep{1981BPT}. Since their introduction to the astronomical community, BPT diagrams, have been enhanced by theoretical limits and empirical divisions to distinguish between AGN-dominated and star-forming sources, as well as between Seyfert and low-ionization nuclear emission-line region (LINER) galaxies (\citealt{2003Kauf,2006Kewley,Schawinski_2007}).

Modeling the emission-line fluxes obtained from spectra is typically carried out using photoionization codes such as \textsc{Cloudy} \citep{Ferland_2013, Ferland_2017} and \textsc{MAPPINGS} \citep{Binette_1985, Sutherland_1993, Dopita_1995}, which provide detailed constraints on the physical properties of ionized gas. These codes incorporate various excitation mechanisms, including ultraviolet (UV) and X-ray radiation, and in the case of \textsc{MAPPINGS}, shocks are also included \citep[e.g.][]{Allen_2008,Ferland_2009,Feltre_2016,Ferland_2017,Feltre_2023,Chatzikos_2023,Zhu_2023}. \eliza{Such models are used to interpret observed emission lines across BPT and other diagnostic diagrams. For example, photoionization models can reproduce line fluxes in individual objects, providing detailed physical constraints, as shown by \cite{PerezMontero_Diaz_2007} for Mrk 209 and by \cite{Oliveira_2024} for nitrogen and oxygen abundances in LINER galaxies.}
The modeling of the specific position of the observations, is affected by the physical characteristics of the emitting source—such as stellar mass, star formation rate (SFR), metallicity, and  dust-to-metal mass ratio \citep{Hirschmann_2017,Feltre_2016,Feltre_2023}, \eliza{as well as density, shape and intensity of the radiation field \citep{Feltre_2016,Zhu_2023}}, making the modeling process quite challenging, especially when examining large samples of galaxies.
\eliza{ A common approach is to compare observational data with precomputed grids of models covering a range of physical parameters. This method forms the basis of widely used tools such as \textsc{HII-CHI-Mistry} \citep{Perez_2014}, \textsc{IZI} \citep{Blanc_2015}, \textsc{NebulaBayes} \citep{Thomas_2018}, and \textsc{HOMERUN} \citep{Marconi_2024}.}


Moreover, cosmic rays (CRs)—highly energetic particles primarily generated by supernova remnants (SNRs) and black hole jets— represent another potential but less explored excitation source \citep{Blasi_2013, Padovani_2017, Veilleux_2020, Wolfire_2022, Kantzas_2023}. Their ability to induce excitation and ionization via secondary electrons deep within molecular clouds significantly impacts nebular emission lines \citep{Spitzer_1968, McKee_1989, Padovani_2018,Gabici_2022}, and has similar effects to supersolar metallicity on the $[\ion{N}{ii}]\lambda6584\rm \mathring{A}$ BPT diagram \citep{2025K}, creating a degeneracy that complicates the interpretation of metallicity and the ionization source. 

Infrared emission-line ratio diagnostics have become essential for examining ionization mechanisms in dusty or obscured regions of galaxies, and have been used to disentangle photoionization due to AGN activity from young stars \citep[see][]{Spinoglio_1992}.
 Diagrams based on high- to low-ionization line ratios—such as $[\ion{O}{IV}]\lambda25.9\,\rm \mu m\,/\, [\ion{Ne}{II}]\lambda12.8\,\rm \mu m$ and $[\ion{Ne}{V}]\lambda14.3\,\rm \mu m\,/\, [\ion{Ne}{II}]\lambda12.8\,\rm \mu m$— are typically used to estimate AGN activity, as they trace ionizing photons with energies beyond those produced by standard stellar populations
 \citep{Genzel_1998, Sturm_2002, Groves_2006}. Yet notably, $[\ion{Ne}{V}]$ emission has also been observed in galaxies lacking clear AGN signatures \citep[e.g.][]{Thuan_2005,Hatano_2023,Hatano_2024,Mingozzi_2025}. Moreover, ratios such as $[\ion{Ne}{III}]/[\ion{Ne}{II}]$,involving lower ionization lines serve as indicators of the ionizing spectrum hardness and have been used to characterize stellar populations in star-forming regions \citep{Thornley_2000,PerezMontero_2024}. With the development of new state-of-the-art facilities, these diagnostics can now be spatially resolved, even in fainter or more embedded sources, and extended with lines such as $[\ion{Ne}{VI}]$ and $[\ion{Ar}{V}]$ \citep{Richardson_2022, Feltre_2023}. These line ratios complement optical diagnostics by providing a dust-insensitive view of excitation conditions, revealing obscured or composite sources, and enabling a more complete interpretation of the ionization mechanisms in galaxies that could also be used to identify ionization due to CRs.

In this study, we extended our previous work \cite{2025K}, hereafter referred to as KFDS25, in order to examine whether CRs are capable of producing the observed emission in both the optical and mid-infrared (MIR) wavelength range.
Specifically, we integrated MIR data from the Mid-Infrared Instrument (MIRI) aboard the \textit{James Webb Space Telescope} (\textit{JWST}), with optical data from the Multi Unit Spectroscopic Explorer (MUSE) on the Very Large Telescope (VLT) to explore the dual impact of photoionization and CR-induced ionization, in a single source, NGC 5728. We tested the effectiveness of the proposed diagnostics on NGC 5728, as it is a jetted and extensively studied AGN, with both optical and MIR observations available for comparison. Section \ref{GALAXIES} describes the main features and the observational data for NGC 5728. Section \ref{methods} outlines our modeling procedure and the steps taken to treat data from certain areas of interest within NGC 5728. Section \ref{results} outlines the findings of the present analysis, and Section \ref{discussion} explores possibilities for further research. We conclude the present study with the summary in Section \ref{summary}.


\section{Observational data}\label{GALAXIES}

\begin{figure*}[!ht!!!!!!!!!!!!!!]
    \centering
    \subfigure[]{\includegraphics[width=0.33\textwidth]{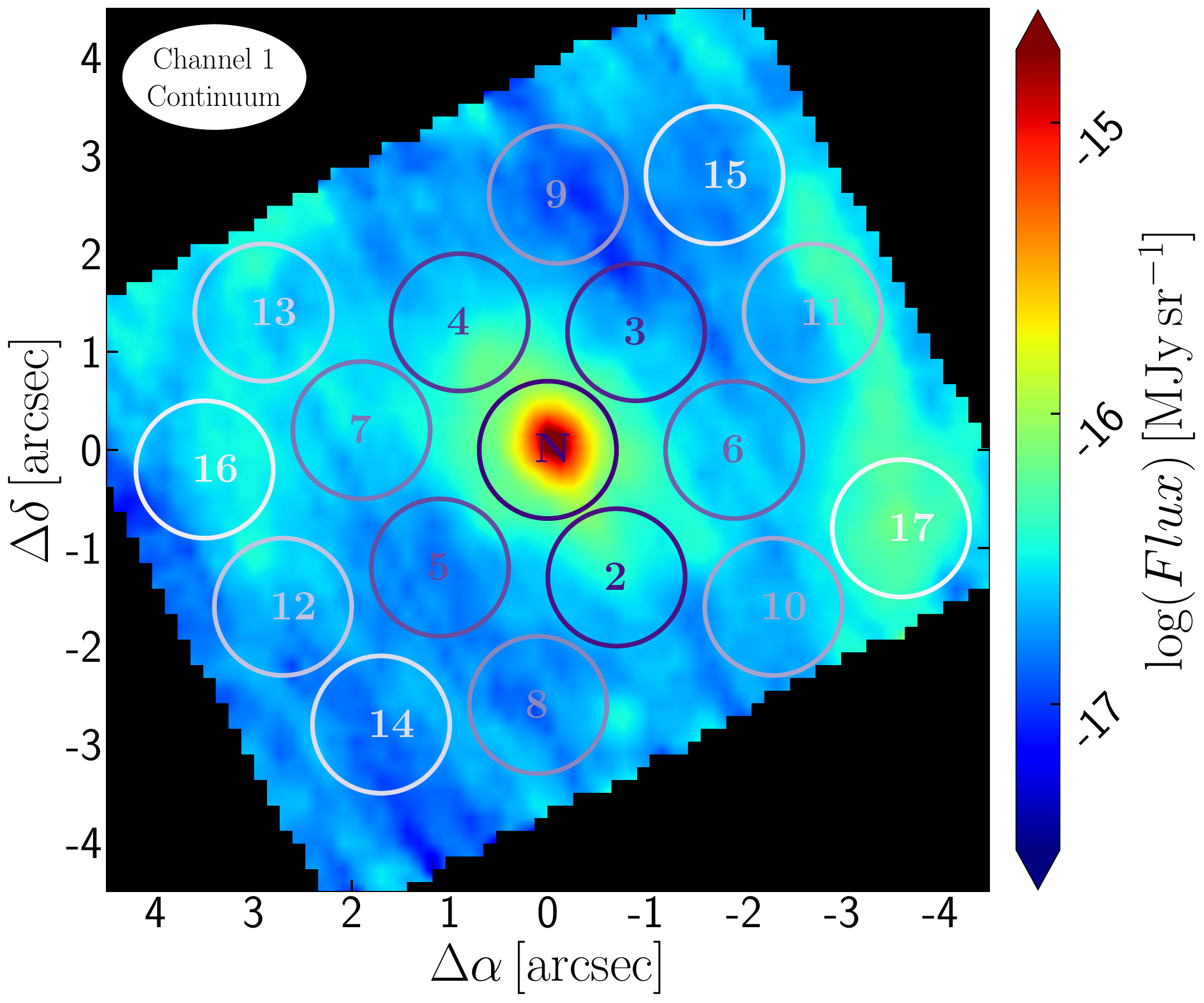}\label{subfig:cont_ch1}}~
    \subfigure[]{\includegraphics[width=0.33\textwidth]{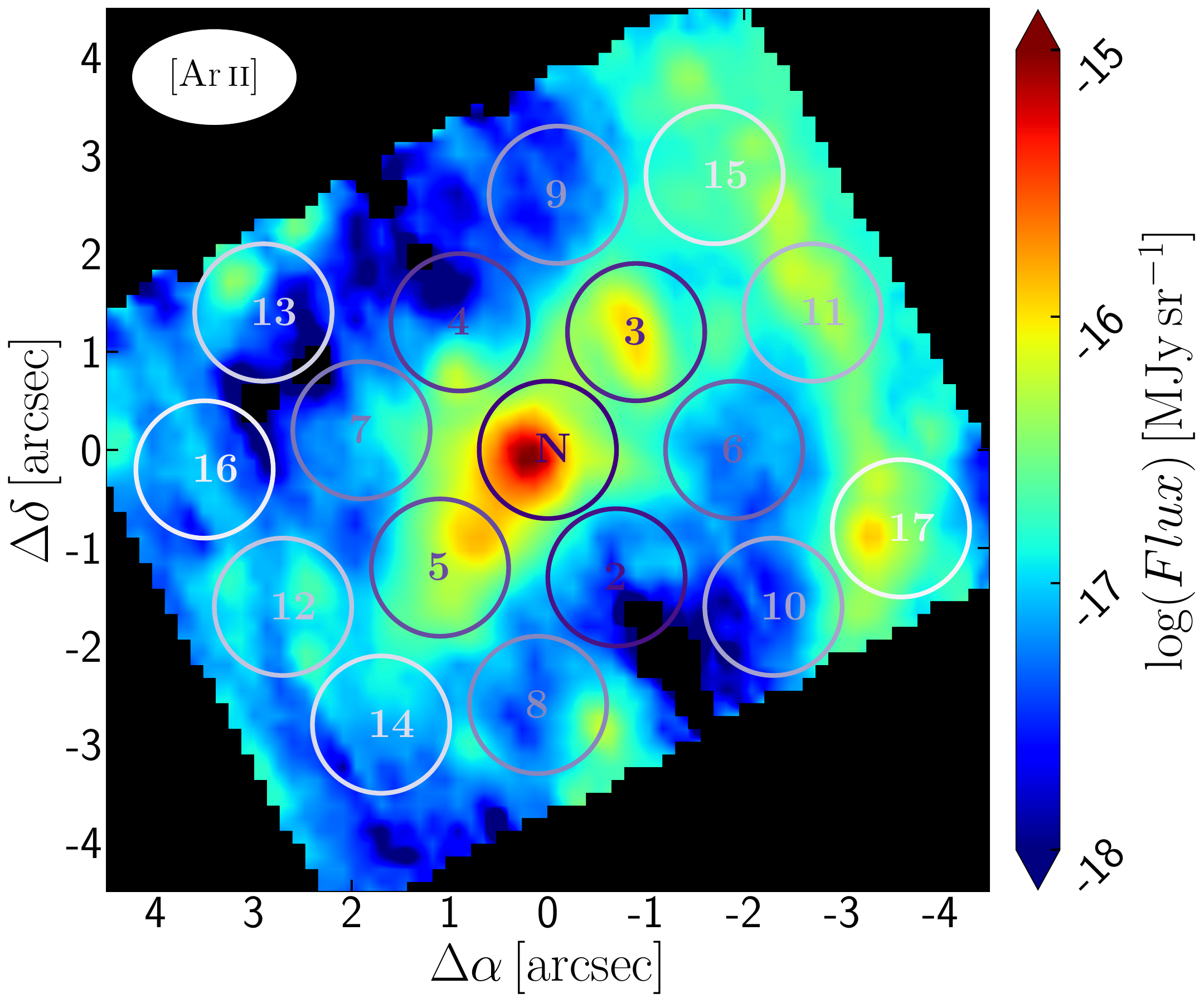}\label{subfig:ar2_ch1}}~
    \subfigure[]{\includegraphics[width=0.33\textwidth]{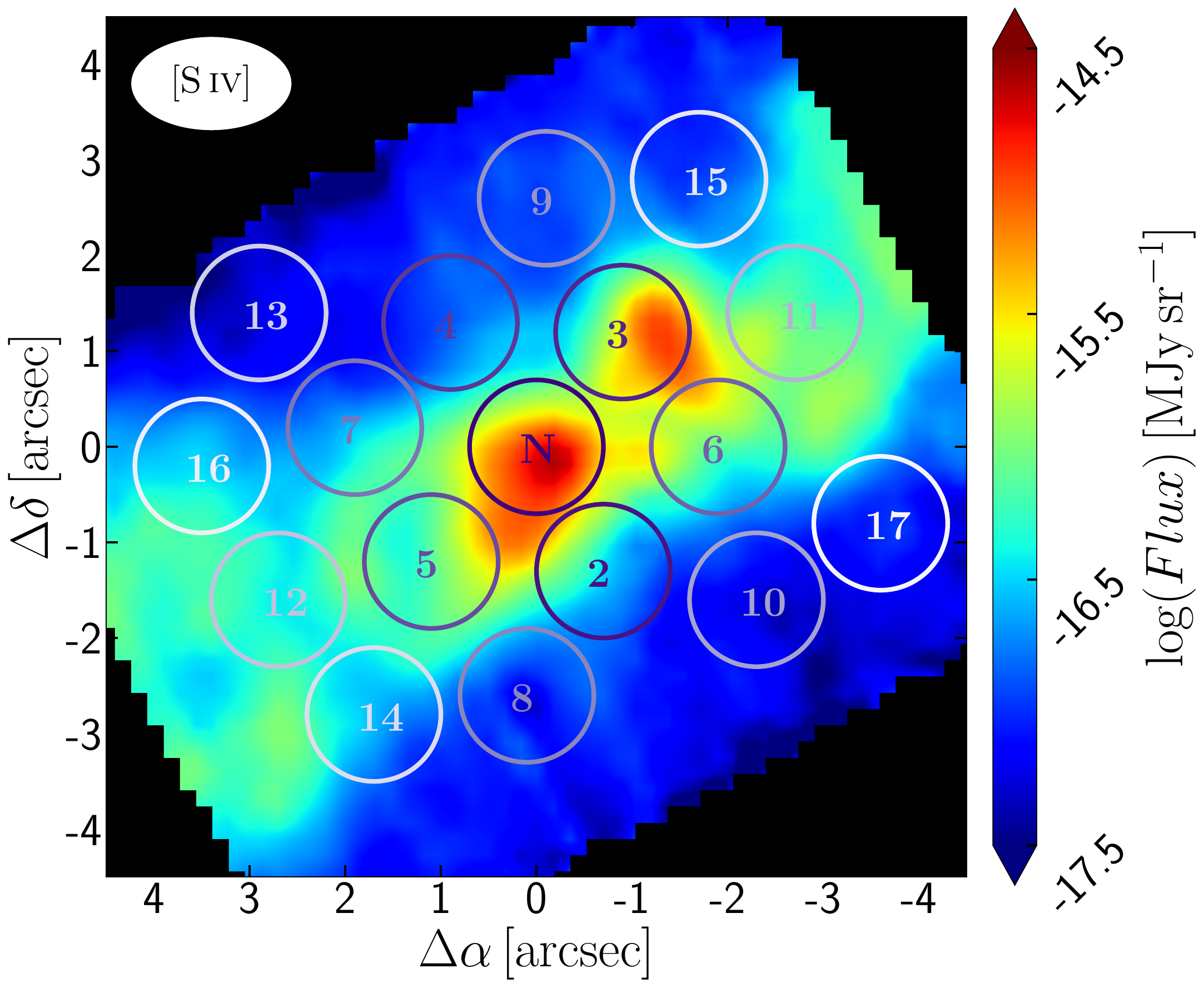}\label{subfig:s4_ch3}}
    \subfigure[]{\includegraphics[width=0.33\textwidth]{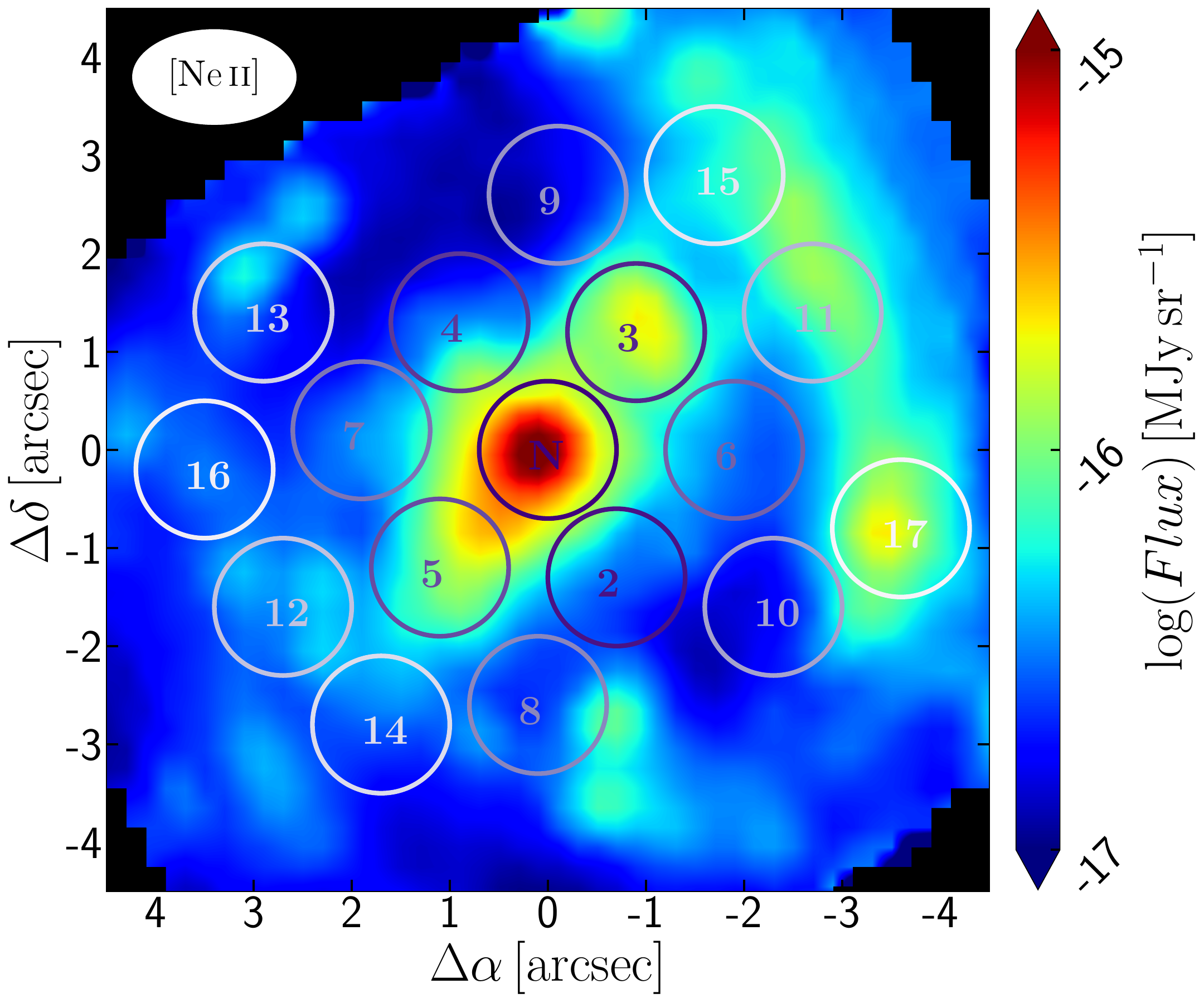}\label{subfig:ne2_ch3}}~
    \subfigure[]{\includegraphics[width=0.33\textwidth]{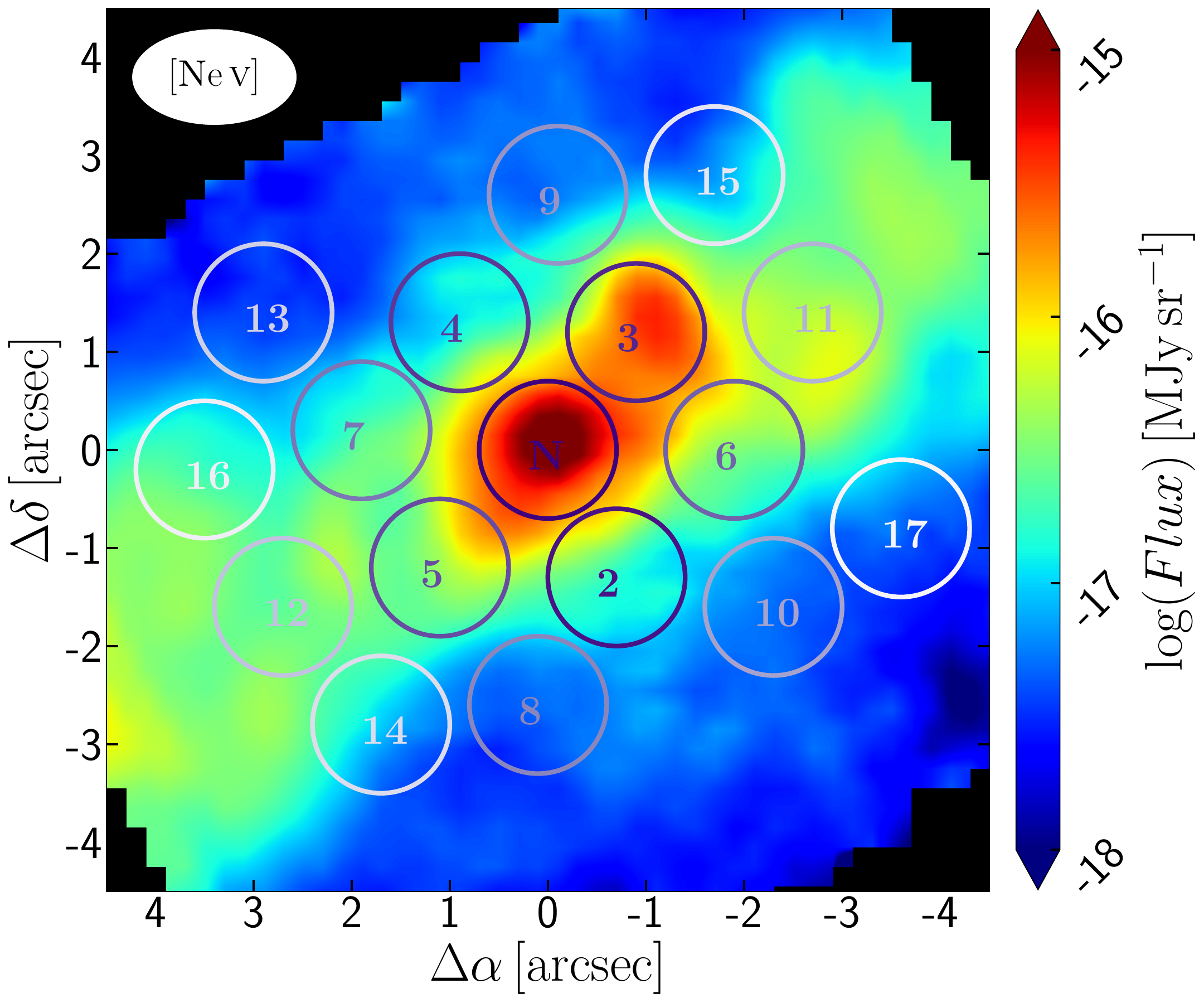}\label{subfig:ne5_ch3}}~
    \subfigure[]{\includegraphics[width=0.33\textwidth]{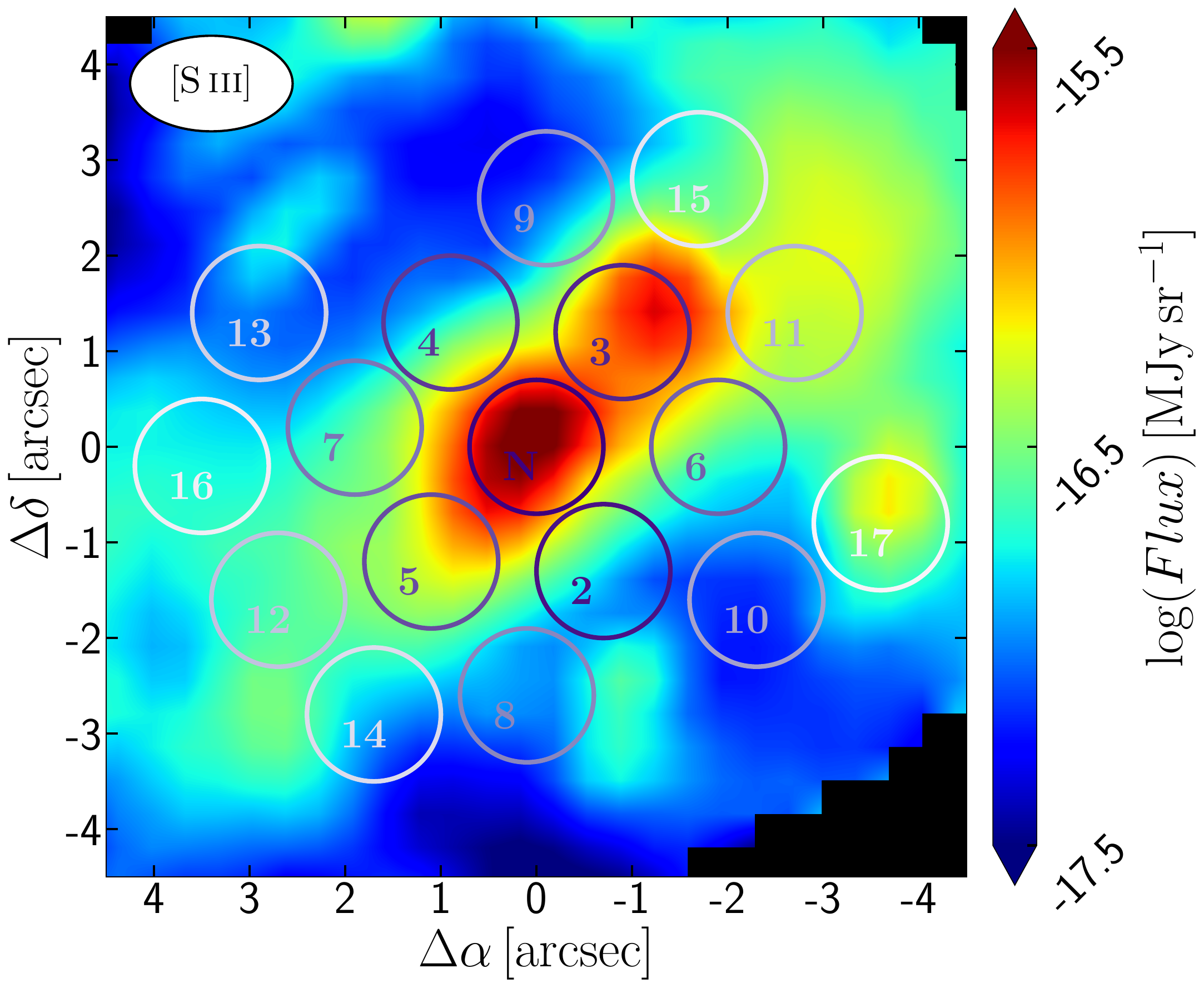}\label{subfig:s3_ch4}}
    \subfigure[]{\includegraphics[width=0.33\textwidth]{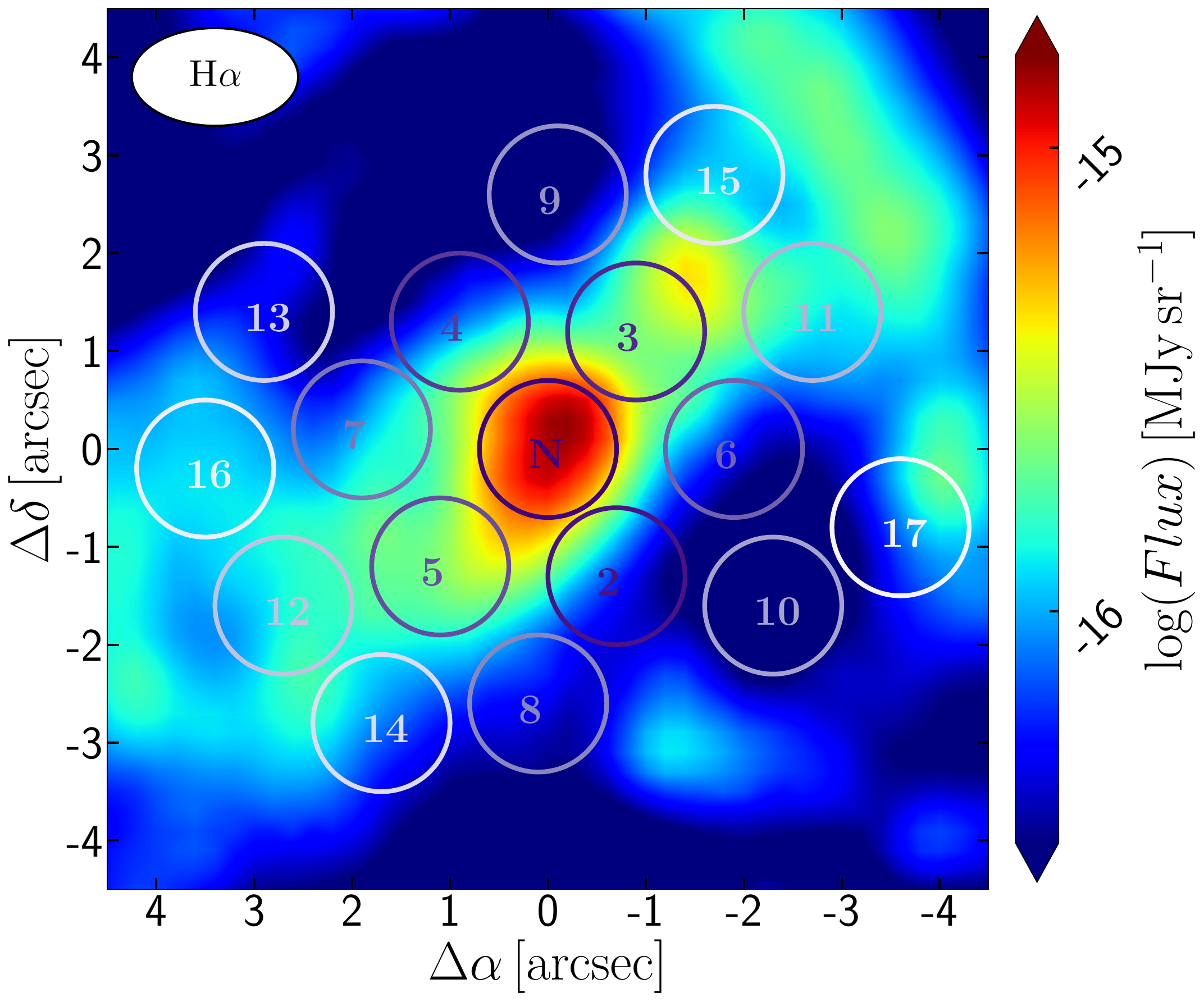}\label{subfig:ha_muse}}~
    \subfigure[]{\includegraphics[width=0.33\textwidth]{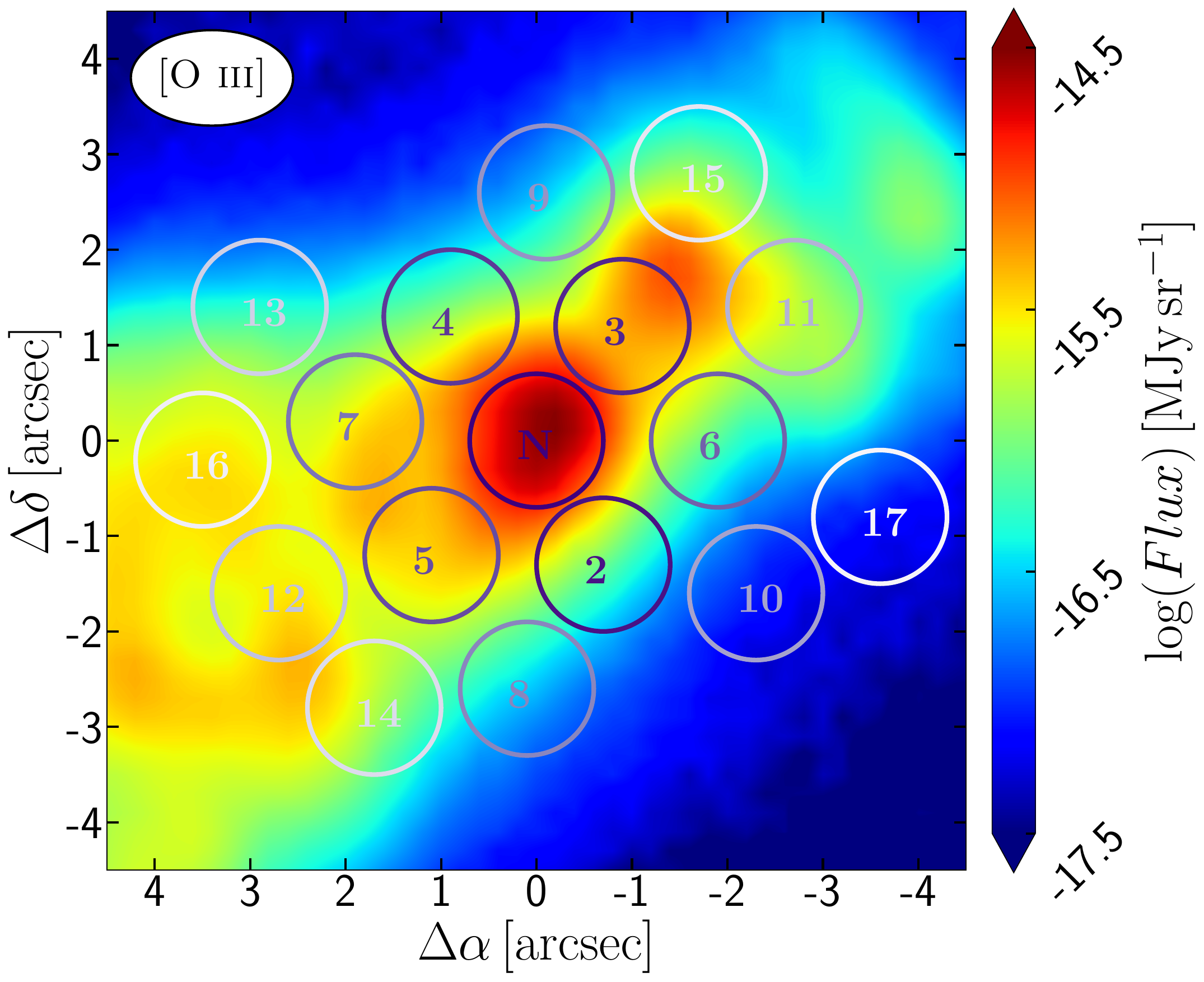}\label{subfig:o3_muse}}~
    \subfigure[]{\includegraphics[width=0.30\textwidth]{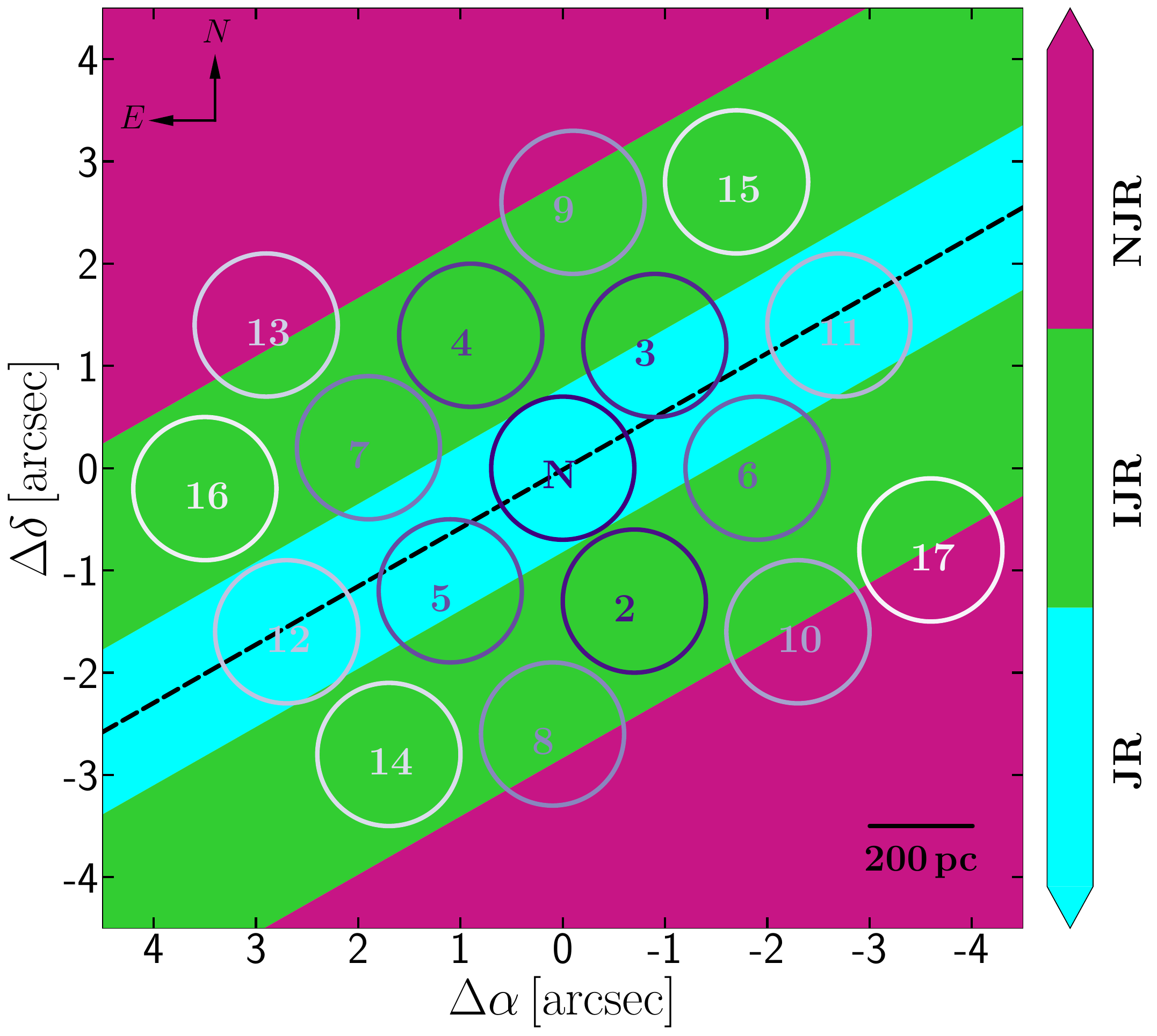}\label{subfig:sketch}}~
    \caption{Apertures chosen to extract spectra from the spectroscopic images of NGC 5728. These are depicted in different shades of purple from deep purple being the nuclear aperture "N" to pale lilac in increasing distance, also noted with numbers. The apertures are drawn over (a) the continuum in Channel 1 of MIRI, extracted in the rest-frame range of 5.90–6.05$\rm\mu$m and over $[\ion{Ar}{II}]{\lambda\rm7\mu m}$, [\ion{S}{IV}]$\lambda 10.5\rm \mu m$, [\ion{Ne}{II}]$\lambda12.8\rm \mu m$, [\ion{Ne}{V}]$\lambda14.9\rm \mu m$, [\ion{S}{III}]$\lambda18.7\rm \mu m$, H$\alpha$, and [\ion{O}{iii}]$\lambda$5007\AA \, (b-h) line emission maps  with the stellar continuum subtracted. Panel~(i) presents a schematic illustration of the adopted jet axis together with the three color-shaded regions used to group the apertures (see Sec.~\ref{apertures}). These regions are defined by the projected vertical distance from the jet axis, relative to the aperture radius: JR (innermost, cyan) includes apertures within one radius (nuclear and/or jet regions); IJR (intermediate, lime green) covers one to two radii (intermediate jet-affected regions); and NJR (outermost, magenta) includes apertures beyond two radii (non-jet or star-forming regions). The multiwavelength data are aligned based on the astrometry of the nucleus.} 
 \label{fig:chosen_apertures}
    \vspace{-0.3cm}
\end{figure*}


We selected NGC 5728, a Seyfert 2 galaxy, which has been observed thoroughly and exhibits clearly apparent jets \citep{Durre_2018,Davies_2024}. Additionally, this source features a star formation ring with a significant number of young stars \citep{Schommer_1988,Shin_2019,Shimizu_2019}. This renders it a perfect case to study the effects of both jet and supernova-related CRs in AGN and starburst-dominated regions.
NGC 5728 proves to be an appropriate target, since it has publicly available VLT/MUSE observations in the European Southern Observatory (ESO) Science Archive Facility (see Section \ref{muse_datacubes}), as well as data from MIRI/\textit{JWST} (see Section \ref{jwst_datacubes}). Lying in close proximity, at a redshift of z=0.00932\footnote{\url{https://ned.ipac.caltech.edu}}, NGC 5728 provides a great opportunity to explore effect of CRs in different galactic environments through a variety of high-quality observational data. 



\subsection{MUSE data}\label{muse_datacubes}

VLT-MUSE raw data are handled by the ESO MUSE pipeline, which includes bias correction, dark subtraction, flat fielding, and background subtraction to improve image quality and reduce noise. Furthermore, the pipeline conducts flux and wavelength calibration for spectral accuracy, and glitch removal from CRs. Subsequently, spectral extraction yields the object's spectra from spatially resolved data cubes. \cite{Weilbacher_2014, Weilbacher_2020} provide a complete overview of these processes in the various operational stages. The MUSE data cube of NGC 5728 is publicly accessible in the ESO Archive\footnote{\url{http://archive.eso.org/cms.html}}, under program ID 097.B-0640 (PI: D. Gadotti). For further details on the data reduction performed on this data cube, see \cite{MING}.




\subsection{\textit{JWST} data}\label{jwst_datacubes}

The data processing pipeline for MIRI Medium Resolution Spectroscopy (MRS) on \textit{JWST} meticulously processes raw observational data through several sophisticated stages to produce scientifically exploitable outputs. Initially, the Detector1 pipeline addresses raw, uncalibrated data from MIRI, correcting instrument-specific imprints and detector anomalies such as dark currents and CRs. This results in slope images that are corrected yet remain uncalibrated, representing the initial rate of signal change. Following this, the Spec2 pipeline performs calibration corrections including flat fielding, wavelength calibration, and spatial adjustments, alongside background subtraction to eliminate sky noise and other irrelevant nonastronomical signals, producing precisely calibrated 2D spectral images. Conclusively, the Spec3 pipeline combines data from multiple exposures and dither configurations, thereby enhancing the signal-to-noise ratio (S/N) and expanding both the spatial and spectral coverage. This process produces fully calibrated 3D data cubes for each observed target, meticulously prepared for comprehensive scientific exploration.

In our study, we employed uncalibrated data of NGC~5728 obtained under the \textit{JWST} Proposal ID~1670 (PI: T. Taro Shimizu), which are publicly available. These data were retrieved from the ESA \textit{JWST} Science Archive\footnote{\url{https://jwst.esac.esa.int/archive/}}, where they were downloaded and processed through all three stages of the \textit{JWST} pipeline using version~1.16.0 and the Calibration References Data System (CRDS) context "jwst\_1295.pmap". The resulting data, spanning the continuous wavelength range of $4.9$--$27.5\,\mu$m across all MRS channels and sub-bands show no substantial differences from the Level~3 products available in the \textit{JWST} archive or from recent studies that initially used different versions of the pipeline and CRDS contexts \citep{Pereira_2022,Davies_2024}. The corresponding datasets are also available through the Mikulski Archive for Space Telescopes (MAST)\footnote{\url{https://mast.stsci.edu}; DOI: \url{https://doi.org/10.17909/fbb5-d563}}.


\subsection{Spectrum extraction}\label{apertures}

The primary objective of this study was to investigate the parameters influencing the excitation of gas across different regions of NGC 5728, and specifically the combined effects of CRs with photoionization. Our approach involved selecting specific regions within NGC 5728 and extracting spectra from the MUSE and \textit{JWST} data cubes using circular apertures. Each aperture was carefully positioned to assess the impact of CRs in both star-forming and jet-influenced areas.

We selected apertures that encompass gas interacting directly with jets and/or outflows, as well as regions indicative of ongoing star formation in the vicinity of NGC 5728. This was achieved by identifying areas exhibiting strong emission across multiple emission lines, as illustrated in Fig.~\ref{fig:chosen_apertures}. To ensure a robust comparison between the optical and MIR data, we adopted circular apertures with a radius of $0.7''$ ($\sim 138\mathrm{pc}$), which closely matches the full width at half maximum (FWHM) of the point spread function (PSF) in the reddest channels of the MIRI and provides a suitable compromise between spatial resolution and S/N.
Moreover, in Fig.~\ref{fig:chosen_apertures}, the selected apertures are depicted in varying shades of purple to indicate their distance from the nucleus. The nucleus itself (“N”) is shown in deep purple, while apertures at larger distances are labeled with increasing numbers and displayed in progressively lighter shades, ending with pale lilac for the most distant regions.
 

In Panel (i) of Fig.~\ref{fig:chosen_apertures}, aperture observations are grouped into three categories based on their projected vertical distance from the jet axis, which is defined by the dichotomy line of the ionization cones shown in the middle panel of fig.~12 in \citet{Shimizu_2019}. Regions within one aperture radius from the axis are considered nuclear-jet regions (JR), shaded cyan in panel (i) and marked with squares in subsequent figures. Apertures located beyond three and a half radii are classified as non-jet regions (NJR), more likely dominated by star formation, and are shaded magenta in panel (i) and represented by stars in other plots. Those situated between these two borders are regarded as intermediately jet-affected regions (IJR), shaded lime green in panel (i) and shown as diamonds elsewhere. If an aperture overlaps multiple regions, it is classified according to where the majority of its area lies.

Furthermore, from each aperture, we extracted the convolved spectra using MPDAF \citep{MPDAF} for the optical and \textsc{Photutils} \citep{phot_2024} for the infrared range.
\eliz{Both the continuum subtraction and the modeling of emission and absorption lines in the optical convolved spectra were performed using \textsc{Pyplatefit} \citep{pypla,Tremonti_2004, Brinchmann_2004}, which fits and subtracts simple stellar population models \citep{Bruzual_2003, Brinchmann_2013} as part of its integrated fitting procedure.}
Infrared emission lines, extracted from the same regions, were fitted with a single Gaussian component using the \textsc{lmfit} package \citep{newville_2015_11813}, while the continuum subtraction involved fitting a polynomial. Additionally, the spectra were smoothed using a median filter prior to fitting to reduce noise and eliminate substantial stellar and AGN contributions. \eliza{A summary of the aperture properties—including positions, minimum S/N values, projected distances to the nucleus, and classifications based on Fig.\ref{subfig:sketch}—is provided in Appendix \ref{ap_prop}. All emission lines were initially selected for fitting by our routines using a $3\sigma$ detection threshold, with a few visually confirmed faint detections included in Table~\ref{tab:aperture_sn}.} Finally, all emission lines detected were fitted in NGC 5728's rest-frame wavelengths.

\section{Methods}\label{methods}
\subsection{Constraining the parameter space}\label{subsec:param_space}

We constrained the ionization parameter, O/H, and N/O ratios for regions within NGC\,5728 using the code \textsc{HII-CHI-Mistry}\footnote{Publicly available at \url{https://home.iaa.csic.es/~epm/HII-CHI-mistry.html}}, developed by \cite{Perez_2014}. This tool applies Bayesian-like statistics to compare observed emission-line ratios against extensive grids of photoionization models. We specifically used the optical and infrared subroutines optimized for AGN \citep{Perez_2019,Perez_Diaz_2022}.

In the optical regime, \textsc{HII-CHI-Mistry} relies on emission lines such as [\ion{O}{ii}]$\lambda3727$\AA, [\ion{Ne}{iii}]$\lambda3868$\AA, [\ion{O}{iii}]$\lambda4363$\AA, [\ion{O}{iii}]$\lambda4959$\AA, [\ion{O}{iii}]$\lambda5007$\AA, [\ion{N}{ii}]$\lambda6584$\AA, and [\ion{S}{ii}]$\lambda\lambda6716,6731$\AA. However, the MUSE spectral coverage (4800–9300\AA) excludes [\ion{O}{ii}]$\lambda3727$\AA~and [\ion{Ne}{iii}]$\lambda3868$\AA. \eliz{Thus, given the spectral coverage of MUSE, our analysis is necessarily limited to all the optical emission lines that fall within its wavelength range—specifically, [\ion{O}{iii}]$\lambda4959$\AA, [\ion{O}{iii}]$\lambda5007$\AA, [\ion{N}{ii}]$\lambda6584$\AA, and [\ion{S}{ii}]$\lambda\lambda6716,6731$\AA.}

In the infrared regime, \textsc{HII-CHI-Mistry} \citep{Fernandez2021,Perez_Diaz_2022} employs emission lines including HI$\lambda 4.05\rm \mu m$, HI$\lambda 7.46 \rm \mu m$, [\ion{S}{IV}]$\lambda 10.5\rm \mu m$, HI$\lambda12.4\rm \mu m$, [\ion{Ne}{II}]$\lambda12.8\rm \mu m$, [\ion{Ne}{V}]$\lambda14.9\rm \mu m$, [\ion{Ne}{III}]$\lambda15.5\rm \mu m$, [\ion{S}{III}]$\lambda18.7\rm \mu m$, [\ion{Ne}{V}]$\lambda24.3\rm \mu m$, [\ion{O}{IV}]$\lambda25.9\rm \mu m$, [\ion{S}{III}]$\lambda33.7\rm \mu m$, [\ion{O}{III}]$\lambda52\rm \mu m$, [\ion{N}{III}]$\lambda57\rm \mu m$, [\ion{O}{III}]$\lambda88\rm \mu m$, [\ion{N}{II}]$\lambda122\rm \mu m$, and [\ion{N}{II}]$\lambda205\rm \mu m$. 
\eliz{While MIRI does not extend to the far-infrared lines included in the full \textsc{HII-CHI-Mistry} grid, such as [\ion{O}{III}]$\lambda88\rm \mu m$ and [\ion{N}{II}]$\lambda122\rm \mu m$, it still provides access to all the MIR emission lines needed for our analysis.}

\begin{table}[!ht]
\centering
\caption{Range of \textsc{HII-CHI-Mistry} estimated values for the ionization parameter, the oxygen abundance, and the N/O relative abundance of the extracted regions, using both the optical and infrared routines.}
\label{tab:metal}
\begin{tabular}{|c|c|c|c|}
\hline
\textbf{NGC 5728} & $\mathbf{\log U}$ & $\mathbf{Z/Z_{\odot}}$ & $\mathbf{\log (N/O)}$ \\
\hline
Optical   & [$-3.4$, $-2.9$] & [$0.4$, $1.0$] & [$-1.1$, $-0.8$] \\
Infrared  & [$-2.7$, $-2.4$] & [$0.4$, $0.7$] & -- \\
\hline
\end{tabular}
\end{table}



\eliza{We use \textsc{HII-CHI-Mistry} (v5.22 optical, v3.01 IR) with a double composite AGN continuum ($\alpha_{\mathrm{UV}} = -1$, $\alpha_{\mathrm{OX}} = -0.8$) and no ionization parameter constraint. The resulting parameter space, summarized in Table~\ref{tab:metal}, guides our simulations. In the IR, the $\mathrm{N}/\mathrm{O}$ ratio is unconstrained due to the lack of far-IR lines. See Section~\ref{Cloudy_models} and Table~\ref{tab:AGN_Model_Grid} for details.} Furthermore, these results support our initial assumption of solar metallicity ($1\, Z_{\odot}$), ensuring consistency with our previous analysis in KFDS25, and align with independent estimates reported by \citet{Perez_Diaz_2021}.
Finally, in Table~\ref{tab:AGN_Model_Grid} we present the relevant elemental abundances adopted as well.


\subsection{{\sc Cloudy} models}\label{Cloudy_models}

Initially, we employed the \eliza{dust-free} AGN models detailed in section 3.3 of KFDS25. However, these models struggled to accurately reproduce high-ionization MIR emission lines such as [\ion{Ar}{V}]$\lambda13.1\mu$m, [\ion{Ne}{V}]$\lambda14.3\rm\mu$m, [\ion{Ne}{V}]$\lambda24.3\rm\mu$m, [\ion{Ne}{VI}]$\lambda7.7\rm\mu$m, and [\ion{O}{IV}]$\lambda25.9\rm\mu$m. To resolve this issue, we created a new grid of AGN models using \textsc{Cloudy} v23.01 \citep{Ferland_2013,Ferland_2017,Chatzikos_2023} with the intermediate AGN spectral energy distribution (SED), "AGN\_Jin12\_Eddr\_mid.sed", described by \cite{Ferland_2020}, \eliz{also shown in teal color in Fig. \ref{fig:SEDs}}. This intermediate SED is a new option introduced in \textsc{Cloudy} 23.01, while previous versions of \textsc{Cloudy} (version 22.01 was employed in KFDS25) did not include this SED. We selected this intermediate SED because it provides a balanced and observationally validated scenario between lower Eddington ratio SEDs, which exhibit weaker far-UV emission and harder X-ray spectra, and higher Eddington ratio SEDs, marked by enhanced far-UV emission and softer X-ray spectra.

As in KFDS25, our \textsc{Cloudy} simulations use a fixed hydrogen column density of $N_{\rm H} = 10^{24}\mathrm{cm}^{-2}$, with the default temperature-based stopping criterion disabled. This ensures each model extends into cooler, deeper layers—crucial for capturing emission in strongly CR-heated models, where higher CR ionization rates produce more extended ionization structures. Using a smaller column or the default stopping criterion would prematurely truncate the emission from cooler regions.


Since \textsc{HII-CHI-Mistry} utilizes AGN photoionization templates generated with \textsc{Cloudy} 17.01 (see \citealt{Fernandez2021,Perez_Diaz_2022}), this is not identical to the new AGN SED described here, which provides improved modeling of high-ionization MIR lines. We tested \textsc{HII-CHI-Mistry} in the MIR, by explicitly excluding the high-ionization lines [\ion{Ne}{V}], and [\ion{O}{IV}], to assess the impact on the derived metallicity and ionization parameter ranges. The exclusion of high-ionization lines does not introduce significant systematic shifts in the derived abundances or ionization parameter beyond the initial uncertainties, confirming that the AGN templates used in the \textsc{HII-CHI-Mistry} routines remain reliable for metallicity and ionization parameter estimates. This robustness is consistent with the findings of \cite{Fernandez2021} and \cite{Perez_Diaz_2022}, who demonstrate that \textsc{HII-CHI-Mistry} produces stable metallicity and ionization parameter estimates even when certain diagnostic lines are absent or when only a subset of the MIR emission lines is available. Finally, in Section \ref{app_sed}, we show that the intermediate AGN template yields results that are identical to those obtained with the previous template in KFDS25.

The AGN model grid used is detailed in Table~\ref{tab:AGN_Model_Grid} and covers the parameter range previously outlined in section 3.3 of KFDS25, while the adopted abundances correspond to solar values \citep{Asplund}. These elemental abundances align with abundances derived via \textsc{HII-CHI-Mistry} for NGC 5728 and are consistent with abundance estimates for AGN narrow line regions (NLR) in the local Universe \citep[e.g.][]{Perez_Diaz_2021,Dors_alone_2021,Dors_2022,Perez_Diaz_2022,Perez_Diaz_2024} Finally, since \textsc{Cloudy} simulations do not include extinction, we applied Calzetti’s law \citep{calzetti} to attenuate the modeled fluxes of the optical emission lines, enabling a direct comparison with observations.


\begin{table}[ht]
\centering
\caption{Summary of AGN model grid parameters and adopted solar elemental abundances.}
\label{tab:AGN_Model_Grid}
\begin{tabular}{|cc|cc|}
\hline
\multicolumn{2}{|c|}{\textbf{Model Parameters}} & \multicolumn{2}{c|}{\textbf{Solar Abundances}} \\
Parameter & Range & Parameter & Value \\
\hline\hline
$\zeta_{\rm CR}$ $[\mathrm{s}^{-1}]$ & $10^{-16}$ to $10^{-12}$ & $\log(\mathrm{O}/\mathrm{H})_\odot + 12$ & 8.69 \\
$\log U$ & $-3.5$ to $-1.5$ & $\log(\mathrm{N}/\mathrm{O})_\odot$ & $-0.86$ \\
$n_{\rm H}$ $[\mathrm{cm}^{-3}]$ & $1$ to $10^4$ & $\log(\mathrm{C}/\mathrm{O})_\odot$ & $-0.3$ \\
&& $\log(\mathrm{He}/\mathrm{H})_\odot$ & $0.085$ \\
\hline
\end{tabular}
\vspace{-0.8cm}
\end{table}

\section{Results}\label{results}

\subsection{Optical line diagnostic diagrams}\label{subsec:bpts}


 With the adoption of the intermediate SED, our models consistently reproduce the low-ionization optical lines—[\ion{N}{ii}], [\ion{S}{ii}], and [\ion{O}{i}]—as effectively as in our previous work (KFDS25), across the same parameter space as shown in Table \ref{tab:AGN_Model_Grid}.
 Furthermore, the intermediate SED shows enhanced consistency with MIR diagnostics, discussed in detail in Section~\ref{subsec:ir_bpts}, and thus represents a more complete modeling approach.


In Fig.~\ref{fig:5728_BPTS_U}, it is evident that all jet-affected regions—whether strongly or moderately influenced-fall above the Kewley line, mainly in the Seyfert domain. CR ionization rates around $10^{-13}\,\rm s^{-1}$ better represent observations from nuclear and jet affected areas in the [\ion{S}{ii}] and [\ion{O}{i}] BPT diagrams, while a rate of approximately $10^{-12}\,\rm s^{-1}$ best models the parameter space of the observed [\ion{N}{ii}] ratios. These findings are in agreement with the results we acquired in KFDS25.

In KFDS25, we demonstrate that regions traditionally characterized as Seyfert and/or LINER sources can also be explained using the star-forming models incorporating CR ionization which led to a new parameter space demarcation line, the SF$\zeta$ line. This boundary captures the upper limit of line ratios expected from star-forming regions influenced by elevated CR ionization rates, offering a more physically motivated and complementary perspective to classical starburst boundaries. Applying this framework allowed us to reassess excitation mechanisms in galaxies without the need to immediately attribute elevated line ratios to AGN activity or extreme metallicities \cite{2004aGroves}. 

\begin{figure*}[!htbp]
    \centering
    \subfigure[]{\includegraphics[width=0.33\textwidth]{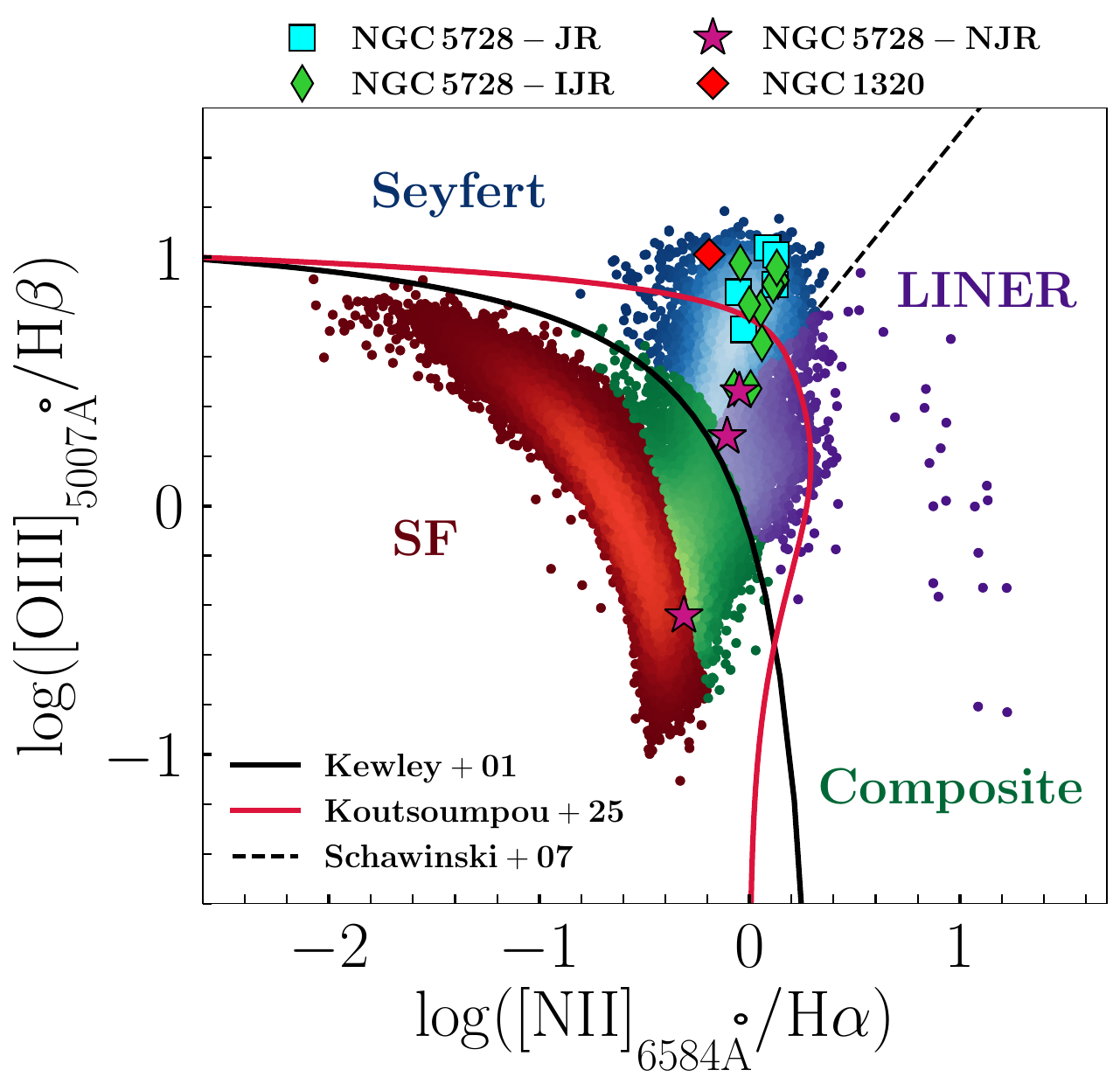}\label{subfig:sfzeta_line_N2}}~
    \subfigure[]{\includegraphics[width=0.33\textwidth]{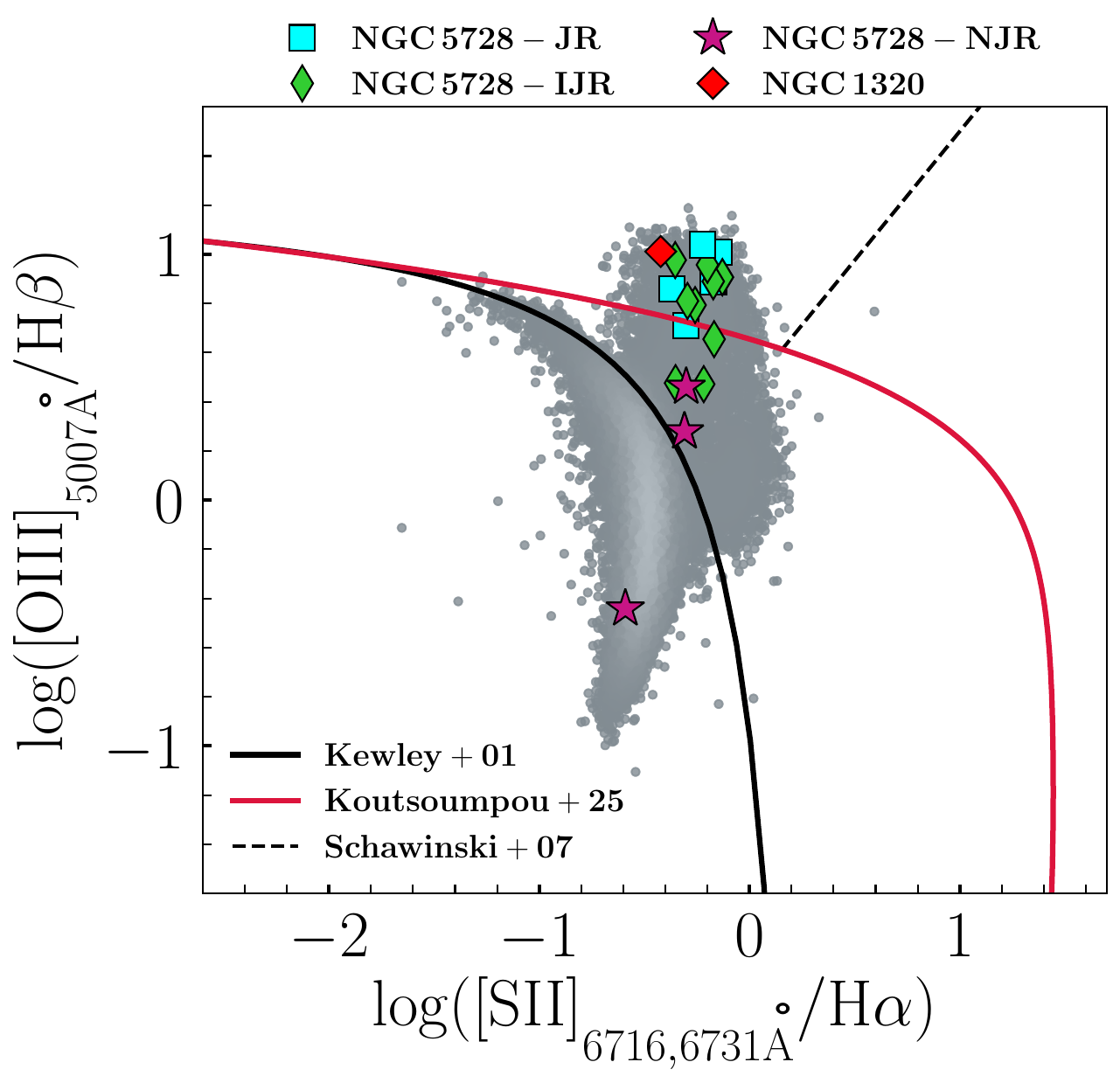}\label{subfig:sfzeta_line_S2}}~
    \subfigure[]{\includegraphics[width=0.33\textwidth]{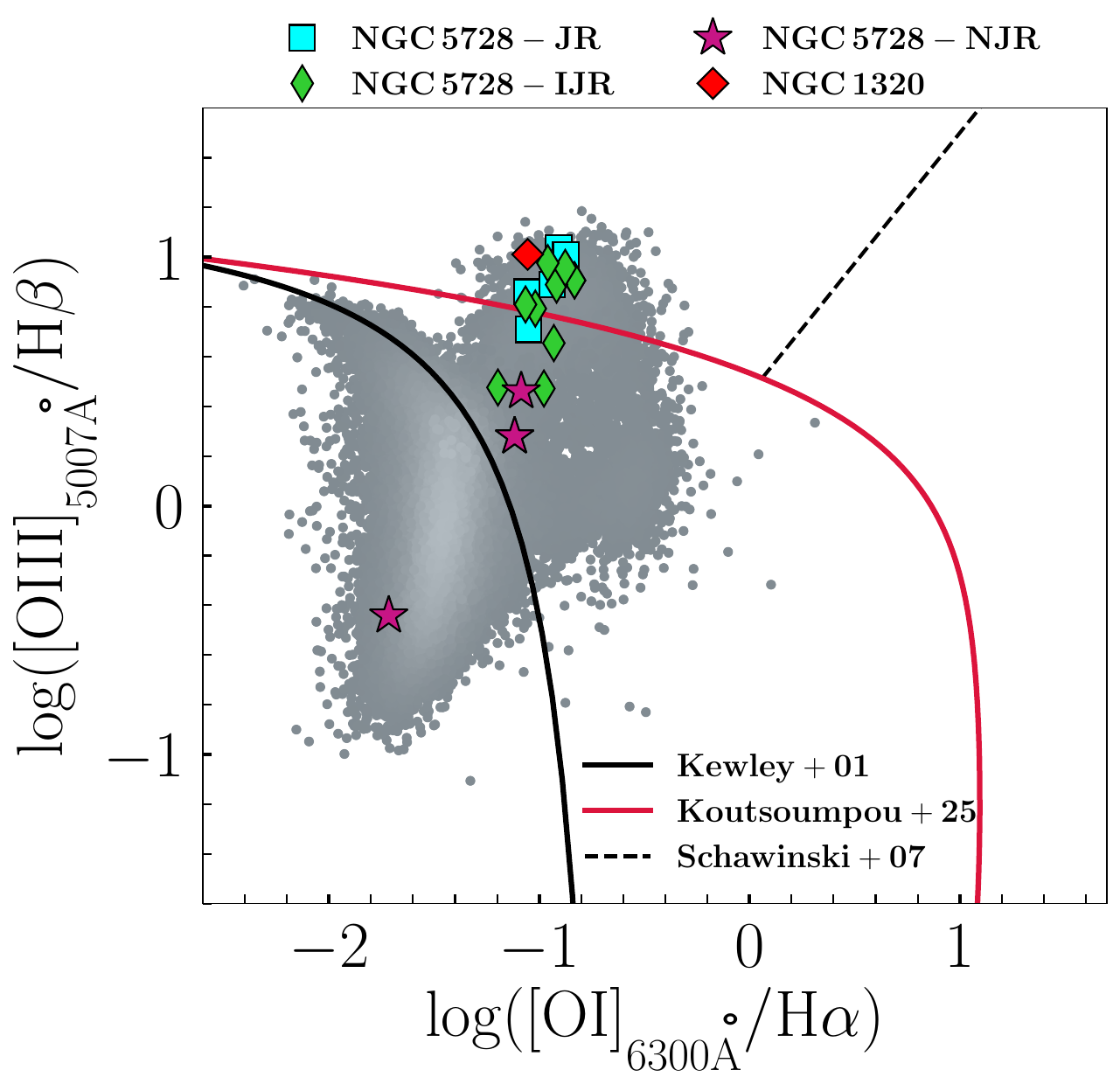}\label{subfig:sfzeta_lineL_O1}}
    \caption{{BPT diagrams depicting [\ion{N}{ii}]/H$\alpha$, [\ion{S}{ii}]/H$\alpha$, and [\ion{O}{i}]/H$\alpha$ ratios. The observations from NGC 5728, and NGC 1320 are marked with stars, and a red diamond, respectively. The cyan squares, lime green diamonds, and magenta stars represent jet-affected (JR), intermediate jet-affected (IJR), and non-jet-affected (NJR) regions, respectively, and correspond to the shaded areas in Fig.~\ref{fig:chosen_apertures}. The Kewley and Schawinski lines are indicated with solid and dashed black lines, respectively, while the Koutsoumpou (SF\texorpdfstring{$\zeta$}{zeta}) line, is depicted by the solid red line. In the background we show the line ratios measured for nearby galaxies from the Sloan Digital Sky Survey (SDSS) data release 7 \citep{Abazajian_2009}.}}
    \label{fig:sfzeta_line}
    \vspace{-0.2cm}
\end{figure*}

Region 17 of NGC 5728 is indicated as a white star in Figure \ref{fig:5728_BPTS_U}, with the white color denoting its location at the greatest distance from the nucleus and the star symbol highlighting its star-forming nature. In the BPT diagrams, this region is located well within the star-forming zone, below the Kewley line and even further below the SF$\zeta$ line, suggesting minimal influence from high CR ionization. This interpretation is further supported by its position away from the jets (magenta-shaded area in Figure \ref{fig:chosen_apertures}i). As such, region 17 is important, as it exhibits line ratios that are consistent with pure star formation and minimal CR impact and provides a baseline for contrasting with jet-impacted or CR-enhanced regions within NGC 5728. Subsequently, regions 10 and 13, marked in lilac and almost white stars in Figure \ref{fig:5728_BPTS_U}, occupy the in-between zone above the Kewley line, while still falling bellow the SF$\zeta$ boundary shown in Figure \ref{fig:sfzeta_line}. These regions originate from areas closer to the jets and are therefore possibly affected by CRs. This confirms that regions such as 10 and 13 are representative of how CR-enhanced star-formation models can explain elevated line ratios in certain parts of NGC 5728, bridging the gap between classical star-forming excitation and AGN-like signatures without requiring a dominant AGN contribution.


Finally, regions closer to the nucleus and the jets, displayed as cyan squares in Fig. ~\ref{fig:sfzeta_line}, lie above the SF$\zeta$ line, making these areas consistent with CR-driven and AGN ionization scenario. Moreover, thin lime green diamonds in Fig.~\ref{fig:sfzeta_line}, representing intermediate regions far from the jets, lie near the SF$\zeta$ line and can be reproduced either by AGN or star-formation models with CRs.
These boundaries are useful in characterizing the intermediate region that is consistent with both the AGN-dominated models and those involving star formation combined with CR ionization ($\zeta_\text{CR} \sim 10^{-13},\rm{s^{-1}}$). Our analysis of NGC 5728 underscores the value of such revised boundaries for interpreting emission-line diagnostics, particularly in spatially resolved regions influenced by CRs.

\subsection{MIR and optical–MIR hybrid line diagnostic diagrams}\label{subsec:ir_bpts}

\begin{figure*}[htbp]
    \centering
    \includegraphics[width=0.91\textwidth]{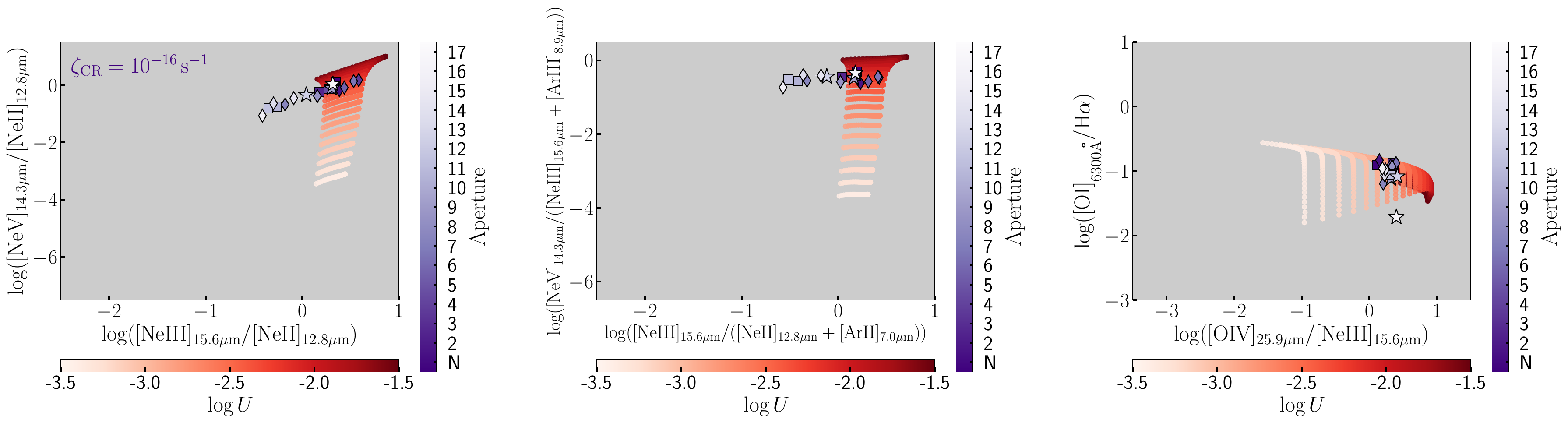}
    \includegraphics[width=0.91\textwidth]{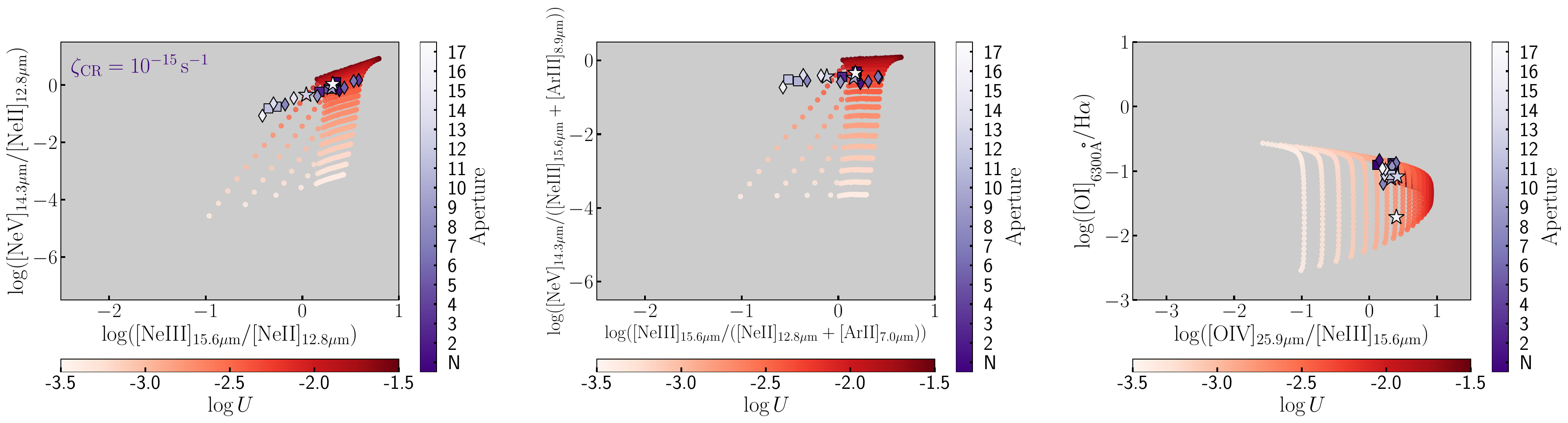}
    \includegraphics[width=0.91\textwidth]{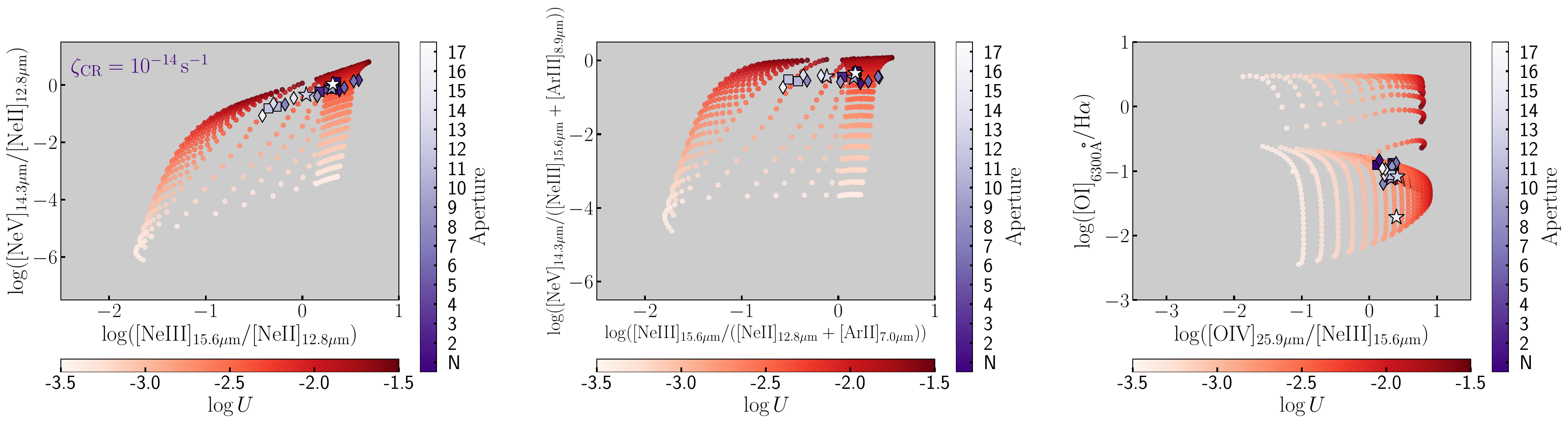}
    \includegraphics[width=0.91\textwidth]{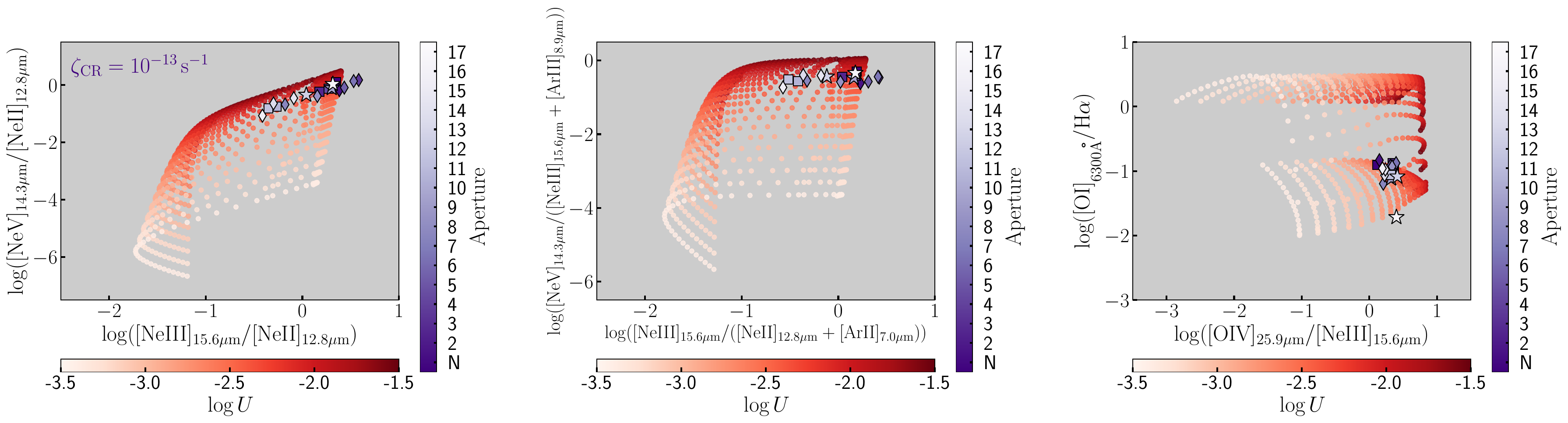}
    \includegraphics[width=0.91\textwidth]{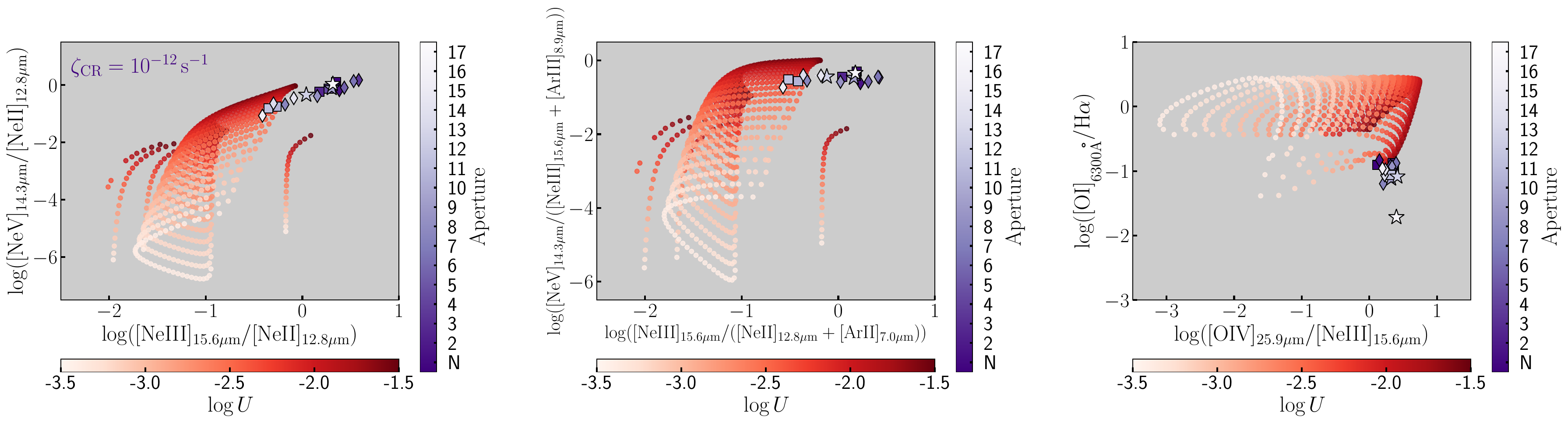}
  \caption{Diagrams with the AGN photoionization models compared with the observations from the selected apertures in NGC 5728 (Fig.\ref{fig:chosen_apertures}). 
  The different shades of purple ranging from deep purple to pale lilac and/or white, represent increasing distance from the nucleus, as also denoted by numbers, with "N" being the nuclear aperture. The different shapes-square, thin diamond, and star-represent the nucleus and/or jet impacted, intermediate and distant areas, respectively. The different shades from white to deep red represent the range of ionization parameter values, $-3.5\leq \log U\leq -1.5$. All the models have solar abundances. The panels from top to bottom correspond to  $\zeta_\mathrm{CR}=10^{-16}\,\rm s^{-1},\,10^{-15}\,\rm s^{-1},\,10^{-14}\,\rm s^{-1},\,10^{-13}\,\rm s^{-1}$, and $10^{-12}\,\rm s^{-1}$, respectively.}\label{fig:5728_new}
\end{figure*}

The first diagnostic plot we selected is [\ion{Ne}{v}]$\lambda14.3\mu$m/ [\ion{Ne}{ii}]$\lambda12.8\mu$m versus [\ion{Ne}{iii}]$\lambda15.6\mu$m/[\ion{Ne}{ii}]$\lambda12.8\mu$m, depicted in the first column of Fig. \ref{fig:5728_new}. This frequently used diagnostic tool has been employed to compare observations with various models, such as pure photoionization models \citep{Feltre_2023} or photoionization and shock models \citep{Zhang_2024}. Given the value of this plot, we reproduced it to verify that our models sufficiently described the parameter space. Our analysis reveals that as $\zeta_{\rm CR}$ increases from $10^{-16}\rm s^{-1}$ in the top row to $10^{-12}\rm s^{-1}$ in the bottom row, the observational data are best represented by models with $\zeta_{\rm CR}$ values  of $\sim 10^{-14}\rm s^{-1}$. These model cases spatially coincide with those presented in figures such as fig. 5 in \cite{Feltre_2023} and fig. 15 in \cite{Zhang_2024}, rendering our models good for reproducing the same parameter space. 
Subsequently, the observations in this plot are effectively captured by models with $\zeta_{\rm CR} \sim 10^{-14}-10^{-13}\rm s^{-1}$. Some regions more directly (square) or indirectly (thin diamonds) influenced by the jets, are adequately modeled at $\zeta_{\rm CR} \sim 10^{-13}\rm s^{-1}$ or even $10^{-12}\rm s^{-1}$.

The diagnostic plot of [\ion{Ne}{iii}]$\lambda15.6\mu$m/ /([\ion{Ne}{ii}]$\lambda12.8\mu$m+[\ion{Ar}{ii}]$\lambda7.0\mu$m) against [\ion{Ne}{v}]$\lambda14.3\mu$m/
([\ion{Ne}{iii}]$\lambda15.6\mu$m+[\ion{Ar}{iii}]$\lambda8.9\mu$m) displayed in the second column of Fig. \ref{fig:5728_new}, was chosen to highlight the substantial shift in the positions of the models within the parameter space as the CR ionization rate increases from $10^{-16}\rm s^{-1}$ in the top row to $10^{-12}\rm s^{-1}$ in the bottom row. This pronounced change makes the selected ratios particularly effective for observing the maximal influence of CRs. 
The shift of the models toward the lower left corner of the plot leads us to conclude that the low ionization lines, namely [\ion{Ne}{ii}], [\ion{Ar}{ii}], [\ion{Ne}{iii}], and [\ion{Ar}{iii}], are significantly impacted by CRs. \eliza{For this diagnostic, we note that models deviating from the folding trend (rightmost, bottom row in Fig. \ref{fig:5728_new}) correspond to the highest CR ionization rate ($10^{-12}\mathrm{s}^{-1}$) and the lowest densities ($\log n_\mathrm{H} \leq 0.1$). At such low densities, strong CR ionization can suppress cooling and low-ionization lines (e.g., [\ion{Ar}{II}], [\ion{Ne}{II}]), disrupting the trend observed at higher densities in this diagram.}


Combining optical and MIR emission-line ratios could represent a promising approach for disentangling the contributions of AGN, star formation, CRs, and shocks in composite galaxies.
Optical diagnostics such as [\ion{O}{i}]/H$\alpha$ are sensitive to the hardness of the ionizing radiation and can trace shocks, while MIR ratios such as [\ion{O}{iv}]/[\ion{Ne}{iii}] probe deeply embedded, high-ionization regions inaccessible in the optical due to dust. As shown by \cite{Feltre_2023}, diagnostic diagrams that combine these regimes---notably [\ion{O}{i}]/H$\alpha$ versus [\ion{O}{iv}]/[\ion{Ne}{iii}]---offer enhanced diagnostic power, allowing degeneracies present in single-wavelength analyses to be resolved. In the third column of Fig.  \ref{fig:5728_new}, we examine the impact of the CR ionization rate on this combination of line ratios. We find that increased CR ionization rates enhance [\ion{Ne}{iii}] and [\ion{O}{i}] emission, resulting in trends that differ from those produced by shocks (see fig. 7 of \cite{Feltre_2023}). While both shocks and CRs can elevate [\ion{O}{i}]/H$\alpha$, CR-driven models yield a broader distribution that better matches the observed parameter space (see Fig. \ref{fig:5728_new}). In contrast, AGN combined with star-formation models tend to underpredict, and AGN combined with shock models mostly overpredict, the observed ratios, providing less overlap with the data (fig. 7 of \citealt{Feltre_2023}). Nonetheless, this approach has limitations, as optical and MIR lines may trace different regions within the ionized clouds, sampling different temperatures and densities. Additionally, uncertainties in extinction corrections can affect the reliability of a combined diagnostic.
\eliza{The [\ion{O}{I}]/H$\alpha$ vs [\ion{O}{IV}]/[\ion{Ne}{III}] diagnostic is promising for distinguishing CR from shock excitation although it is strongly metallicity-dependent, as [\ion{O}{I}]/H$\alpha$ directly traces O/H. Since both our models and those of \cite{Feltre_2023} use fixed metallicity, we likely underestimate the true spread in this ratio and may overlook CR-shock degeneracies. Exploring additional } 
mixed optical-MIR diagnostics \eliza{in the future will provide} powerful tools for investigating the interplay between star formation, AGN activity, shocks, and CRs in complex galactic environments where these mechanisms coexist.

Additionally, the majority of observed emission-line ratios are accurately represented by models at $\zeta_{\rm CR} \sim 10^{-14}\rm s^{-1}$. Notably, star-forming regions, depicted with white and/or lilac stars based on their proximity to the nucleus, are well modeled without the influence of CRs, a behavior that is also present in the position of these stars in the BPT diagrams (Fig. \ref{fig:5728_BPTS_U}). However, data represented by squares and thin diamonds, which are either directly or indirectly affected by jets, are accurately described by even the highest CR ionization models, as illustrated in the second column and last row of Fig. \ref{fig:5728_new}. Finally, in Appendix~\ref{appendix_gal}, we present additional MIR emission lines, including mid-ionization species such as [\ion{S}{iv}]$\lambda10.5\,\mu$m and [\ion{S}{iii}]$\lambda18.7\,\mu$m, as well as high-ionization lines such as [\ion{Ar}{v}]$\lambda7.9\,\mu$m, [\ion{Ar}{v}]$\lambda13.1\,\mu$m, [\ion{Ne}{v}]$\lambda14.3\,\mu$m, [\ion{Ne}{v}]$\lambda24.3\,\mu$m, [\ion{Ne}{vi}]$\lambda7.7\,\mu$m, and [\ion{O}{iv}]$\lambda25.9\,\mu$m. The diagnostic plots presented across Fig. \ref{fig:5728_new} and Fig. \ref{fig:5728_APPENDIX} effectively lay out the varying impacts of CRs on emission lines with different ionization potentials. We confirm that while high-ionization lines are mainly influenced by photoionization and are thus not good indicators of CR presence, low-ionization lines show significant sensitivity to CRs. We further study the link between CRs and MIR emission by examining the temperature of nebular gas and the emissivity of different lines within the depth of an emission region in Section \ref{subsec:structure_plots}.


\subsection{Gas stratification diagrams} \label{subsec:structure_plots}

In our study in KFDS25, we found that photoionization governs the illuminated face of the cloud (i.e. the nebular layers are defined with respect to the ionizing source), whereas CRs play a more critical role in the shielded regions where photoionization is minimal. Specifically, CR rates $\sim 10^{-13} \,\rm{s^{-1}}$ significantly impact the thermal dynamics of the ionized gas, leading to the formation of a secondary, low-ionization layer of warm ionized gas ($\sim 8000\rm K$) well beyond the region dominated by photoionization. This inner layer contributes to the emission of low-ionization lines such as [\ion{N}{ii}]$\lambda$6584\AA, [\ion{S}{ii}]$\lambda\lambda$6716,6731\AA, and [\ion{O}{i}]$\lambda$6300\AA. In contrast, the emissivities of higher ionization lines such as [\ion{O}{iii}]$\lambda$5007\AA, H$\alpha$, and H$\beta$, are only slightly increased (see figs. 8, 9, and 10 in KFDS25). This analysis was replicated for different emission lines at MIR wavelengths to determine if low-ionization MIR emission lines were similarly affected as presented in Section \ref{subsec:ir_bpts}.

\begin{figure*}[!htp]
    \centering
   \subfigure[]{\includegraphics[width=0.33\textwidth]{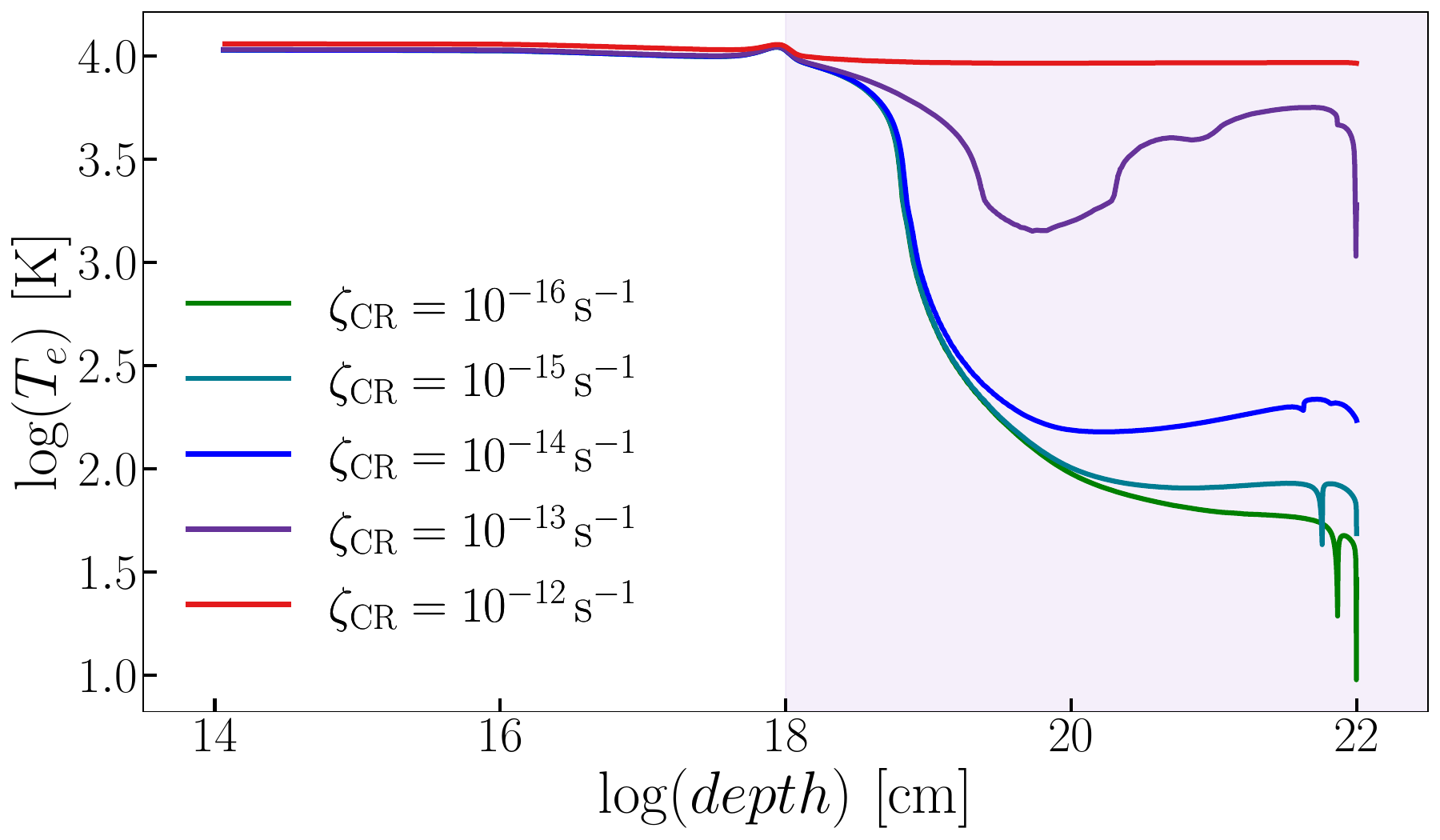}}\label{subfig:temp_nh2_agn}~    
   \subfigure[]{\includegraphics[width=0.33\textwidth]{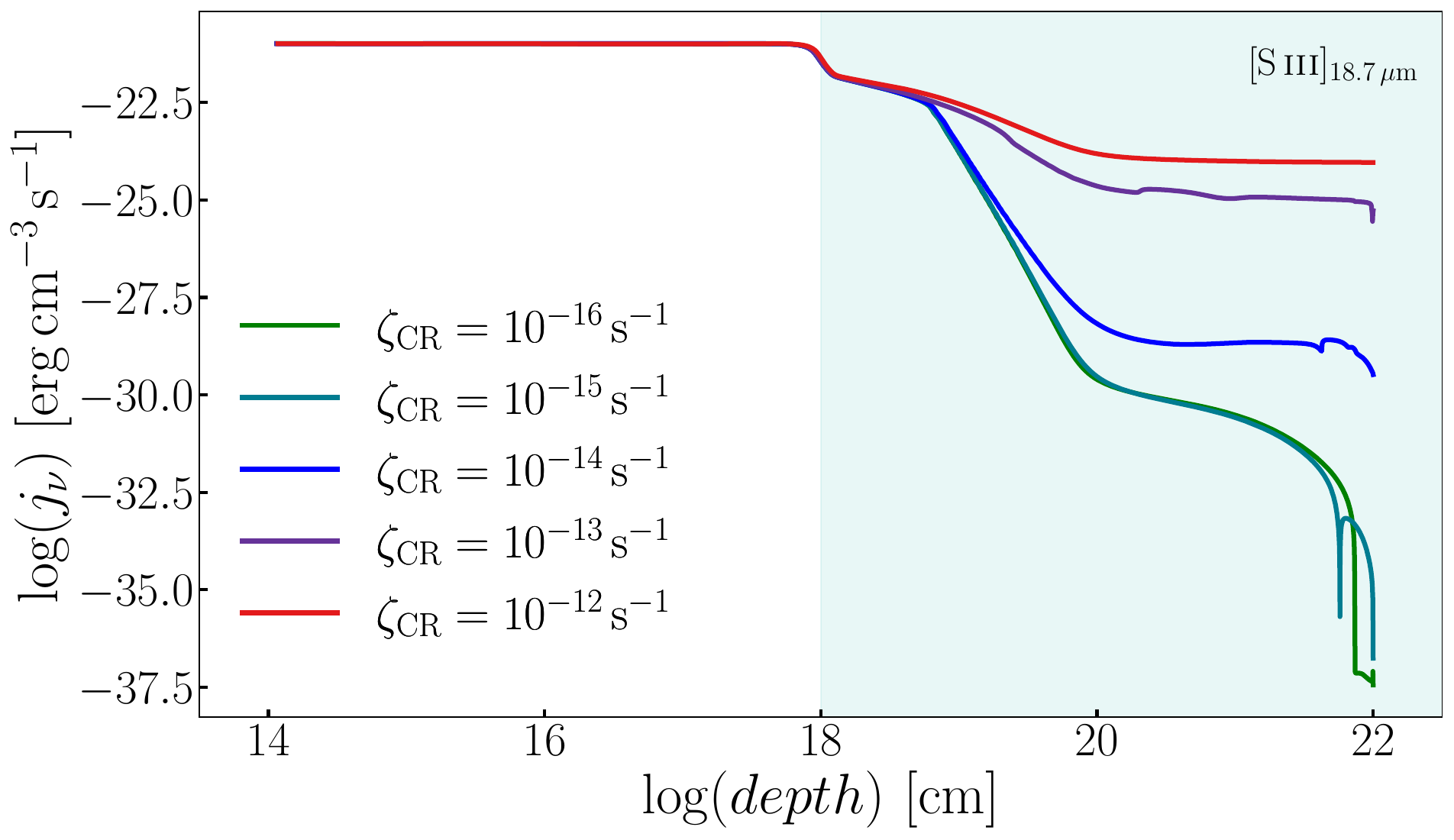}}\label{subfig:s3_nh2_agn}~
    \subfigure[]{\includegraphics[width=0.33\textwidth]{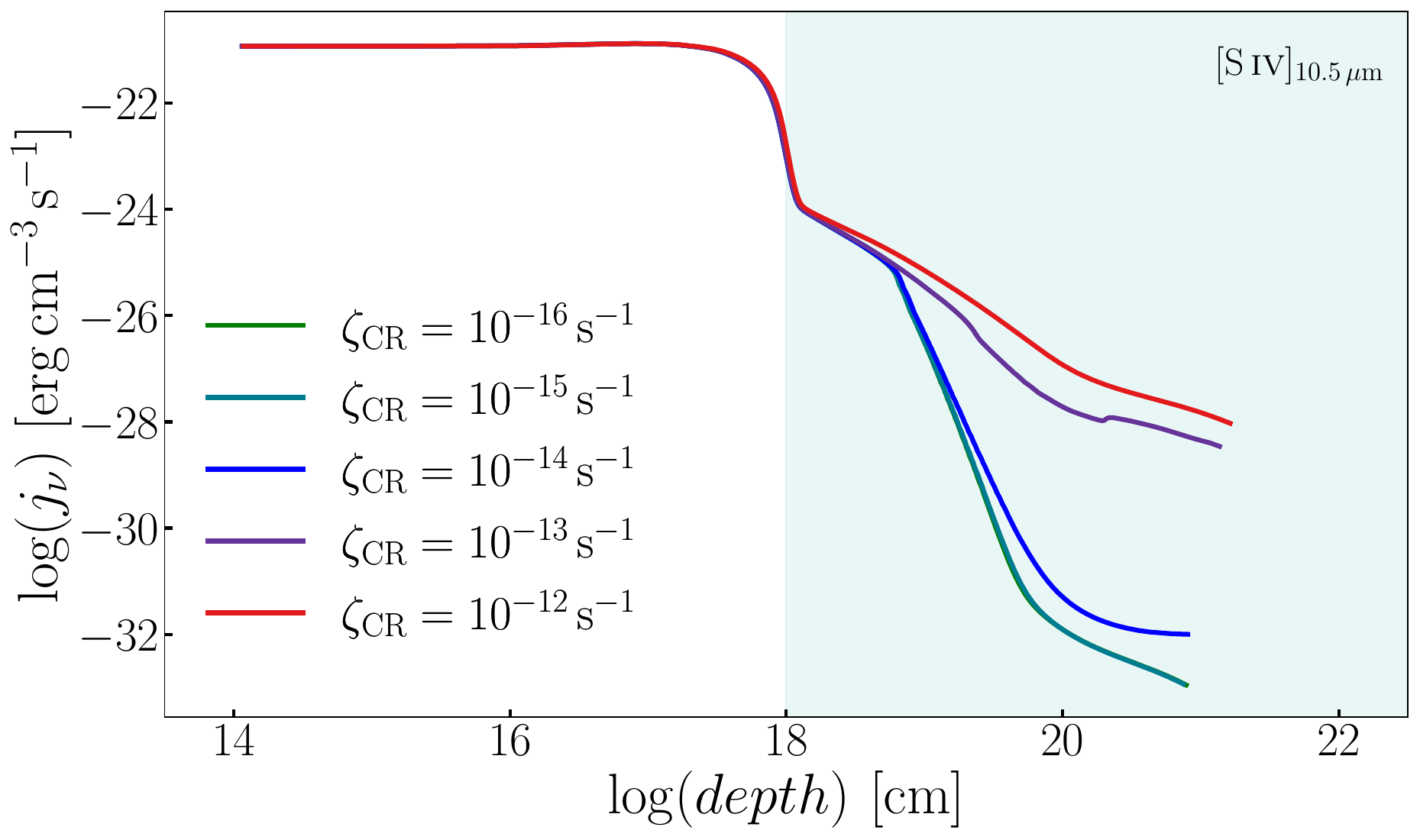}}\label{subfig:s4_nh2_agn}
     \subfigure[]{\includegraphics[width=0.33\textwidth]{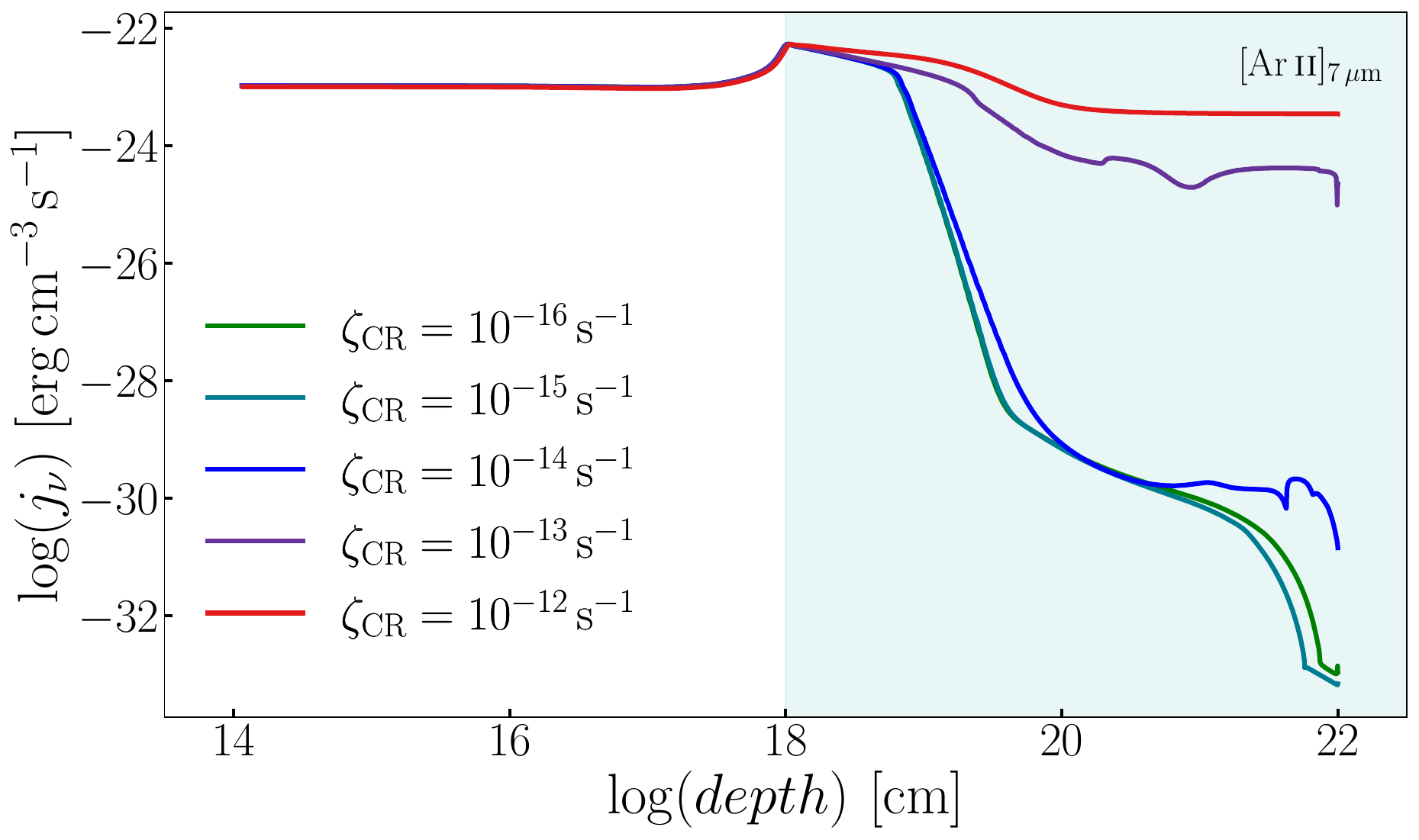}}\label{subfig:ar2_nh2_agn}~
    \subfigure[]{\includegraphics[width=0.33\textwidth]{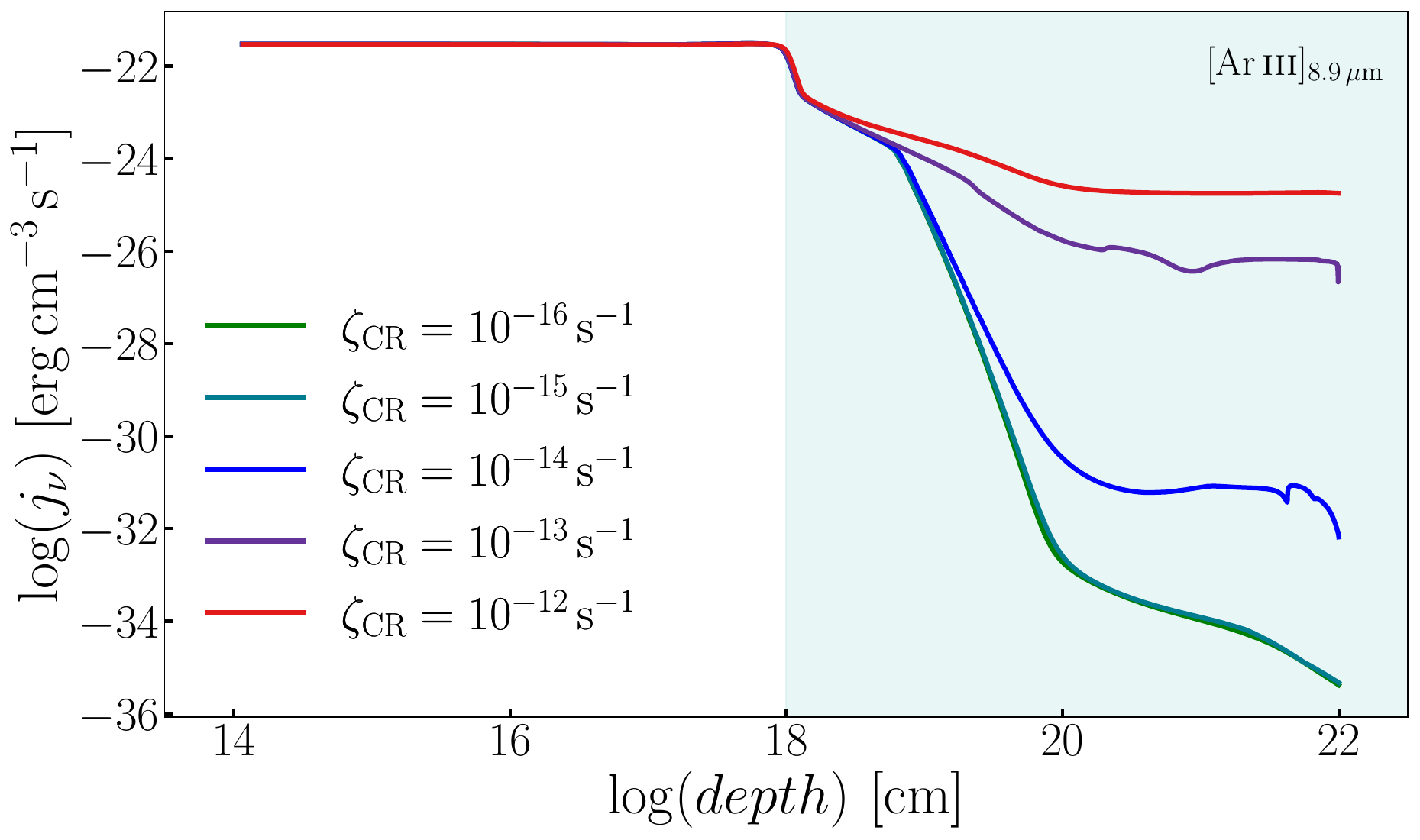}}\label{subfig:ar3_nh2_agn}~
    \subfigure[]{\includegraphics[width=0.33\textwidth]{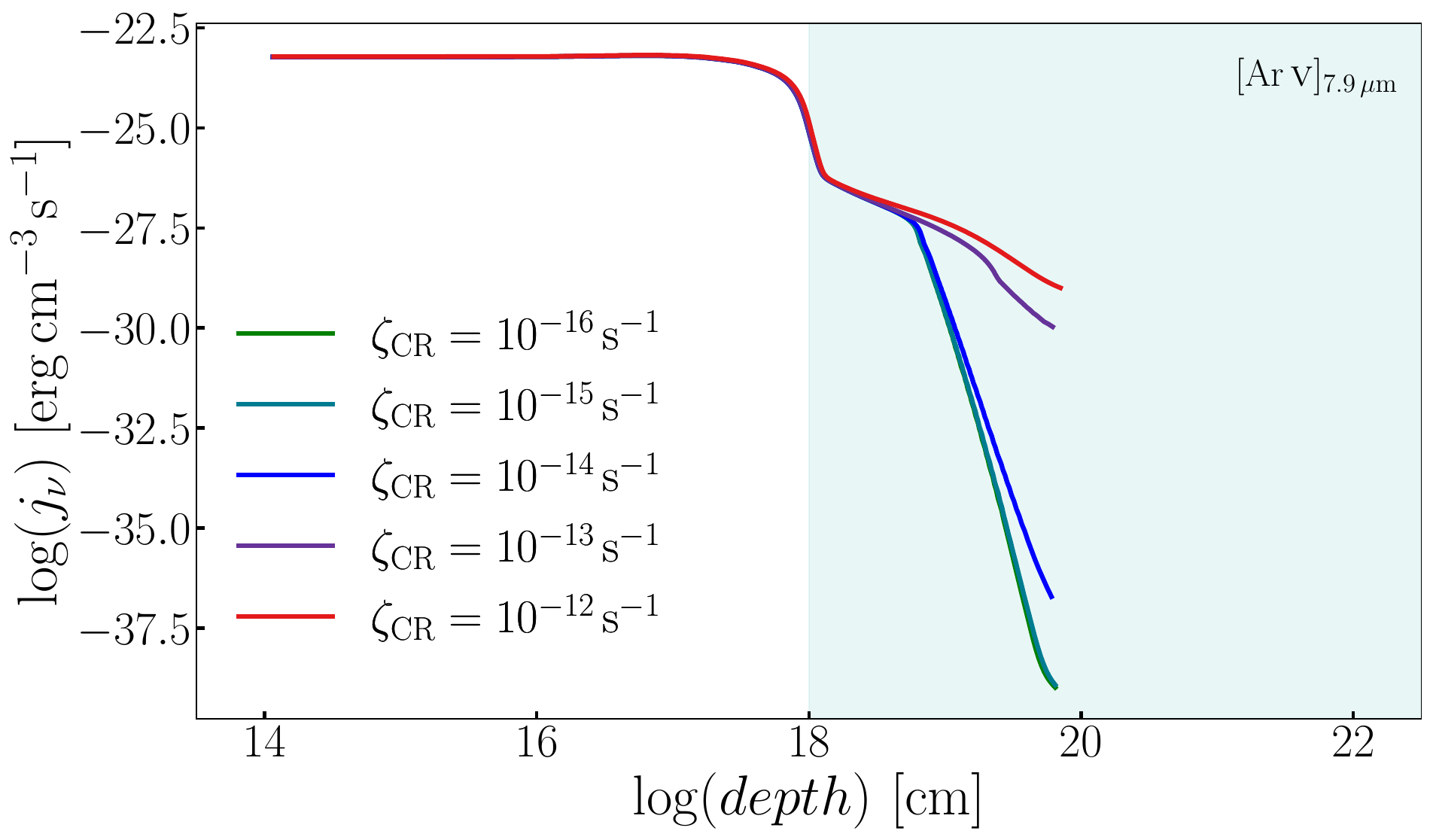}}\label{subfig:ar5_nh2_agn}
    \subfigure[]{\includegraphics[width=0.33\textwidth]{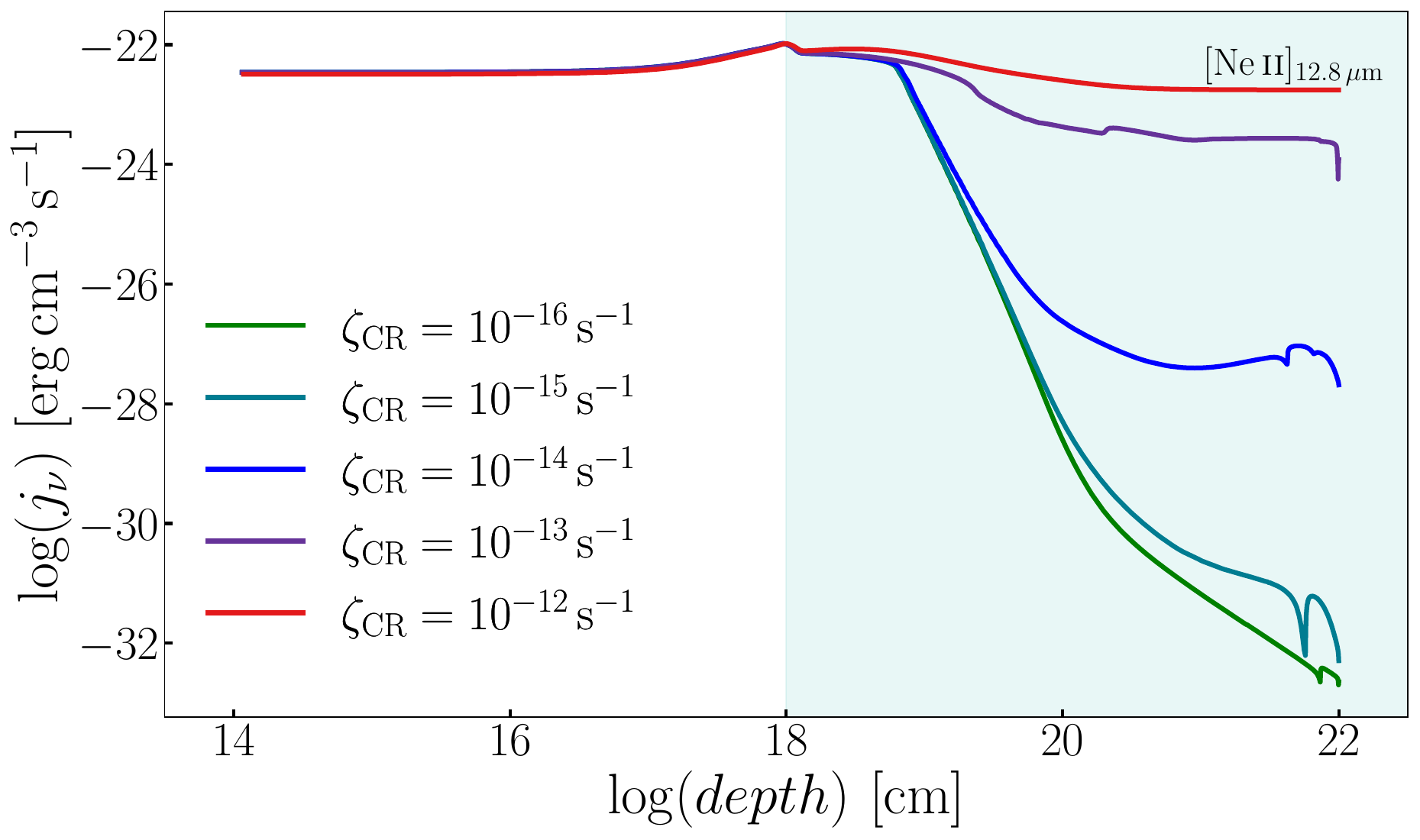}}\label{subfig:ne2_nh2_agn}~
    \subfigure[]{\includegraphics[width=0.33\textwidth]{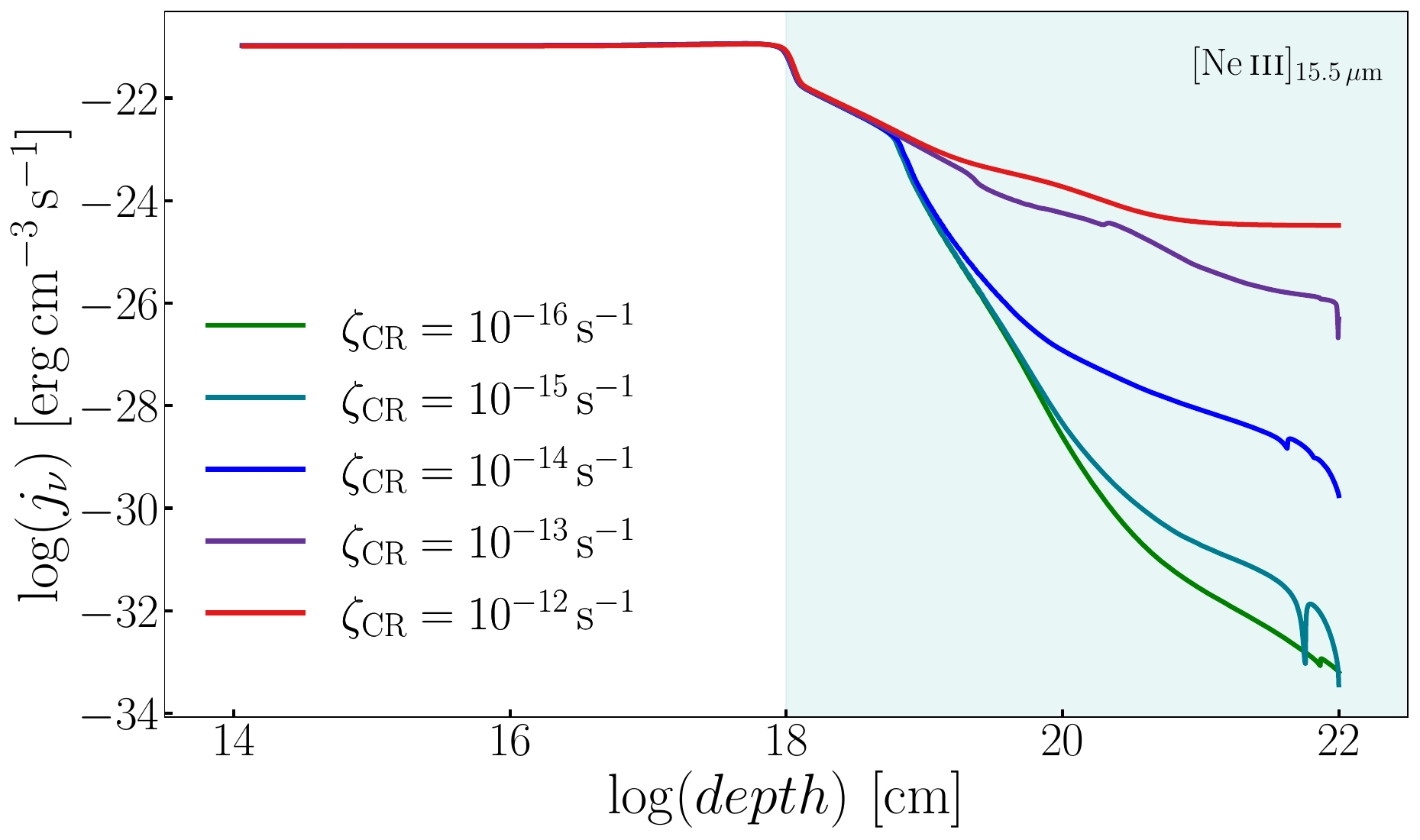}}\label{subfig:ne3_nh2_agn}~
    \subfigure[]{\includegraphics[width=0.33\textwidth]{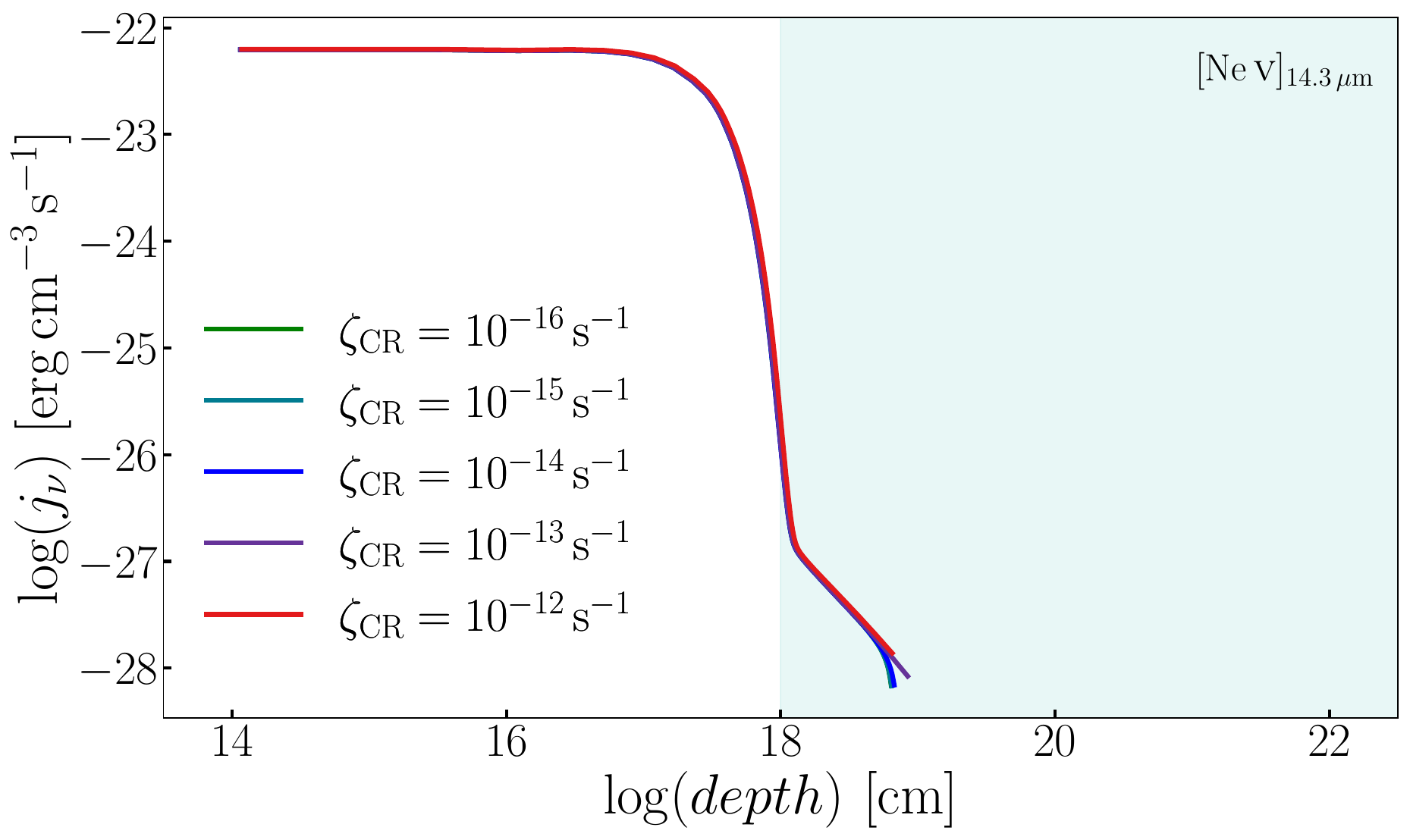}}\label{subfig:ne5_nh2_agn}
     \caption{Temperature and line emissivity vs. depth in the simulated cloud for AGN models, for an initial density of $n_{\rm H}=100\,\rm{cm^{-3}}$, and for $\zeta_\mathrm{CR}=10^{-16}\,\rm s^{-1},10^{-15}\,\rm s^{-1},\,10^{-14}\,\rm s^{-1},\,10^{-13}\,\rm s^{-1}$, and $10^{-12}\,\rm s^{-1}$, and $\log U=-3.0$. The different panels a-f correspond to kinetic temperature and the emissivity of $[\ion{S}{III}]_{\rm18.7\mu m},\, [\ion{S}{IV}]_{\rm10.5\mu m},\,[\ion{Ar}{II}]_{\rm7\mu m},\,[\ion{Ar}{III}]_{\rm8.9\mu m},\,[\ion{Ar}{V}]_{\rm7.9\mu m},\,[\ion{Ne}{II}]_{\rm12.8\mu m},\,[\ion{Ne}{III}]_{\rm15.5\mu m}$, and $[\ion{Ne}{V}]_{\rm14.3\mu m}$,
     respectively. The shaded area 
    indicates the approximate region in which CR heating becomes dominant. The CR-dominated area is depicted in lilac in the temperature plot and in teal in the emissivity plots. }\label{fig:mir_lines_struc}
\end{figure*}

As illustrated in Fig. \ref{fig:mir_lines_struc}, the interaction of CRs with gas significantly influences the MIR emission lines. For an initial hydrogen density of $n_{\rm H}=100\,\rm{cm^{-3}}$ and an ionization parameter $\log U=-3.0$, we observe varied sensitivities across different MIR emission lines under increasing CR ionization rates from $10^{-16}\,\rm{s^{-1}}$ to $10^{-12}\,\rm{s^{-1}}$. We find that while photoionization influences the electron temperature at the cloud's illuminated face, high CR rates $(\gtrsim 10^{-14}\,\rm{s^{-1}})$ dictate the excitation in deeper layers. Specifically, in the second and third row of the first column of Fig. \ref{fig:mir_lines_struc}, $[\ion{Ar}{II}]{\lambda\rm7\mu m}$ and $[\ion{Ne}{II}]{\lambda\rm12.8\mu m}$, with ionization potentials of 15.76 eV and 21.56 eV respectively, exhibit the most pronounced changes in emissivity due to higher $\zeta_{\rm CR}$, highlighting their strong sensitivity to CRs. The subsequent panels of the second column within Fig. \ref{fig:mir_lines_struc} reveal that  $[\ion{S}{III}]{\lambda\rm18.7\mu m}$, $[\ion{Ar}{III}]{\lambda\rm8.9\mu m}$, and $[\ion{Ne}{III}]{\lambda\rm15.5\mu m}$, having ionization potentials of 23.33eV, 27.63 eV, and 40.96 eV respectively, also experience considerable emissivity enhancement, though to a lesser degree than $[\ion{Ar}{II}]$ and $[\ion{Ne}{II}]$. Conversely, displaying a gradient of decreasing responsiveness to CRs, $[\ion{S}{IV}]{\lambda\rm10.5\mu m}$ with an ionization potential of 34.79 eV, shows some effects; while $[\ion{Ar}{V}]$, with an ionization potential of 59.81 eV, exhibits only minimal changes; and $[\ion{Ne}{V}]$, with an ionization potential of 97.12 eV, shows no sensitivity to CRs. These results indicate that the highest-ionization emission lines are largely unaffected by CR-driven ionization within the cloud.

This differentiation in sensitivity to CR rates among different emission lines and different ionization states of the same line is deeply rooted in the ionization energy required for each particular emission. The behavior of low-ionization lines in the MIR spectrum exhibits similar trends found in the low-ionization lines in the optical spectrum, as documented in our findings for KFDS25. Specifically, $[\ion{N}{ii}]{\rm\lambda6584\AA}$, $[\ion{S}{ii}]{\rm\lambda\lambda6716,6731\AA}$, and $[\ion{O}{i}]{\rm\lambda6300\AA}$ in the optical spectrum, with ionization potentials of 14.53 eV, 10.36 eV, and 13.62 eV, respectively, exhibit increased emissivity under higher CR ionization rates, paralleling the effects seen in MIR lines such as $[\ion{Ne}{II}]{\rm\lambda12.8\mu m}$ and $[\ion{S}{III}]{\rm\lambda18.7\mu m}$. This finding reinforces the interconnected nature of optical and MIR spectral behavior, demonstrating a unified response to CR ionization, which could be exploited in astrophysical environments not optically thin in the optical, such as dust-obscured nuclei of galaxies, star-forming regions, dense molecular clouds, or circumstellar disks.




\section{Discussion}\label{discussion}
\subsection{CR effect on MIR low ionization lines}\label{CR_MIR}


CRs significantly influence the ionization and the heating of the ISM in environments such as AGN and starburst galaxies \citep{McKee_1989,Padovani_2009,Padovani_2018,Gabici_2022}. 
Our analysis in KFDS25, demonstrates that the impact of CRs on low-ionization optical lines such as [\ion{N}{ii}], [\ion{S}{ii}], and [\ion{O}{i}], is more evident than on higher-ionization lines such as H$\alpha$, H$\beta$, and [\ion{O}{iii}], due to a secondary low-ionization CR-induced layer beyond photoionization-dominated regions. We analyze that the effect of CRs on optical emission lines introduces new limits in the classical line-ratio BPT diagnostic diagrams and, as showcased in section 5.4 of KFDS25, can also affect metallicity and ionization parameter estimates.



\begin{figure*}[!htp]
    \centering
    \includegraphics[width=\textwidth]{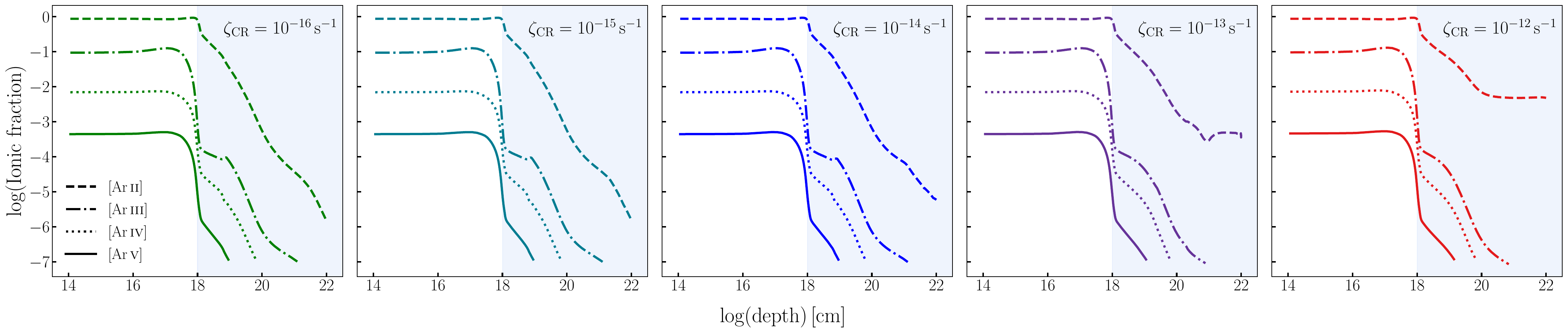}\label{subfig:ionic_Ar}
   \includegraphics[width=\textwidth]{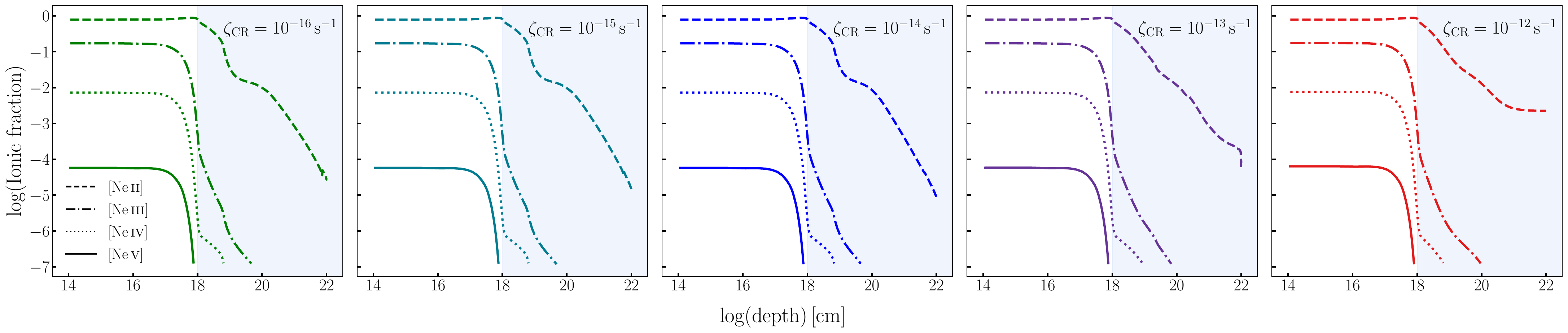}\label{subfig:ionic_Ne}
     \caption{Ionic fraction vs. depth in the simulated cloud for AGN models, for an initial density of $n_{\rm H}=100\,\rm{cm^{-3}}$, and for $\zeta_\mathrm{CR}=10^{-16}\,\rm s^{-1},10^{-15}\,\rm s^{-1},\,10^{-14}\,\rm s^{-1},\,10^{-13}\,\rm s^{-1}$, and $10^{-12}\,\rm s^{-1}$, from left to right. Top row: Ar ionic fractions. Bottom row: Ne ionic fractions. The blue-shaded area 
    indicates the approximate region where CR heating becomes dominant.}\label{fig:ionic_struc}
\end{figure*}

Based on Section \ref{subsec:ir_bpts}, the MIR emission lines most sensitive to CR ionization are [\ion{Ar}{ii}] and [\ion{Ne}{ii}]. This conclusion is supported by both their relatively low ionization potentials, 15.76eV, and 21.56eV, respectively, and their emissivity profiles in Fig.~\ref{fig:mir_lines_struc}, which exhibit a pronounced enhancement with increasing $\zeta_{\rm CR}$. In contrast, lines from higher-ionization species such as [\ion{Ne}{iii}], [\ion{Ar}{iii}], [\ion{S}{iv}], and even more in [\ion{Ar}{v}], [\ion{Ne}{v}] display minimal changes, remaining largely governed by photoionization (see Fig. \ref{fig:mir_lines_struc}). 
This differential response arises from the physical stratification within the cloud. In the \eliza{layers closer to the ionizing source}, where photoionization from the AGN dominates, the intense radiation field enhances the emission of high-ionization lines such as [\ion{Ar}{v}] and [\ion{Ne}{v}], while the ionic fractions of low-ionization species \eliza{such as [\ion{Ar}{ii}] and [\ion{Ne}{ii}]} remain \eliza{low due to their efficient ionization to higher states}.
Deeper into the cloud, however, the radiation field becomes increasingly attenuated, allowing high-ionization ions to recombine and \eliza{the ionic fractions of low-ionization species, like [\ion{Ar}{ii}] and [\ion{Ne}{ii}], to increase}. In these shielded regions, CRs begin to dominate the ionization balance, sustaining a partially ionized phase and efficiently boosting emission from these low-ionization lines. In Fig. \ref{fig:ionic_struc}, Ar and Ne ionic fractions versus depth, clearly show that [\ion{Ar}{ii}] and [\ion{Ne}{ii}] increase in abundance in regions where CR heating is dominant and the higher ionization stages are no longer sustained. Hence, the combined ratio [\ion{Ar}{ii}] + [\ion{Ne}{ii}] offers a powerful diagnostic, effectively isolating the CR-influenced component of the ionized gas and distinguishing it from photoionized regions. This makes it a particularly suitable tool for probing environments where CRs are expected to be present.


Additionally, we tested the effect of dust by running extra \textsc{Cloudy} simulations using standard prescriptions that include either ISM grains or polycyclic aromatic hydrocarbons (PAHs) (\texttt{grains ISM} and \texttt{grains PAH}, respectively). These dusty models produced MIR line ratios nearly identical to dust-free models. Low-ionization lines ([\ion{Ne}{ii}]$\lambda12.8\,\mu$m, [\ion{Ar}{ii}]$\lambda7.0\,\mu$m, [\ion{S}{iii}]$\lambda18.7\,\mu$m) showed comparable emissivities overall, but their emission decreased faster with depth due to more efficient cooling when dust is present. Additionally, dust slightly increase the electron temperatures near the illuminated face of the cloud, where hydrogen recombination dominates, particularly at lower $\zeta_{\rm CR}$. In contrast, the impact of dust in highly-excited lines ([\ion{Ne}{v}]$\lambda14.3\,\mu$m, [\ion{Ar}{v}]$\lambda7.9\,\mu$m) is negligible, since these lines originate predominantly in regions close to the ionizing source where AGN photoionization strongly dominates.

Following the discussion in KFDS25, we estimated the expected CR ionization rate for the central region of NGC~5728 using the measured radio continuum fluxes at 1.4\,GHz and 5\,GHz \cite{Singh_2013} and assuming equipartition between the magnetic field and relativistic electron energy densities. For a spectral index of $\alpha \sim 1.09$, and a $\sim 20''$ beam at 5\,GHz, corresponding to a size of $\sim 1.85\,\rm{kpc}$, we obtain an equipartition magnetic field of $\sim 4\,\rm{\mu G}$.
Adopting $p \sim 3.2$ for the electron energy distribution spectrum, derived from the synchrotron slope via $p=2\alpha+1$ \citep{Rybicki_Lightman}, and following the methodology described in \cite{Gabici_2022} and KFDS25, we obtain a CR ionization rate of $\sim 10^{-11}\,\rm{s^{-1}}$ for the nuclear region of NGC~5728. A lower value of $p = 2.4$, corresponding to the standard slope in optically thin synchrotron sources, would result in $\sim 10^{-14}\,\rm{s^{-1}}$. These values are consistent with the CR rates inferred from our optical and MIR models, which reproduce the observed ratios in those regions of NGC~5728 most affected by the jet. Deeper radio observations, at higher angular resolution and particularly targeting optically thin emission, would improve CR ionization rate estimates.



\subsection{Diagnostic power of MIR versus optical}\label{MIR_vs_opt}


MIR emission lines offer a significant observational advantage over their optical counterparts due to their much lower sensitivity to dust extinction and temperature variations. As shown in table 3 of \citet{Wang_2019}, the relative extinction decreases from $A_\lambda/A_V = 1.0$ at 0.5525$\rm\mu$m (Johnson \textit{V}) to $\sim 0.025$ at 8$\rm \mu$m (\textit{Spitzer} [8.0]). Thus, an optical extinction of $A_V = 10$~mag results in $\sim 0.25$~mag of extinction at 8$\rm \mu$m, representing a reduction by a factor of 40. Conversely, producing just $0.1$~mag of extinction in the MIR requires an optical extinction of about $A_V \sim 4$~mag. This steep attenuation gradient highlights the power of MIR diagnostics for investigating heavily obscured environments, where optical line emission is significantly diminished.

In addition to differences in dust extinction, another key distinction between optical and MIR diagnostics lies in their sensitivity to the electron temperature, $T_{\rm e}$. As shaded in pale pink in the first panel of Fig.~\ref{fig:mir_lines_struc}, our models exhibit a pronounced drop in $T_{\rm e}$ at a depth of approximately $10^{18}\, \rm{cm}$ into the cloud. This temperature decline coincides with the onset of recombination, where the decreasing number of free electrons reduces the overall thermal energy, leading to substantial cooling of the gas. Optical lines, such as [\ion{S}{ii}] and [\ion{O}{i}], are particularly sensitive to $T_{\rm e}$, as their emissivities rely heavily on collisional excitation, as argued in KFDS25. Consequently, this sharp drop in $T_{\rm e}$ leads to a corresponding decline in their emissivity at greater depths. This behavior is evident in the predictions of our AGN models with $\zeta_{\rm CR} \gtrsim 10^{-13}\,\rm{s^{-1}}$ for $\log U = -3$, where the optical line intensities diminish steeply beyond the photoionized surface layers (see fig. 9, and 10 in KFDS25). In contrast, the MIR lines such as [\ion{Ar}{ii}]$\lambda$7.0$\mu$m and [\ion{Ne}{ii}]$\lambda$12.8$\mu$m (see Fig. \ref{fig:mir_lines_struc}) exhibit a much flatter emissivity profile across the same depth range. This difference arises because IR fine-structure transitions have much lower excitation energies and are less sensitive to the thermal electron population (see also fig. 1 in \citealt{Fernandez2021}). 
Therefore, even as the gas cools and recombines, MIR lines can continue to be emitted efficiently in the deeper, CR-affected layers, making them reliable probes of low-ionization gas in regions where optical diagnostics are suppressed. 

Moreover, recent studies of large, diffuse ionized nebulae with uncertain origin \citep{Lumb_Calle_2024}, highlight the importance of resolving ionization mechanisms in faint emission-line structures. These findings support the need for diagnostics capable of identifying the excitation mechanisms in spatially extended regions. In this context, hybrid optical–MIR diagnostics offer valuable potential for resolving such degeneracies. In particular, the [\ion{O}{iv}]/\eliza{[\ion{Ne}{iii}]} versus [\ion{O}{i}]/H$\alpha$ diagram emerges as a key tool.

\subsection{CR excitation in AGN MIR diagnostics}\label{llagn}


\begin{figure*}[!htp]
    \centering
  \centering
    \subfigure[]{\includegraphics[width=0.33\textwidth]{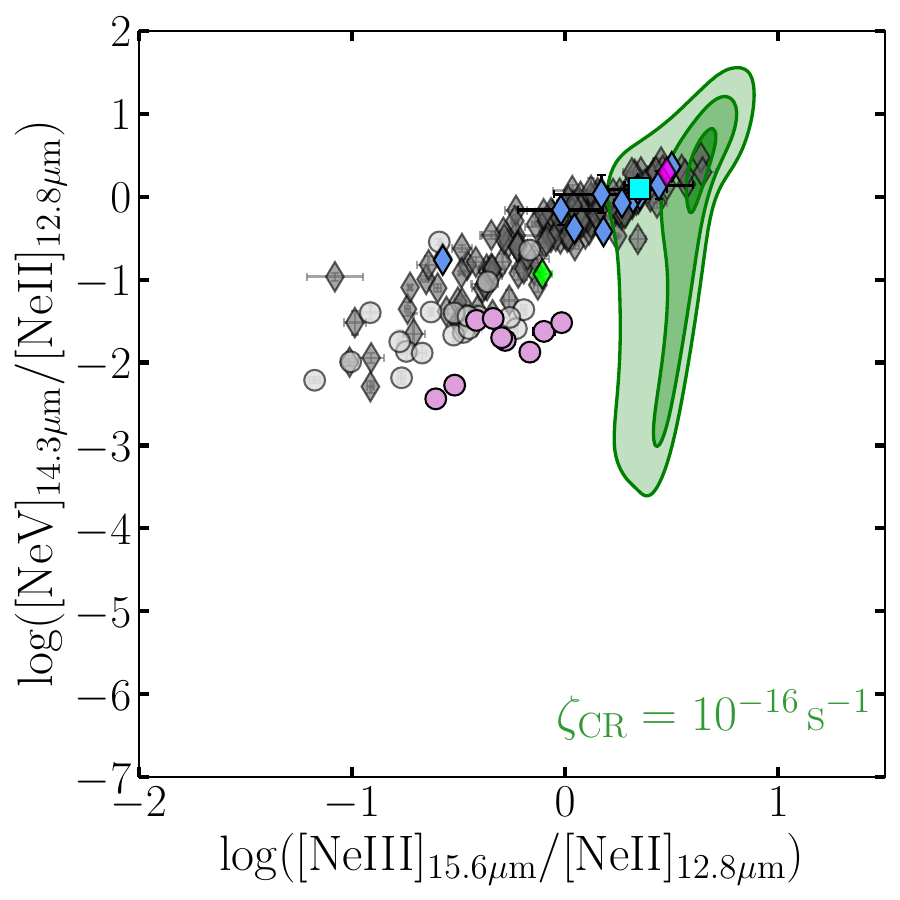}\label{subfig:kde_zeta0__16}}~
    ~\subfigure[]{\includegraphics[width=0.33\textwidth]{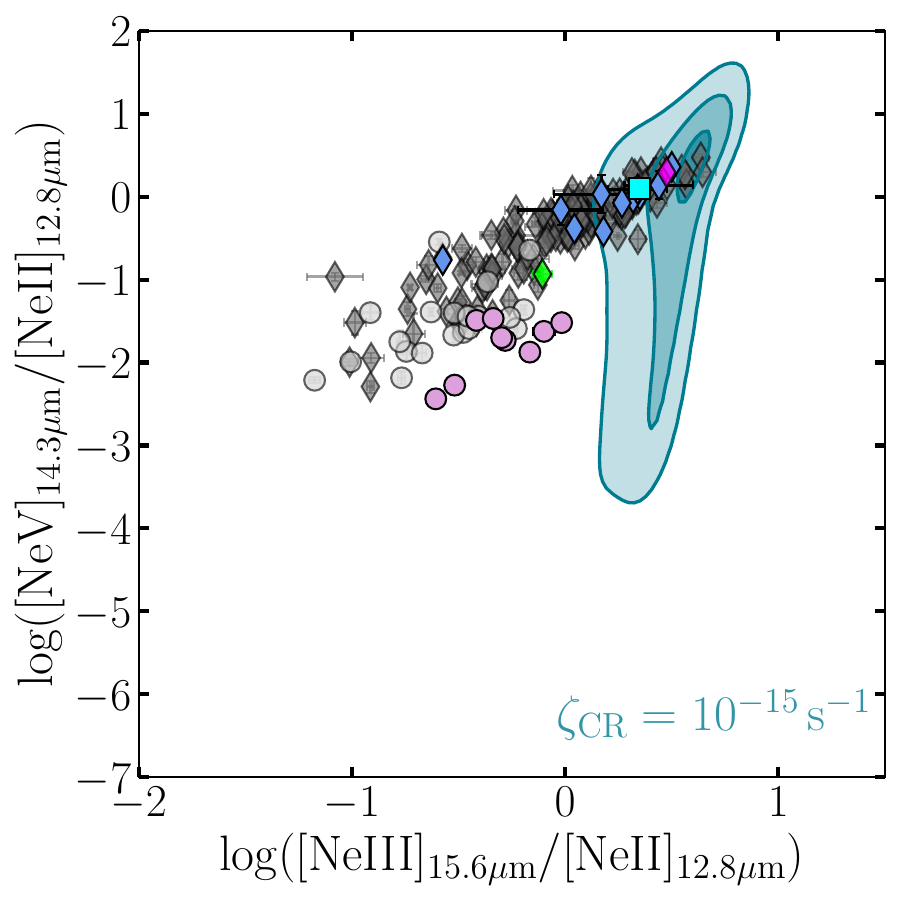}}\label{subfig:kde_zeta1__15}~\subfigure[]{\includegraphics[width=0.33\textwidth]{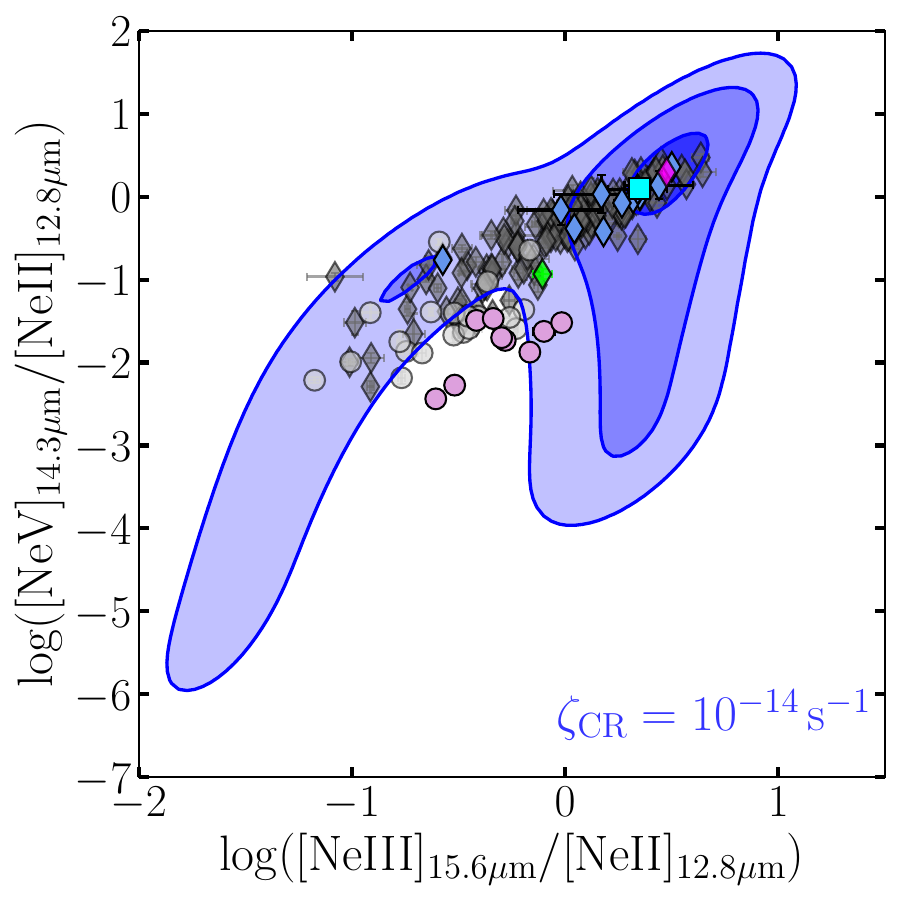}\label{subfig:kde_zeta2__14}}
    \subfigure[]{\includegraphics[width=0.33\textwidth]{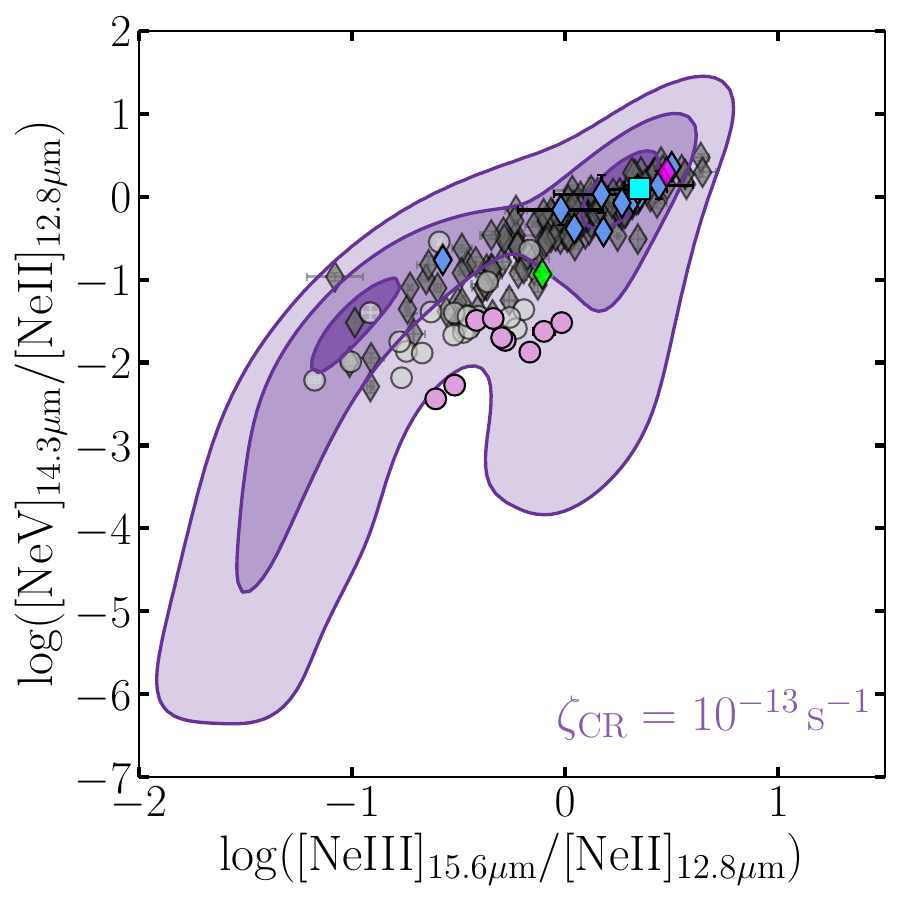}\label{subfig:kde_zeta3__13}}~\subfigure[]{\includegraphics[width=0.33\textwidth]{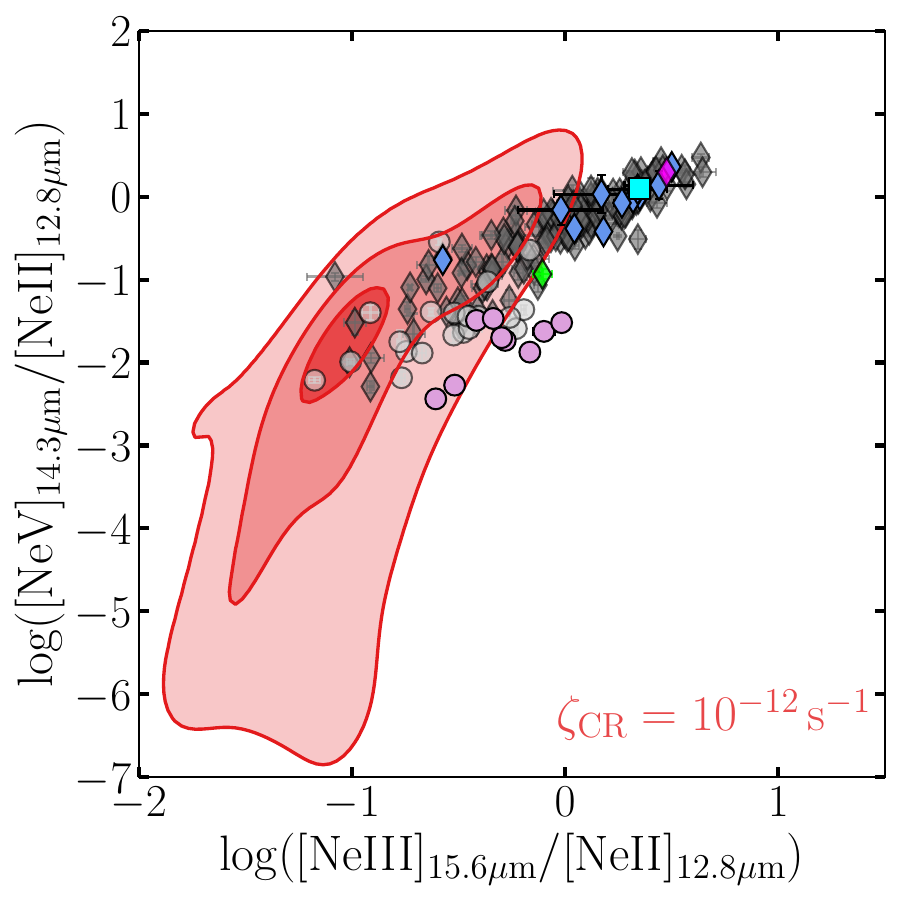}\label{subfig:kde_zeta4__12}} 
    ~\subfigure[]{\includegraphics[width=0.31\textwidth]{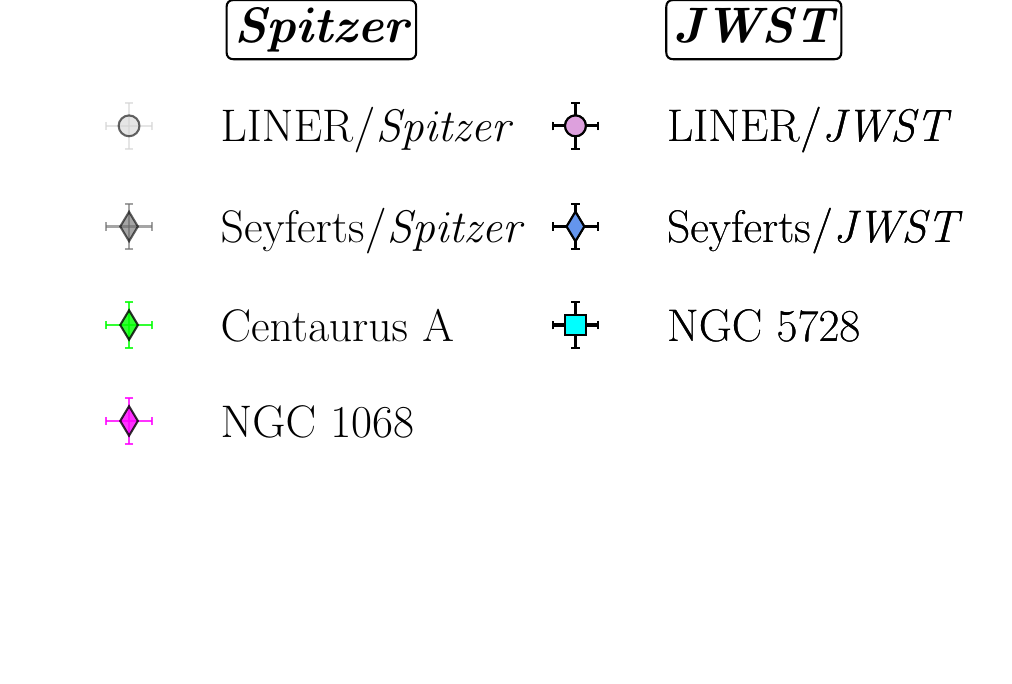}\label{subfig:legend}}
     \caption{Diagnostic diagrams depicting the area covered by AGN models with solar abundances for  $-3.5 \leq \log U \leq -1.5$ and for $0 \leq \log n_\text{H} \leq 4$. The solid contour lines map regions containing 10\%, 50\%, and 90\% of the models for $\zeta_\mathrm{CR}=10^{-16}\,\rm s^{-1},10^{-15}\,\rm s^{-1},\,10^{-14}\,\rm s^{-1},\,10^{-13}\,\rm s^{-1}$, and $10^{-12}\,\rm s^{-1}$, from left to right in green-, teal-, blue-, purple-, and red-colored contours, respectively. In the background are archival \textit{Spitzer}/IRS measurements of Seyfert and LINER nuclei \citep{Dudik_2007,Dudik_2009,Goulding_2009,Tommasin2008,Tommasin2010,Pereira_Santaella_2010,Fernandez_2016,Fernandez2021}, along with recent nuclear line ratios from \textit{JWST}/MIRI observations from the literature  (\citealt{Pereira_2022, Alvarez_2023, Armus_2023, Herrero_2024, Zhang_2024, Hernandez_2025, Goold_2024, Hermosa_2025,Goold2026,2026Fernandez})}

     
     \label{fig:kde}
\end{figure*}


CRs enhance the 
emission-line fluxes from low-ionization species relative to the pure photoionization case, leading to a systematic decrease in the 
high-to-low ionization line ratios. This trend arises from the unique ability of CRs to penetrate far deeper into cloud structures than ionizing photons, maintaining the ionization and excitation of low-ionization species in partially shielded regions where the radiation becomes ineffective.

Figure~\ref{fig:kde} illustrates this effect for the diagnostic diagram of [\ion{Ne}{v}]/[\ion{Ne}{ii}] versus [\ion{Ne}{iii}]/[\ion{Ne}{ii}] through our models, which assume solar abundances and span a wide range in the ionization parameter ($-3.5 \leq \log U \leq -1.5$) and hydrogen density ($0 \leq \log n_\mathrm{H} \leq 4$). The model predictions are presented as solid contours enclosing 10\%, 50\%, and 90\% of all the predicted ratios for these input parameters, thus excluding outlier values. The contours colored in green, teal, blue, purple, and red, represent the increasing CR ionization rate, $\zeta_\mathrm{CR} = 10^{-16},10^{-15},10^{-14},10^{-13},$ and $10^{-12}\,\mathrm{s}^{-1}$, respectively. These models are compared with the distribution of emission-line ratios measured for nearby AGN. The gray data points in Fig.~\ref{fig:kde} correspond to \textit{Spitzer}/IRS spectroscopic observations for a sample of nearby Seyfert galaxies and LINERs, compiled by \citep{Fernandez2021} from the literature \citep{Dudik_2007,Dudik_2009,Goulding_2009,Tommasin2008,Tommasin2010,Pereira_Santaella_2010,Fernandez_2016}. This full dataset most likely captures the diversity of MIR line ratios across the local population of active nuclei, including AGN photoionization, shocks, AGN and/or starburst composites, and jet-dominated nuclei.

Photoionization-dominated models with $\log U \gtrsim -2.5$ are in agreement with the highest [\ion{Ne}{v}]/[\ion{Ne}{ii}] and [\ion{Ne}{iii}]/[\ion{Ne}{ii}] ratios observed in the AGN population (Fig.~\ref{subfig:kde_zeta0__16}). As the CR ionization rate increases, the models extend toward progressively lower [\ion{Ne}{iii}]/[\ion{Ne}{ii}] values, highlighting the 
increasing contribution of CRs to 
low-ionization lines (Figs.~\ref{subfig:kde_zeta2__14}--\ref{subfig:kde_zeta4__12}). This contrasts with the behavior in pure AGN photoionization models at decreasing ionization parameter values ($\log U \lesssim -3$), which primarily affect the [\ion{Ne}{v}]/[\ion{Ne}{ii}] ratio and deviate from the observed ratio (e.g. Fig.~\ref{subfig:kde_zeta0__16}). 
Thus, CR ionization may be a relevant ionization mechanism, together with shocks and star-formation contribution \citep[e.g.][]{Feltre_2023}, to explain the lower-excitation end of the observed distribution. The latter is mostly populated by LINER galaxies, which are expected to exhibit a higher incidence of compact jets compared to the Seyfert population \citep{Baldi2023}. 
In particular, the distribution of observed ratios is well reproduced by models with CR ionization rates in the $(\zeta_\mathrm{CR} \gtrsim 10^{-14}$--$10^{-13}\,\mathrm{s}^{-1})$ range, in agreement with measurements obtained in the central region of active galaxies \citep[e.g.][]{Gonz_2013,Gonz_Alf_2018}. This suggests that CRs may provide a dominant or complementary excitation mechanism in low-luminosity AGN where the radiation field alone fails to account for MIR diagnostics.

The nucleus of NGC~5728, consistent with AGN photoionization-dominated ratios (Figs. \ref{fig:sfzeta_line}, \ref{fig:5728_new}, and \ref{fig:5728_BPTS_U}), is marked by a cyan square in Fig.~\ref{fig:kde}. The locations of NGC~1068 and Centaurus~A, whose optical line ratios were previously analyzed in KFDS25, are also highlighted by thin diamonds in magenta and lime green, respectively. Additional Seyfert and LINER galaxies with published line fluxes from nuclear measurements in recent \textit{JWST}/MIRI observations (\citealt{Pereira_2022, Alvarez_2023, Armus_2023, Herrero_2024, Zhang_2024, Hernandez_2025, Goold_2024, Hermosa_2025,Goold2026,2026Fernandez})

are indicated in blue and pink, respectively. The nucleus of Centaurus~A, which exhibits prominent jet activity, is far from the AGN photoionization locus and can be described by models with CR ionization rates of $\zeta_\mathrm{CR} \sim 10^{-14}$--$10^{-13}\,\mathrm{s}^{-1}$.
\eliz{Finally, it is noteworthy that Mrk231, a Seyfert galaxy exhibiting both jets and starburst activity \citep{Aalto_2015,Silpa_2021}, falls within the region of Fig. \ref{fig:kde} that can be modeled with $\zeta_\mathrm{CR} \gtrsim 10^{-14},\mathrm{s}^{-1}$. It is depicted as a blue thin diamond, and corresponds to the lowest [\ion{Ne}{iii}]/[\ion{Ne}{ii}] ratio among the \textit{JWST}-observed Seyfert galaxies considered here, as reported by \citet{Herrero_2024}, supporting the idea that CRs play a role in shaping the emission-line properties of such galaxies.}

\section{Summary}\label{summary}




We conducted a comprehensive study of NGC~5728 by integrating VLT/MUSE optical and \textit{JWST}/MIRI MIR observations to investigate the impact of CRs on the ISM in AGN. Using updated \textsc{Cloudy} models that include CR ionization  and a more adequate AGN ionizing spectrum, we explored a broad parameter space in the ionization parameter, gas density, and CR ionization rate. Our models, assuming solar abundances, successfully reproduce both optical and MIR emission-line ratios without requiring supersolar metallicities.

We find that low-ionization lines such as [\ion{Ne}{ii}] and [\ion{Ar}{ii}] are particularly sensitive to CRs, while high-ionization lines such as [\ion{Ne}{v}] and [\ion{Ar}{v}] remain governed by photoionization. This allows high-ionization MIR diagnostics to serve as robust tracers of AGN activity, whereas low-ionization MIR lines provide valuable insight into the CR-dominated regime of the ISM. CR ionization rates ranging from $\zeta_{\rm CR} \sim 10^{-14}$--$10^{-13}\mathrm{s}^{-1}$ are found to reproduce the observed MIR emission in regions affected by jets and outflows in NGC~5728, consistent with evidence of nonthermal activity from the synchrotron fitting of radio originating from the nucleus. Additionally, by combining optical and MIR diagnostics, we show that diagnostic diagrams involving [\ion{O}{i}]/H$\alpha$ and [\ion{O}{iv}]/[\ion{Ne}{iii}] may offer a viable route toward disentangling the degeneracy between CR- and shock-induced excitation. 

Finally, our models successfully reproduced the MIR line ratio parameter space of Centaurus A and NGC 1068, two benchmark systems with elevated CR activity previously examined in KFDS25. This demonstrates that photoionization along with CRs can account for the excitation conditions observed across both optical and MIR wavelengths. Centaurus A is consistent with ionization rates as high as $\zeta_{\rm CR} \sim 10^{-12}\mathrm{s}^{-1}$, while NGC 5728 and NGC 1068 correspond to rates ranging from $\zeta_{\rm CR} \sim 10^{-14}$ to $10^{-13}\mathrm{s}^{-1}$. These results extend the conclusions of KFDS25 by demonstrating  that AGN models with elevated CR ionization rates provide a plausible explanation for the excitation observed in low-luminosity AGN.

Notably, AGN-like emission-line ratios are consistent with high CR ionization rates ($\zeta_{\mathrm{CR}} \gtrsim 10^{-14}\mathrm{s}^{-1}$), suggesting that CRs serve as a complementary excitation mechanism in such environments. Altogether, the consistency across wavelengths supports the scenario proposed in KFDS25, in which CRs, possibly originating from jets or SNRs, significantly contribute to both the ionization and heating of the gas in AGN.

\begin{acknowledgements}
We are grateful to the anonymous referee for their constructive comments for overall improving this work. We also sincerely thank Kameron Goold and the ReveaLLAGN collaboration for kindly providing their data for comparison with our models. EK gratefully acknowledges Prof. A. Mastichiadis for his early supervision and constructive support during the initial stages of this work, and Prof. D. Hatzidimitriou for her insightful feedback during the preparation of this manuscript. EK acknowledges full financial support by the State Scholarships Foundation (IKY) from the proceeds of the "N. D. Chrysovergis" bequest. JAFO acknowledges financial support by the Spanish Ministry of Science and Innovation (MCIN/AEI/10.13039/501100011033), by "ERDF A way of making Europe'' and by "European Union NextGenerationEU/PRTR'' through the grants PID2021-124918NB-C44 and CNS2023-145339; MCIN and the European Union -- NextGenerationEU through the Recovery and Resilience Facility project ICTS-MRR-2021-03-CEFCA. 

This work is based on observations made with the NASA/ESA/CSA \textit{JWST}, associated with program ID~1670 (PI: T.~Taro~Shimizu). The corresponding datasets are publicly available in MAST at the Space Telescope Science Institute, operated by the Association of Universities for Research in Astronomy, Inc., under NASA contract NAS~5-03127.
\end{acknowledgements}

\bibliographystyle{aa.bst} 
\bibliography{name,biblio,jwst} 

@ARTICLE{2025K,
       author = {{Koutsoumpou}, E. and {Fern{\'a}ndez-Ontiveros}, J.~A. and {Dasyra}, K.~M. and {Spinoglio}, L.},
        title = "{Cosmic-ray ionization of low-excitation lines in active galactic nuclei and starburst galaxies}",
      journal = {\aap},
         year = 2025,
        month = jan,
       volume = {693},
          eid = {A215},
        pages = {A215},
          doi = {10.1051/0004-6361/202452232},
archivePrefix = {arXiv},
       eprint = {2411.17811},
 primaryClass = {astro-ph.GA},
       adsurl = {https://ui.adsabs.harvard.edu/abs/2025A&A...693A.215K}
}

@ARTICLE{Jin_2012,
       author = {{Jin}, Chichuan and {Ward}, Martin and {Done}, Chris},
        title = "{A combined optical and X-ray study of unobscured type 1 active galactic nuclei - III. Broad-band SED properties}",
      journal = {\mnras},
         year = 2012,
        month = sep,
       volume = {425},
       number = {2},
        pages = {907-929},
          doi = {10.1111/j.1365-2966.2012.21272.x},
archivePrefix = {arXiv},
       eprint = {1205.1846},
 primaryClass = {astro-ph.HE},
       adsurl = {https://ui.adsabs.harvard.edu/abs/2012MNRAS.425..907J}
}

@ARTICLE{Binette_1985,
       author = {{Binette}, L.},
        title = "{Photoionization models for liners : gas distribution abundances.}",
      journal = {\aap},
         year = 1985,
        month = feb,
       volume = {143},
        pages = {334-346},
       adsurl = {https://ui.adsabs.harvard.edu/abs/1985A&A...143..334B}
}

@ARTICLE{Sutherland_1993,
       author = {{Sutherland}, Ralph S. and {Dopita}, M.~A.},
        title = "{Cooling Functions for Low-Density Astrophysical Plasmas}",
      journal = {\apjs},
         year = 1993,
        month = sep,
       volume = {88},
        pages = {253},
          doi = {10.1086/191823},
       adsurl = {https://ui.adsabs.harvard.edu/abs/1993ApJS...88..253S}
}

@article{Hirschmann_2017,
    author = {Hirschmann, Michaela and Charlot, Stephane and Feltre, Anna and Naab, Thorsten and Choi, Ena and Ostriker, Jeremiah P. and Somerville, Rachel S.},
    title = {Synthetic nebular emission from massive galaxies – I: origin of the cosmic evolution of optical emission-line ratios},
    journal = {Monthly Notices of the Royal Astronomical Society},
    volume = {472},
    number = {2},
    pages = {2468-2495},
    year = {2017},
    month = {08},
    issn = {0035-8711},
    doi = {10.1093/mnras/stx2180},
    url = {https://doi.org/10.1093/mnras/stx2180},
    eprint = {https://academic.oup.com/mnras/article-pdf/472/2/2468/19943145/stx2180.pdf},
}

@ARTICLE{Feltre_2023,
       author = {{Feltre}, A. and {Gruppioni}, C. and {Marchetti}, L. and {Mahoro}, A. and {Salvestrini}, F. and {Mignoli}, M. and {Bisigello}, L. and {Calura}, F. and {Charlot}, S. and {Chevallard}, J. and {Romero-Colmenero}, E. and {Curtis-Lake}, E. and {Delvecchio}, I. and {Dors}, O.~L. and {Hirschmann}, M. and {Jarrett}, T. and {Marchesi}, S. and {Moloko}, M.~E. and {Plat}, A. and {Pozzi}, F. and {Sefako}, R. and {Traina}, A. and {Vaccari}, M. and {V{\"a}is{\"a}nen}, P. and {Vallini}, L. and {Vidal-Garc{\'\i}a}, A. and {Vignali}, C.},
        title = "{Optical and mid-infrared line emission in nearby Seyfert galaxies}",
      journal = {\aap},
         year = 2023,
        month = jul,
       volume = {675},
          eid = {A74},
        pages = {A74},
          doi = {10.1051/0004-6361/202245516},
archivePrefix = {arXiv},
       eprint = {2301.02252},
 primaryClass = {astro-ph.GA},
       adsurl = {https://ui.adsabs.harvard.edu/abs/2023A&A...675A..74F}
}

@ARTICLE{Fernandez_2016,
       author = {{Fern{\'a}ndez-Ontiveros}, Juan Antonio and {Spinoglio}, Luigi and {Pereira-Santaella}, Miguel and {Malkan}, Matthew A. and {Andreani}, Paola and {Dasyra}, Kalliopi M.},
        title = "{Far-infrared Line Spectra of Active Galaxies from the Herschel/PACS Spectrometer: The Complete Database}",
      journal = {\apjs},
         year = 2016,
        month = oct,
       volume = {226},
       number = {2},
          eid = {19},
        pages = {19},
          doi = {10.3847/0067-0049/226/2/19},
archivePrefix = {arXiv},
       eprint = {1607.02511},
 primaryClass = {astro-ph.GA},
       adsurl = {https://ui.adsabs.harvard.edu/abs/2016ApJS..226...19F}
}

@article{Ferland_2020,
    author = {Ferland, G J and Done, C and Jin, C and Landt, H and Ward, M J},
    title = {State-of-the-art AGN SEDs for photoionization models: BLR predictions confront the observations},
    journal = {Monthly Notices of the Royal Astronomical Society},
    volume = {494},
    number = {4},
    pages = {5917-5922},
    year = {2020},
    month = {05},
    issn = {0035-8711},
    doi = {10.1093/mnras/staa1207},
    url = {https://doi.org/10.1093/mnras/staa1207},
    eprint = {https://academic.oup.com/mnras/article-pdf/494/4/5917/33209134/staa1207.pdf},
}

@ARTICLE{Shin_2019,
       author = {{Shin}, Jaejin and {Woo}, Jong-Hak and {Chung}, Aeree and {Baek}, Junhyun and {Cho}, Kyuhyoun and {Kang}, Daeun and {Bae}, Hyun-Jin},
        title = "{Positive and Negative Feedback of AGN Outflows in NGC 5728}",
      journal = {\apj},
         year = 2019,
        month = aug,
       volume = {881},
       number = {2},
          eid = {147},
        pages = {147},
          doi = {10.3847/1538-4357/ab2e72},
archivePrefix = {arXiv},
       eprint = {1907.00982},
 primaryClass = {astro-ph.GA},
       adsurl = {https://ui.adsabs.harvard.edu/abs/2019ApJ...881..147S}
}

@ARTICLE{Shimizu_2019,
       author = {{Shimizu}, T. Taro and {Davies}, R.~I. and {Lutz}, D. and {Burtscher}, L. and {Lin}, M. and {Baron}, D. and {Davies}, R.~L. and {Genzel}, R. and {Hicks}, E.~K.~S. and {Koss}, M. and {Maciejewski}, W. and {M{\"u}ller-S{\'a}nchez}, F. and {Orban de Xivry}, G. and {Price}, S.~H. and {Ricci}, C. and {Riffel}, R. and {Riffel}, R.~A. and {Rosario}, D. and {Schartmann}, M. and {Schnorr-M{\"u}ller}, A. and {Sternberg}, A. and {Sturm}, E. and {Storchi-Bergmann}, T. and {Tacconi}, L. and {Veilleux}, S.},
        title = "{The multiphase gas structure and kinematics in the circumnuclear region of NGC 5728}",
      journal = {\mnras},
         year = 2019,
        month = dec,
       volume = {490},
       number = {4},
        pages = {5860-5887},
          doi = {10.1093/mnras/stz2802},
archivePrefix = {arXiv},
       eprint = {1907.03801},
 primaryClass = {astro-ph.GA},
       adsurl = {https://ui.adsabs.harvard.edu/abs/2019MNRAS.490.5860S}
}

@ARTICLE{Thornley_2000,
       author = {{Thornley}, Michele D. and {F{\"o}rster Schreiber}, Natascha M. and {Lutz}, Dieter and {Genzel}, Reinhard and {Spoon}, Henrik W.~W. and {Kunze}, Dietmar and {Sternberg}, Amiel},
        title = "{Massive Star Formation and Evolution in Starburst Galaxies: Mid-infrared Spectroscopy with the ISO Short Wavelength Spectrometer}",
      journal = {\apj},
         year = 2000,
        month = aug,
       volume = {539},
       number = {2},
        pages = {641-657},
          doi = {10.1086/309261},
archivePrefix = {arXiv},
       eprint = {astro-ph/0003334},
 primaryClass = {astro-ph},
       adsurl = {https://ui.adsabs.harvard.edu/abs/2000ApJ...539..641T}
}

@ARTICLE{Richardson_2022,
       author = {{Richardson}, Chris T. and {Simpson}, Connor and {Polimera}, Mugdha S. and {Kannappan}, Sheila J. and {Bellovary}, Jillian M. and {Greene}, Christopher and {Jenkins}, Sam},
        title = "{Optical and JWST Mid-IR Emission Line Diagnostics for Simultaneous IMBH and Stellar Excitation in z 0 Dwarf Galaxies}",
      journal = {\apj},
         year = 2022,
        month = mar,
       volume = {927},
       number = {2},
          eid = {165},
        pages = {165},
          doi = {10.3847/1538-4357/ac510c},
archivePrefix = {arXiv},
       eprint = {2202.01330},
 primaryClass = {astro-ph.GA},
       adsurl = {https://ui.adsabs.harvard.edu/abs/2022ApJ...927..165R}
}

@ARTICLE{Durre_2018,
       author = {{Durr{\'e}}, Mark and {Mould}, Jeremy},
        title = "{The AGN Ionization Cones of NGC 5728. I. Excitation and Nuclear Structure}",
      journal = {\apj},
         year = 2018,
        month = nov,
       volume = {867},
       number = {2},
          eid = {149},
        pages = {149},
          doi = {10.3847/1538-4357/aae68e},
archivePrefix = {arXiv},
       eprint = {1810.03258},
 primaryClass = {astro-ph.GA},
       adsurl = {https://ui.adsabs.harvard.edu/abs/2018ApJ...867..149D}
}

@ARTICLE{Davies_2024,
       author = {{Davies}, R. and {Shimizu}, T. and {Pereira-Santaella}, M. and {Alonso-Herrero}, A. and {Audibert}, A. and {Bellocchi}, E. and {Boorman}, P. and {Campbell}, S. and {Cao}, Y. and {Combes}, F. and {Delaney}, D. and {D{\'\i}az-Santos}, T. and {Eisenhauer}, F. and {Esparza Arredondo}, D. and {Feuchtgruber}, H. and {F{\"o}rster Schreiber}, N.~M. and {Fuller}, L. and {Gandhi}, P. and {Garc{\'\i}a-Bernete}, I. and {Garc{\'\i}a-Burillo}, S. and {Garc{\'\i}a-Lorenzo}, B. and {Genzel}, R. and {Gillessen}, S. and {Gonz{\'a}lez Mart{\'\i}n}, O. and {Haidar}, H. and {Hermosa Mu{\~n}oz}, L. and {Hicks}, E.~K.~S. and {H{\"o}nig}, S. and {Imanishi}, M. and {Izumi}, T. and {Labiano}, A. and {Leist}, M. and {Levenson}, N.~A. and {Lopez-Rodriguez}, E. and {Lutz}, D. and {Ott}, T. and {Packham}, C. and {Rabien}, S. and {Ramos Almeida}, C. and {Ricci}, C. and {Rigopoulou}, D. and {Rosario}, D. and {Rouan}, D. and {Santos}, D.~J.~D. and {Shangguan}, J. and {Stalevski}, M. and {Sternberg}, A. and {Sturm}, E. and {Tacconi}, L. and {Villar Mart{\'\i}n}, M. and {Ward}, M. and {Zhang}, L.},
        title = "{GATOS: missing molecular gas in the outflow of NGC 5728 revealed by JWST}",
      journal = {\aap},
         year = 2024,
        month = sep,
       volume = {689},
          eid = {A263},
        pages = {A263},
          doi = {10.1051/0004-6361/202449875},
archivePrefix = {arXiv},
       eprint = {2406.17072},
 primaryClass = {astro-ph.GA},
       adsurl = {https://ui.adsabs.harvard.edu/abs/2024A&A...689A.263D}
}

@ARTICLE{Schommer_1988,
       author = {{Schommer}, Robert A. and {Caldwell}, Nelson and {Wilson}, A.~S. and {Baldwin}, J.~A. and {Phillips}, M.~M. and {Williams}, T.~B. and {Turtle}, A.~J.},
        title = "{Ionized Gas and Radio Emission in the Barred Seyfert Galaxy NGC 5728}",
      journal = {\apj},
         year = 1988,
        month = jan,
       volume = {324},
        pages = {154},
          doi = {10.1086/165887},
       adsurl = {https://ui.adsabs.harvard.edu/abs/1988ApJ...324..154S}
}

@software{phot_2024,
  author       = {Larry Bradley and
                  Brigitta Sip{\H o}cz and
                  Thomas Robitaille and
                  Erik Tollerud and
                  Z\`e Vin{\'{\i}}cius and
                  Christoph Deil and
                  Kyle Barbary and
                  Tom J Wilson and
                  Ivo Busko and
                  Axel Donath and
                  Hans Moritz G{\"u}nther and
                  Mihai Cara and
                  P. L. Lim and
                  Sebastian Me{\ss}linger and
                  Simon Conseil and
                  Zach Burnett and
                  Azalee Bostroem and
                  Michael Droettboom and
                  E. M. Bray and
                  Lars Andersen Bratholm and
                  Adam Ginsburg and
                  William Jamieson and
                  Geert Barentsen and
                  Matt Craig and
                  Brett M. Morris and
                  Marshall Perrin and
                  Shivangee Rathi and
                  Sergio Pascual and
                  Iskren Y. Georgiev},
  title        = {astropy/photutils: 2.0.2},
  month        = oct,
  year         = 2024,
  publisher    = {Zenodo},
  version      = {2.0.2},
  doi          = {10.5281/zenodo.13989456},
  url          = {https://doi.org/10.5281/zenodo.13989456},
}

@ARTICLE{Pereira_Santaella_2010,
       author = {{Pereira-Santaella}, Miguel and {Diamond-Stanic}, Aleksandar M. and {Alonso-Herrero}, Almudena and {Rieke}, George H.},
        title = "{The Mid-infrared High-ionization Lines from Active Galactic Nuclei and Star-forming Galaxies}",
      journal = {\apj},
         year = 2010,
        month = dec,
       volume = {725},
       number = {2},
        pages = {2270-2280},
          doi = {10.1088/0004-637X/725/2/2270},
archivePrefix = {arXiv},
       eprint = {1010.5129},
 primaryClass = {astro-ph.CO},
       adsurl = {https://ui.adsabs.harvard.edu/abs/2010ApJ...725.2270P}
}

@ARTICLE{Sturm_2002,
       author = {{Sturm}, E. and {Lutz}, D. and {Verma}, A. and {Netzer}, H. and {Sternberg}, A. and {Moorwood}, A.~F.~M. and {Oliva}, E. and {Genzel}, R.},
        title = "{Mid-Infrared line diagnostics of active galaxies. A spectroscopic AGN survey with ISO-SWS}",
      journal = {\aap},
         year = 2002,
        month = oct,
       volume = {393},
        pages = {821-841},
          doi = {10.1051/0004-6361:20021043},
archivePrefix = {arXiv},
       eprint = {astro-ph/0207381},
 primaryClass = {astro-ph},
       adsurl = {https://ui.adsabs.harvard.edu/abs/2002A&A...393..821S}
}

@article{Genzel_1998,
doi = {10.1086/305576},
url = {https://dx.doi.org/10.1086/305576},
year = {1998},
month = {may},
publisher = {},
volume = {498},
number = {2},
pages = {579},
author = {Genzel, R. and Lutz, D. and Sturm, E. and Egami, E. and Kunze, D. and Moorwood, A. F. M. and Rigopoulou, D. and Spoon, H. W. W. and Sternberg, A. and Tacconi-Garman, L. E. and Tacconi, L. and Thatte, N.},
title = {What Powers Ultraluminous IRAS Galaxies?*},
journal = {The Astrophysical Journal}
}

@ARTICLE{PerezMontero_Diaz_2007,
       author = {{P{\'e}rez-Montero}, Enrique and {D{\'\i}az}, {\'A}ngeles I.},
        title = "{The nature of the Wolf-Rayet galaxy Mrk 209 from photoionization models}",
      journal = {\mnras},
         year = 2007,
        month = may,
       volume = {377},
       number = {3},
        pages = {1195-1205},
          doi = {10.1111/j.1365-2966.2007.11670.x},
archivePrefix = {arXiv},
       eprint = {astro-ph/0703319},
 primaryClass = {astro-ph},
       adsurl = {https://ui.adsabs.harvard.edu/abs/2007MNRAS.377.1195P}
}

@ARTICLE{Mingozzi_2025,
       author = {{Mingozzi}, Matilde and {Garcia del Valle-Espinosa}, Macarena and {James}, Bethan L. and {Rickards Vaught}, Ryan J. and {Hayes}, Matthew and {Amor{\'\i}n}, Ricardo O. and {Leitherer}, Claus and {Aloisi}, Alessandra and {Hunt}, Leslie and {Law}, David and {Richardson}, Chris T. and {Pidgeon}, Aidan and {Arellano-C{\'o}rdova}, Karla Z. and {Berg}, Danielle A. and {Chisholm}, John and {Hernandez}, Svea and {Jones}, Logan and {Kumari}, Nimisha and {Martin}, Crystal L. and {Ravindranath}, Swara and {Vallini}, Livia and {Xu}, Xinfeng},
        title = "{Exploring the Mysterious High-ionization Source Powering [Ne V] in High-z Analog SBS0335-052 E with JWST/MIRI}",
      journal = {\apj},
         year = 2025,
        month = jun,
       volume = {985},
       number = {2},
          eid = {253},
        pages = {253},
          doi = {10.3847/1538-4357/adc996},
archivePrefix = {arXiv},
       eprint = {2502.07662},
 primaryClass = {astro-ph.GA},
       adsurl = {https://ui.adsabs.harvard.edu/abs/2025ApJ...985..253M}
}

@ARTICLE{PerezMontero_2024,
       author = {{P{\'e}rez-Montero}, E. and {Fern{\'a}ndez-Ontiveros}, J.~A. and {P{\'e}rez-D{\'\i}az}, B. and {V{\'\i}lchez}, J.~M. and {Kumari}, N. and {Amor{\'\i}n}, R.},
        title = "{Exploring the hardness of the ionising radiation with the infrared softness diagram. I. Similar effective temperature scales for starbursts and (ultra)luminous infrared galaxies}",
      journal = {\aap},
         year = 2024,
        month = apr,
       volume = {684},
          eid = {A40},
        pages = {A40},
          doi = {10.1051/0004-6361/202348089},
archivePrefix = {arXiv},
       eprint = {2401.09765},
 primaryClass = {astro-ph.GA},
       adsurl = {https://ui.adsabs.harvard.edu/abs/2024A&A...684A..40P}
}

@ARTICLE{Hermosa_2025,
       author = {{Hermosa Mu{\~n}oz}, L. and {Alonso-Herrero}, A. and {Labiano}, A. and {Guillard}, P. and {Pantoni}, L. and {Buiten}, V. and {Dicken}, D. and {Baes}, M. and {B{\"o}ker}, T. and {Colina}, L. and {Donnan}, F. and {Garc{\'\i}a-Bernete}, I. and {{\"O}stlin}, G. and {van der Werf}, P. and {Ward}, M.~J. and {Brandl}, B.~R. and {Walter}, F. and {Wright}, G. and {G{\"u}del}, M. and {Henning}, Th. and {Lagage}, P. -O. and {Ray}, T.},
        title = "{MICONIC: Dual active galactic nuclei, star formation, and ionised gas outflows in NGC 6240 seen with MIRI/JWST}",
      journal = {\aap},
         year = 2025,
        month = jan,
       volume = {693},
          eid = {A321},
        pages = {A321},
          doi = {10.1051/0004-6361/202452437},
archivePrefix = {arXiv},
       eprint = {2412.14707},
 primaryClass = {astro-ph.GA},
       adsurl = {https://ui.adsabs.harvard.edu/abs/2025A&A...693A.321H}
}

@ARTICLE{Zhang_2024,
       author = {{Zhang}, Lulu and {Packham}, Chris and {Hicks}, Erin K.~S. and {Davies}, Ric I. and {Shimizu}, Taro T. and {Alonso-Herrero}, Almudena and {Hermosa Mu{\~n}oz}, Laura and {Garc{\'\i}a-Bernete}, Ismael and {Pereira-Santaella}, Miguel and {Audibert}, Anelise and {L{\'o}pez-Rodr{\'\i}guez}, Enrique and {Bellocchi}, Enrica and {Bunker}, Andrew J. and {Combes}, Francoise and {D{\'\i}az-Santos}, Tanio and {Gandhi}, Poshak and {Garc{\'\i}a-Burillo}, Santiago and {Garc{\'\i}a-Lorenzo}, Bego{\~n}a and {Gonz{\'a}lez-Mart{\'\i}n}, Omaira and {Imanishi}, Masatoshi and {Labiano}, Alvaro and {Leist}, Mason T. and {Levenson}, Nancy A. and {Ramos Almeida}, Cristina and {Ricci}, Claudio and {Rigopoulou}, Dimitra and {Rosario}, David J. and {Stalevski}, Marko and {Ward}, Martin J. and {Esparza-Arredondo}, Donaji and {Delaney}, Dan and {Fuller}, Lindsay and {Haidar}, Houda and {H{\"o}nig}, Sebastian and {Izumi}, Takuma and {Rouan}, Daniel},
        title = "{The Galaxy Activity, Torus, and Outflow Survey (GATOS). IV. Exploring Ionized Gas Outflows in Central Kiloparsec Regions of GATOS Seyferts}",
      journal = {\apj},
         year = 2024,
        month = oct,
       volume = {974},
       number = {2},
          eid = {195},
        pages = {195},
          doi = {10.3847/1538-4357/ad6a4b},
archivePrefix = {arXiv},
       eprint = {2409.09771},
 primaryClass = {astro-ph.GA},
       adsurl = {https://ui.adsabs.harvard.edu/abs/2024ApJ...974..195Z}
}

@article{Goold_2024,
doi = {10.3847/1538-4357/ad3065},
url = {https://dx.doi.org/10.3847/1538-4357/ad3065},
year = {2024},
month = {may},
publisher = {The American Astronomical Society},
volume = {966},
number = {2},
pages = {204},
author = {Goold, Kameron and Seth, Anil and Molina, Mallory and Ohlson, David and Runnoe, Jessie C. and Böker, Torsten and Davis, Timothy A. and Dumont, Antoine and Eracleous, Michael and Fernández-Ontiveros, Juan Antonio and Gallo, Elena and Goulding, Andy D. and Greene, Jenny E. and Ho, Luis C. and Markoff, Sera B. and Neumayer, Nadine and Plotkin, Richard M. and Prieto, Almudena and Satyapal, Shobita and van de Ven, Glenn and Walsh, Jonelle L. and Yuan, Feng and Feldmeier-Krause, Anja and Gültekin, Kayhan and Hönig, Sebastian and Kirkpatrick, Allison and Lützgendorf, Nora and Reines, Amy E. and Strader, Jay and Trump, Jonathan R. and Voggel, Karina T.},
title = {ReveaLLAGN 0: First Look at JWST MIRI Data of Sombrero and NGC 1052},
journal = {The Astrophysical Journal}
}

@ARTICLE{Hernandez_2025,
       author = {{Hernandez}, Svea and {Smith}, Linda J. and {Jones}, Logan H. and {Togi}, Aditya and {Mel{\'e}ndez}, Marcio B. and {Abril-Melgarejo}, Valentina and {Adamo}, Angela and {Alonso Herrero}, Almudena and {D{\'\i}az-Santos}, Tanio and {Fischer}, Travis C. and {Garc{\'\i}a-Burillo}, Santiago and {Hirschauer}, Alec S. and {Hunt}, Leslie K. and {James}, Bethan and {Lebouteiller}, Vianney and {Long}, Knox S. and {Mingozzi}, Matilde and {Ramambason}, Lise and {Ramos Almeida}, Cristina},
        title = "{JWST/MIRI Detection of [Ne v] and [Ne VI] in M83: Evidence for the Long Sought-after Active Galactic Nucleus?}",
      journal = {\apj},
         year = 2025,
        month = apr,
       volume = {983},
       number = {2},
          eid = {154},
        pages = {154},
          doi = {10.3847/1538-4357/adba5d},
archivePrefix = {arXiv},
       eprint = {2502.17621},
 primaryClass = {astro-ph.GA},
       adsurl = {https://ui.adsabs.harvard.edu/abs/2025ApJ...983..154H}
}

@ARTICLE{Herrero_2024,
       author = {{Alonso Herrero}, A. and {Hermosa Mu{\~n}oz}, L. and {Labiano}, A. and {Guillard}, P. and {Buiten}, V.~A. and {Dicken}, D. and {van der Werf}, P. and {{\'A}lvarez-M{\'a}rquez}, J. and {B{\"o}ker}, T. and {Colina}, L. and {Eckart}, A. and {Garc{\'\i}a-Mar{\'\i}n}, M. and {Jones}, O.~C. and {Pantoni}, L. and {P{\'e}rez-Gonz{\'a}lez}, P.~G. and {Rouan}, D. and {Ward}, M.~J. and {Baes}, M. and {{\"O}stlin}, G. and {Royer}, P. and {Wright}, G.~S. and {G{\"u}del}, M. and {Henning}, Th. and {Lagage}, P. -O. and {van Dishoeck}, E.~F.},
        title = "{MICONIC: JWST/MIRI MRS observations of the nuclear and circumnuclear regions of Mrk 231}",
      journal = {\aap},
         year = 2024,
        month = oct,
       volume = {690},
          eid = {A95},
        pages = {A95},
          doi = {10.1051/0004-6361/202450071},
archivePrefix = {arXiv},
       eprint = {2407.02180},
 primaryClass = {astro-ph.GA},
       adsurl = {https://ui.adsabs.harvard.edu/abs/2024A&A...690A..95A}
}

@ARTICLE{Pereira_2022,
       author = {{Pereira-Santaella}, M. and {{\'A}lvarez-M{\'a}rquez}, J. and {Garc{\'\i}a-Bernete}, I. and {Labiano}, A. and {Colina}, L. and {Alonso-Herrero}, A. and {Bellocchi}, E. and {Garc{\'\i}a-Burillo}, S. and {H{\"o}nig}, S.~F. and {Ramos Almeida}, C. and {Rosario}, D.},
        title = "{Low-power jet-interstellar medium interaction in NGC 7319 revealed by JWST/MIRI MRS}",
      journal = {\aap},
         year = 2022,
        month = sep,
       volume = {665},
          eid = {L11},
        pages = {L11},
          doi = {10.1051/0004-6361/202244725},
archivePrefix = {arXiv},
       eprint = {2208.04835},
 primaryClass = {astro-ph.GA},
       adsurl = {https://ui.adsabs.harvard.edu/abs/2022A&A...665L..11P}
}

@ARTICLE{Alvarez_2023,
       author = {{{\'A}lvarez-M{\'a}rquez}, J. and {Labiano}, A. and {Guillard}, P. and {Dicken}, D. and {Argyriou}, I. and {Patapis}, P. and {Law}, D.~R. and {Kavanagh}, P.~J. and {Larson}, K.~L. and {Gasman}, D. and {Mueller}, M. and {Alberts}, S. and {Brandl}, B.~R. and {Colina}, L. and {Garc{\'\i}a-Mar{\'\i}n}, M. and {Jones}, O.~C. and {Noriega-Crespo}, A. and {Shivaei}, I. and {Temim}, T. and {Wright}, G.~S.},
        title = "{Nuclear high-ionisation outflow in the Compton-thick AGN NGC 6552 as seen by the JWST mid-infrared instrument}",
      journal = {\aap},
         year = 2023,
        month = apr,
       volume = {672},
          eid = {A108},
        pages = {A108},
          doi = {10.1051/0004-6361/202244880},
archivePrefix = {arXiv},
       eprint = {2209.01695},
 primaryClass = {astro-ph.GA},
       adsurl = {https://ui.adsabs.harvard.edu/abs/2023A&A...672A.108A}
}

@ARTICLE{Armus_2023,
       author = {{Armus}, L. and {Lai}, T. and {U}, V. and {Larson}, K.~L. and {Diaz-Santos}, T. and {Evans}, A.~S. and {Malkan}, M.~A. and {Rich}, J. and {Medling}, A.~M. and {Law}, D.~R. and {Inami}, H. and {Muller-Sanchez}, F. and {Charmandaris}, V. and {van der Werf}, P. and {Stierwalt}, S. and {Linden}, S. and {Privon}, G.~C. and {Barcos-Mu{\~n}oz}, L. and {Hayward}, C. and {Song}, Y. and {Appleton}, P. and {Aalto}, S. and {Bohn}, T. and {B{\"o}ker}, T. and {Brown}, M.~J.~I. and {Finnerty}, L. and {Howell}, J. and {Iwasawa}, K. and {Kemper}, F. and {Marshall}, J. and {Mazzarella}, J.~M. and {McKinney}, J. and {Murphy}, E.~J. and {Sanders}, D. and {Surace}, J.},
        title = "{GOALS-JWST: Mid-infrared Spectroscopy of the Nucleus of NGC 7469}",
      journal = {\apjl},
         year = 2023,
        month = jan,
       volume = {942},
       number = {2},
          eid = {L37},
        pages = {L37},
          doi = {10.3847/2041-8213/acac66},
archivePrefix = {arXiv},
       eprint = {2209.13125},
 primaryClass = {astro-ph.GA},
       adsurl = {https://ui.adsabs.harvard.edu/abs/2023ApJ...942L..37A}
}

@ARTICLE{Thuan_2005,
       author = {{Thuan}, Trinh X. and {Izotov}, Yuri I.},
        title = "{High-Ionization Emission in Metal-deficient Blue Compact Dwarf Galaxies}",
      journal = {\apjs},
         year = 2005,
        month = dec,
       volume = {161},
       number = {2},
        pages = {240-270},
          doi = {10.1086/491657},
archivePrefix = {arXiv},
       eprint = {astro-ph/0507209},
 primaryClass = {astro-ph},
       adsurl = {https://ui.adsabs.harvard.edu/abs/2005ApJS..161..240T}
}

@ARTICLE{Hatano_2024,
       author = {{Hatano}, Shun and {Ouchi}, Masami and {Umeda}, Hiroya and {Nakajima}, Kimihiko and {Kawaguchi}, Toshihiro and {Isobe}, Yuki and {Aoyama}, Shohei and {Watanabe}, Kuria and {Harikane}, Yuichi and {Kusakabe}, Haruka and {Matsumoto}, Akinori and {Moriya}, Takashi J. and {Nishigaki}, Moka and {Ono}, Yoshiaki and {Onodera}, Masato and {Sugahara}, Yuma and {Suzuki}, Akihiro and {Xu}, Yi and {Zhang}, Yechi},
        title = "{EMPRESS. XIV. Strong High-ionization Lines of Young Galaxies at z = 0{\textendash}8: Ionizing Spectra Consistent with the Intermediate-mass Black Holes with M $_{BH}$ {\ensuremath{\sim}} {}10$^{3}${\textendash}{}10$^{6}$ M $_{{\ensuremath{\odot}}}$}",
      journal = {\apj},
         year = 2024,
        month = may,
       volume = {966},
       number = {2},
          eid = {170},
        pages = {170},
          doi = {10.3847/1538-4357/ad335c},
archivePrefix = {arXiv},
       eprint = {2305.02189},
 primaryClass = {astro-ph.GA},
       adsurl = {https://ui.adsabs.harvard.edu/abs/2024ApJ...966..170H}
}

@ARTICLE{Hatano_2023,
       author = {{Hatano}, Shun and {Ouchi}, Masami and {Nakajima}, Kimihiko and {Kawaguchi}, Toshihiro and {Kokubo}, Mitsuru and {Kikuta}, Satoshi and {Tominaga}, Nozomu and {Xu}, Yi and {Watanabe}, Kuria and {Harikane}, Yuichi and {Isobe}, Yuki and {Matsumoto}, Akinori and {Nishigaki}, Moka and {Ono}, Yoshiaki and {Onodera}, Masato and {Sugahara}, Yuma and {Umeda}, Hiroya and {Zhang}, Yechi},
        title = "{Active Massive Black Hole Found in the Young Star-Forming Dwarf Galaxy SBS 0335-052E}",
      journal = {arXiv e-prints},
         year = 2023,
        month = apr,
          eid = {arXiv:2304.03726},
        pages = {arXiv:2304.03726},
          doi = {10.48550/arXiv.2304.03726},
archivePrefix = {arXiv},
       eprint = {2304.03726},
 primaryClass = {astro-ph.GA},
       adsurl = {https://ui.adsabs.harvard.edu/abs/2023arXiv230403726H}
}

@ARTICLE{2026Fernandez,
       author = {{Fern{\'a}ndez-Ontiveros}, J.~A. and {Spinoglio}, L. and {Pereira-Santaella}, M. and {Hern{\'a}n-Caballero}, A. and {Hatziminaoglou}, E. and {P{\'e}rez-Montero}, E. and {V{\'\i}lchez}, J.~M. and {P{\'e}rez-D{\'\i}az}, B. and {Amor{\'\i}n}, R. and {Malkan}, M.~A. and {Dasyra}, K.~M.},
        title = "{Wild is the wind from low-luminosity AGN: a jet-driven gas bubble blowing out a massive CO-dark outflow in ESO 420-G13}",
      journal = {arXiv e-prints},
         year = 2026,
        month = apr,
          eid = {arXiv:2604.06305},
        pages = {arXiv:2604.06305},
          doi = {10.48550/arXiv.2604.06305},
archivePrefix = {arXiv},
       eprint = {2604.06305},
 primaryClass = {astro-ph.GA},
       adsurl = {https://ui.adsabs.harvard.edu/abs/2026arXiv260406305F}
}

@ARTICLE{Lumb_Calle_2024,
       author = {{Lumbreras-Calle}, A. and {Fern{\'a}ndez-Ontiveros}, J.~A. and {Infante-Sainz}, R. and {Akhlaghi}, M. and {Montoro-Molina}, B. and {P{\'e}rez-D{\'\i}az}, B. and {del Pino}, A. and {Vives-Arias}, H. and {Hern{\'a}n-Caballero}, A. and {L{\'o}pez-Sanjuan}, C. and {Mart{\'\i}n-Guerrero}, M.~A. and {Eskandarlou}, S. and {Ederoclite}, A.},
        title = "{Andromeda's tenuous veil: extensive nebular emission near (yet far from) M31}",
      journal = {arXiv e-prints},
         year = 2024,
        month = dec,
          eid = {arXiv:2412.08327},
        pages = {arXiv:2412.08327},
          doi = {10.48550/arXiv.2412.08327},
archivePrefix = {arXiv},
       eprint = {2412.08327},
 primaryClass = {astro-ph.GA},
       adsurl = {https://ui.adsabs.harvard.edu/abs/2024arXiv241208327L}
}

@ARTICLE{Spinoglio_1992,
       author = {{Spinoglio}, Luigi and {Malkan}, Matthew A.},
        title = "{Infrared Line Diagnostics of Active Galactic Nuclei}",
      journal = {\apj},
         year = 1992,
        month = nov,
       volume = {399},
        pages = {504},
          doi = {10.1086/171943},
       adsurl = {https://ui.adsabs.harvard.edu/abs/1992ApJ...399..504S}
}

@ARTICLE{Dudik_2007,
       author = {{Dudik}, R.~P. and {Weingartner}, J.~C. and {Satyapal}, S. and {Fischer}, Jacqueline and {Dudley}, C.~C. and {O'Halloran}, B.},
        title = "{Mid-Infrared Fine-Structure Line Ratios in Active Galactic Nuclei Observed with the Spitzer IRS: Evidence for Extinction by the Torus}",
      journal = {\apj},
         year = 2007,
        month = jul,
       volume = {664},
       number = {1},
        pages = {71-87},
          doi = {10.1086/518685},
archivePrefix = {arXiv},
       eprint = {0704.0547},
 primaryClass = {astro-ph},
       adsurl = {https://ui.adsabs.harvard.edu/abs/2007ApJ...664...71D}
}

@ARTICLE{Dudik_2009,
       author = {{Dudik}, R.~P. and {Satyapal}, S. and {Marcu}, D.},
        title = "{A Spitzer Spectroscopic Survey of Low-Ionization Nuclear Emission-Line Regions: Characterization of the Central Source}",
      journal = {\apj},
         year = 2009,
        month = feb,
       volume = {691},
       number = {2},
        pages = {1501-1524},
          doi = {10.1088/0004-637X/691/2/1501},
archivePrefix = {arXiv},
       eprint = {0811.1252},
 primaryClass = {astro-ph},
       adsurl = {https://ui.adsabs.harvard.edu/abs/2009ApJ...691.1501D}
}

@ARTICLE{Goold2026,
       author = {{Goold}, Kameron and {Seth}, Anil and {Molina}, Mallory and {Ohlson}, David and {Acharya}, Nischal and {B{\"o}ker}, Torsten and {Dumont}, Antoine and {Eracleous}, Michael and {Feldmeier-Krause}, Anja and {Fern{\'a}ndez-Ontiveros}, Juan Antonio and {Gallo}, Elena and {Goulding}, Andy D. and {G{\"u}ltekin}, Kayhan and {Ho}, Luis C. and {Neumayer}, Nadine and {Plotkin}, Richard M. and {Prieto}, Almudena and {Runnoe}, Jessie C. and {Satyapal}, Shobita and {van de Ven}, Glenn and {Walsh}, Jonelle L. and {Yuan}, Feng and {L{\"u}tzgendorf}, Nora},
        title = "{ReveaLLAGN 1: JWST Emission-Line Spectra Reveal Low-Luminosity AGN with UV-Deficient SEDs and Warm Molecular Gas}",
      journal = {arXiv e-prints},
         year = 2026,
        month = jan,
          eid = {arXiv:2601.16977},
        pages = {arXiv:2601.16977},
archivePrefix = {arXiv},
       eprint = {2601.16977},
 primaryClass = {astro-ph.GA},
       adsurl = {https://ui.adsabs.harvard.edu/abs/2026arXiv260116977G}
}

@ARTICLE{Goulding_2009,
       author = {{Goulding}, A.~D. and {Alexander}, D.~M.},
        title = "{Towards a complete census of AGN in nearby Galaxies: a large population of optically unidentified AGN}",
      journal = {\mnras},
         year = 2009,
        month = sep,
       volume = {398},
       number = {3},
        pages = {1165-1193},
          doi = {10.1111/j.1365-2966.2009.15194.x},
archivePrefix = {arXiv},
       eprint = {0906.0772},
 primaryClass = {astro-ph.CO},
       adsurl = {https://ui.adsabs.harvard.edu/abs/2009MNRAS.398.1165G}
}

@ARTICLE{Singh_2013,
       author = {{Singh}, V. and {Shastri}, P. and {Ishwara-Chandra}, C.~H. and {Athreya}, R.},
        title = "{Low-frequency radio observations of Seyfert galaxies: A test of the unification scheme}",
      journal = {\aap},
         year = 2013,
        month = jun,
       volume = {554},
          eid = {A85},
        pages = {A85},
          doi = {10.1051/0004-6361/201221003},
archivePrefix = {arXiv},
       eprint = {1304.0720},
 primaryClass = {astro-ph.CO},
       adsurl = {https://ui.adsabs.harvard.edu/abs/2013A&A...554A..85S}
}

@ARTICLE{Baldi2023,
       author = {{Baldi}, Ranieri D.},
        title = "{The nature of compact radio sources: the case of FR 0 radio galaxies}",
      journal = {\aapr},
         year = 2023,
        month = dec,
       volume = {31},
       number = {1},
          eid = {3},
        pages = {3},
          doi = {10.1007/s00159-023-00148-3},
archivePrefix = {arXiv},
       eprint = {2307.08379},
 primaryClass = {astro-ph.GA},
       adsurl = {https://ui.adsabs.harvard.edu/abs/2023A&ARv..31....3B}
}

@ARTICLE{Wang_2019,
       author = {{Wang}, Shu and {Chen}, Xiaodian},
        title = "{The Optical to Mid-infrared Extinction Law Based on the APOGEE, Gaia DR2, Pan-STARRS1, SDSS, APASS, 2MASS, and WISE Surveys}",
      journal = {\apj},
         year = 2019,
        month = jun,
       volume = {877},
       number = {2},
          eid = {116},
        pages = {116},
          doi = {10.3847/1538-4357/ab1c61},
archivePrefix = {arXiv},
       eprint = {1904.04575},
 primaryClass = {astro-ph.GA},
       adsurl = {https://ui.adsabs.harvard.edu/abs/2019ApJ...877..116W}
}

@article{Veilleux_2020,
	doi = {10.1007/s00159-019-0121-9},
  
	url = {https://doi.org/10.1007%2Fs00159-019-0121-9},
  
	year = 2020,
	month = {April},
  
	publisher = {Springer Science and Business Media {LLC}},
  
	volume = {28},
  
	number = {1},
  
	author = {Sylvain Veilleux and Roberto Maiolino and Alberto D. Bolatto and Susanne Aalto},
  
	title = {Cool outflows in galaxies and their implications},
  
	journal = {The Astronomy and Astrophysics Review}
}

@article{2,
  author = {R.Morganti},
  journal = {Front.Astron.Space Sci.},
  volume = {4},
  year = {2017}
}

@article{9,
  author = {K.Zubovas \& A.R.King},
  journal = {MNRAS},
  volume = {439},
  pages = {400--406},
  year = {2014}
}

@article{10,
  author = {A.Y.Wagner \& G.V.Bicknell},
  journal = {ApJ},
  volume = {728},
  pages = {29},
  year = {2011}
  
}

@article{16,
  author = {K.Sakamoto et al.},
  journal = {ApJ},
  volume = {797},
  pages = {90},
  year = {2014}
  }

@article{17,
	doi = {10.1086/301358},
  
	url = {https://doi.org/10.1086%2F301358},
  
	year = 2000,
	month = {May},
  
	publisher = {American Astronomical Society},
  
	volume = {119},
  
	number = {5},
  
	pages = {2085--2091},
  
	author = {T. A. Oosterloo and R. Morganti and A. Tzioumis and J. Reynolds and E. King and P. McCulloch and Z. Tsvetanov},
  
	title = {A Strong Jet-Cloud Interaction in the Seyfert Galaxy {IC} 5063: {VLBI} Observations},
  
	journal = {The Astronomical Journal}
}

@ARTICLE{Ferland_2009,
       author = {{Ferland}, G.~J. and {Fabian}, A.~C. and {Hatch}, N.~A. and {Johnstone}, R.~M. and {Porter}, R.~L. and {van Hoof}, P.~A.~M. and {Williams}, R.~J.~R.},
        title = "{Collisional heating as the origin of filament emission in galaxy clusters}",
      journal = {\mnras},
         year = 2009,
        month = feb,
       volume = {392},
       number = {4},
        pages = {1475-1502},
          doi = {10.1111/j.1365-2966.2008.14153.x},
archivePrefix = {arXiv},
       eprint = {0810.5372},
 primaryClass = {astro-ph},
       adsurl = {https://ui.adsabs.harvard.edu/abs/2009MNRAS.392.1475F}
}

@ARTICLE{Ferland_2017,
       author = {{Ferland}, G.~J. and {Chatzikos}, M. and {Guzm{\'a}n}, F. and {Lykins}, M.~L. and {van Hoof}, P.~A.~M. and {Williams}, R.~J.~R. and {Abel}, N.~P. and {Badnell}, N.~R. and {Keenan}, F.~P. and {Porter}, R.~L. and {Stancil}, P.~C.},
        title = "{The 2017 Release Cloudy}",
      journal = {Revista Mexicana de Astronomia y Astrofisica},
         year = 2017,
        month = oct,
       volume = {53},
        pages = {385-438},
          doi = {10.48550/arXiv.1705.10877},
archivePrefix = {arXiv},
       eprint = {1705.10877},
 primaryClass = {astro-ph.GA},
       adsurl = {https://ui.adsabs.harvard.edu/abs/2017RMxAA..53..385F}
}

@ARTICLE{calzetti,
       author = {{Calzetti}, Daniela},
        title = "{The Dust Opacity of Star-forming Galaxies}",
      journal = {Publications of the ASP},
         year = 2001,
        month = dec,
       volume = {113},
       number = {790},
        pages = {1449-1485},
          doi = {10.1086/324269},
archivePrefix = {arXiv},
       eprint = {astro-ph/0109035},
 primaryClass = {astro-ph},
       adsurl = {https://ui.adsabs.harvard.edu/abs/2001PASP..113.1449C}
}

@ARTICLE{Chatzikos_2023,
       author = {{Chatzikos}, M. and {Bianchi}, S. and {Camilloni}, F. and {Chakraborty}, P. and {Gunasekera}, C.~M. and {Guzm{\'a}n}, F. and {Milby}, J.~S. and {Sarkar}, A. and {Shaw}, G. and {van Hoof}, P.~A.~M. and {Ferland}, G.~J.},
        title = "{The 2023 Release of Cloudy}",
      journal = {\rmxaa},
         year = 2023,
        month = oct,
       volume = {59},
        pages = {327-343},
          doi = {10.22201/ia.01851101p.2023.59.02.12},
archivePrefix = {arXiv},
       eprint = {2308.06396},
 primaryClass = {astro-ph.GA},
       adsurl = {https://ui.adsabs.harvard.edu/abs/2023RMxAA..59..327C}
}

@article{Luridiana_2014,
	doi = {10.1051/0004-6361/201323152},
  
	url = {https://doi.org/10.1051%2F0004-6361%2F201323152},
  
	year = 2014,
	month = {December},
  
	publisher = {{EDP} Sciences},
  
	volume = {573},
  
	pages = {A42},
  
	author = {V. Luridiana and C. Morisset and R. A. Shaw},
  
	title = {{PyNeb}: a new tool for analyzing emission lines},
  
	journal = {Astronomy \& Astrophysics}
}

@ARTICLE{Allen_2008,
       author = {{Allen}, Mark G. and {Groves}, Brent A. and {Dopita}, Michael A. and {Sutherland}, Ralph S. and {Kewley}, Lisa J.},
        title = "{The MAPPINGS III Library of Fast Radiative Shock Models}",
      journal = {\apjs},
         year = 2008,
        month = sep,
       volume = {178},
       number = {1},
        pages = {20-55},
          doi = {10.1086/589652},
archivePrefix = {arXiv},
       eprint = {0805.0204},
 primaryClass = {astro-ph},
       adsurl = {https://ui.adsabs.harvard.edu/abs/2008ApJS..178...20A}
}

@ARTICLE{Asplund,
       author = {{Asplund}, Martin and {Grevesse}, Nicolas and {Sauval}, A. Jacques and {Scott}, Pat},
        title = "{The Chemical Composition of the Sun}",
      journal = {\araa},
         year = 2009,
        month = sep,
       volume = {47},
       number = {1},
        pages = {481-522},
          doi = {10.1146/annurev.astro.46.060407.145222},
archivePrefix = {arXiv},
       eprint = {0909.0948},
 primaryClass = {astro-ph.SR},
       adsurl = {https://ui.adsabs.harvard.edu/abs/2009ARA&A..47..481A}
}

@ARTICLE{Aalto_2015,
       author = {{Aalto}, S. and {Garcia-Burillo}, S. and {Muller}, S. and {Winters}, J.~M. and {Gonzalez-Alfonso}, E. and {van der Werf}, P. and {Henkel}, C. and {Costagliola}, F. and {Neri}, R.},
        title = "{High resolution observations of HCN and HCO$^{+}$J = 3-2 in the disk and outflow of Mrk 231. Detection of vibrationally excited HCN in the warped nucleus}",
      journal = {\aap},
         year = 2015,
        month = feb,
       volume = {574},
          eid = {A85},
        pages = {A85},
          doi = {10.1051/0004-6361/201423987},
archivePrefix = {arXiv},
       eprint = {1411.2474},
 primaryClass = {astro-ph.GA},
       adsurl = {https://ui.adsabs.harvard.edu/abs/2015A&A...574A..85A}
}

@ARTICLE{Silpa_2021,
       author = {{Silpa}, S. and {Kharb}, P. and {O'Dea}, C.~P. and {Baum}, S.~A. and {Sebastian}, B. and {Mukherjee}, D. and {Harrison}, C.~M.},
        title = "{AGN jets and winds in polarized light: the case of Mrk 231}",
      journal = {\mnras},
         year = 2021,
        month = oct,
       volume = {507},
       number = {2},
        pages = {2550-2561},
          doi = {10.1093/mnras/stab2110},
archivePrefix = {arXiv},
       eprint = {2107.09466},
 primaryClass = {astro-ph.GA},
       adsurl = {https://ui.adsabs.harvard.edu/abs/2021MNRAS.507.2550S}
}

@article{pypla,
	author = {{Bacon, Roland} and {Brinchmann, Jarle} and {Conseil, Simon} and {Maseda, Michael} and {Nanayakkara, Themiya} and {Wendt, Martin} and {Bacher, Raphael} and {Mary, David} and {Weilbacher, Peter M.} and {Krajnovi\'{}c, Davor} and {Boogaard, Leindert} and {Bouch\'e, Nicolas} and {Contini, Thierry} and {Epinat, Beno\^{\i}t} and {Feltre, Anna} and {Guo, Yucheng} and {Herenz, Christian} and {Kollatschny, Wolfram} and {Kusakabe, Haruka} and {Leclercq, Floriane} and {Michel-Dansac, L\'eo} and {Pello, Roser} and {Richard, Johan} and {Roth, Martin} and {Salvignol, Gregory} and {Schaye, Joop} and {Steinmetz, Matthias} and {Tresse, Laurence} and {Urrutia, Tanya} and {Verhamme, Anne} and {Vitte, Eloise} and {Wisotzki, Lutz} and {Zoutendijk, Sebastiaan L.}},
	title = {The MUSE Hubble Ultra Deep Field surveys: Data release II},
	DOI= "10.1051/0004-6361/202244187",
	url= "https://doi.org/10.1051/0004-6361/202244187",
	journal = {A\&A},
	year = 2023,
	volume = 670,
	pages = "A4",
}

@MISC{MPDAF,
       author = {{Bacon}, Roland and {Piqueras}, Laure and {Conseil}, Simon and {Richard}, Johan and {Shepherd}, Martin},
        title = "{MPDAF: MUSE Python Data Analysis Framework}",
 howpublished = {Astrophysics Source Code Library, record ascl:1611.003},
         year = 2016,
        month = nov,
          eid = {ascl:1611.003},
        pages = {ascl:1611.003},
archivePrefix = {ascl},
       eprint = {1611.003},
       adsurl = {https://ui.adsabs.harvard.edu/abs/2016ascl.soft11003B}
}

@ARTICLE{Blasi_2013,
       author = {{Blasi}, Pasquale},
        title = "{The origin of galactic cosmic rays}",
      journal = {\aapr},
         year = 2013,
        month = nov,
       volume = {21},
          eid = {70},
        pages = {70},
          doi = {10.1007/s00159-013-0070-7},
archivePrefix = {arXiv},
       eprint = {1311.7346},
 primaryClass = {astro-ph.HE},
       adsurl = {https://ui.adsabs.harvard.edu/abs/2013A&ARv..21...70B}
}

@ARTICLE{1981BPT,
	author = {{Baldwin}, J. A. and {Phillips}, M. M. and {Terlevich}, R.},
	title = "{Classification parameters for the emission-line spectra of extragalactic objects.}",
	journal = {\pasp},
	year = 1981,
	month = feb,
	volume = {93},
	pages = {5-19},
	doi = {10.1086/130766},
	adsurl = {https://ui.adsabs.harvard.edu/abs/1981PASP...93....5B}
}

@ARTICLE{Kantzas_2023,
       author = {{Kantzas}, D. and {Markoff}, S. and {Cooper}, A.~J. and {Gaggero}, D. and {Petropoulou}, M. and {De La Torre Luque}, P.},
        title = "{Possible contribution of X-ray binary jets to the Galactic cosmic ray and neutrino flux}",
      journal = {\mnras},
         year = 2023,
        month = sep,
       volume = {524},
       number = {1},
        pages = {1326-1342},
          doi = {10.1093/mnras/stad1909},
archivePrefix = {arXiv},
       eprint = {2306.12715},
 primaryClass = {astro-ph.HE},
       adsurl = {https://ui.adsabs.harvard.edu/abs/2023MNRAS.524.1326K}
}

@ARTICLE{2003Kauf,
	author = {{Kauffmann}, Guinevere and {Heckman}, Timothy M. and {Tremonti}, Christy and {Brinchmann}, Jarle and {Charlot}, St{\'e}phane and {White}, Simon D.~M. and {Ridgway}, Susan E. and {Brinkmann}, Jon and {Fukugita}, Masataka and {Hall}, Patrick B. and {Ivezi{\'c}}, {\v{Z}}eljko and {Richards}, Gordon T. and {Schneider}, Donald P.},
	title = "{The host galaxies of active galactic nuclei}",
	journal = {\mnras},
	year = 2003,
	month = dec,
	volume = {346},
	number = {4},
	pages = {1055-1077},
	doi = {10.1111/j.1365-2966.2003.07154.x},
	archivePrefix = {arXiv},
	eprint = {astro-ph/0304239},
	primaryClass = {astro-ph},
	adsurl = {https://ui.adsabs.harvard.edu/abs/2003MNRAS.346.1055K}
}

@ARTICLE{2006Kewley,
	author = {{Kewley}, Lisa J. and {Groves}, Brent and {Kauffmann}, Guinevere and {Heckman}, Tim},
	title = "{The host galaxies and classification of active galactic nuclei}",
	journal = {\mnras},
	year = 2006,
	month = nov,
	volume = {372},
	number = {3},
	pages = {961-976},
	doi = {10.1111/j.1365-2966.2006.10859.x},
	archivePrefix = {arXiv},
	eprint = {astro-ph/0605681},
	primaryClass = {astro-ph},
	adsurl = {https://ui.adsabs.harvard.edu/abs/2006MNRAS.372..961K}
}

@ARTICLE{Fernandez2021,
       author = {{Fern{\'a}ndez-Ontiveros}, J.~A. and {P{\'e}rez-Montero}, E. and {V{\'\i}lchez}, J.~M. and {Amor{\'\i}n}, R. and {Spinoglio}, L.},
        title = "{Measuring chemical abundances with infrared nebular lines: HII-CHI-MISTRY-IR}",
      journal = {\aap},
         year = 2021,
        month = aug,
       volume = {652},
          eid = {A23},
        pages = {A23},
          doi = {10.1051/0004-6361/202039716},
archivePrefix = {arXiv},
       eprint = {2103.09253},
 primaryClass = {astro-ph.GA},
       adsurl = {https://ui.adsabs.harvard.edu/abs/2021A&A...652A..23F}
}

@ARTICLE{2004aGroves,
       author = {{Groves}, Brent A. and {Dopita}, Michael A. and {Sutherland}, Ralph S.},
        title = "{Dusty, Radiation Pressure-Dominated Photoionization. I. Model Description, Structure, and Grids}",
      journal = {\apjs},
         year = 2004,
        month = jul,
       volume = {153},
       number = {1},
        pages = {9-73},
          doi = {10.1086/421113},
archivePrefix = {arXiv},
       eprint = {astro-ph/0404175},
 primaryClass = {astro-ph},
       adsurl = {https://ui.adsabs.harvard.edu/abs/2004ApJS..153....9G}
}

@ARTICLE{Groves_2006,
       author = {{Groves}, Brent A. and {Heckman}, Timothy M. and {Kauffmann}, Guinevere},
        title = "{Emission-line diagnostics of low-metallicity active galactic nuclei}",
      journal = {\mnras},
         year = 2006,
        month = oct,
       volume = {371},
       number = {4},
        pages = {1559-1569},
          doi = {10.1111/j.1365-2966.2006.10812.x},
archivePrefix = {arXiv},
       eprint = {astro-ph/0607311},
 primaryClass = {astro-ph},
       adsurl = {https://ui.adsabs.harvard.edu/abs/2006MNRAS.371.1559G}
}

@ARTICLE{Gabici_2022,
       author = {{Gabici}, Stefano},
        title = "{Low-energy cosmic rays: regulators of the dense interstellar medium}",
      journal = {\aapr},
         year = 2022,
        month = dec,
       volume = {30},
       number = {1},
          eid = {4},
        pages = {4},
          doi = {10.1007/s00159-022-00141-2},
archivePrefix = {arXiv},
       eprint = {2203.14620},
 primaryClass = {astro-ph.HE},
       adsurl = {https://ui.adsabs.harvard.edu/abs/2022A&ARv..30....4G}
}

@article{MING,
title={The MAGNUM survey: different gas properties in the outflowing and disc components in nearby active galaxies with MUSE},
volume={622},
ISSN={1432-0746},
url={http://dx.doi.org/10.1051/0004-6361/201834372},
DOI={10.1051/0004-6361/201834372},
journal = {\aap},
publisher={EDP Sciences},
author={Mingozzi, M. and Cresci, G. and Venturi, G. and Marconi, A. and Mannucci, F. and Perna, M. and Belfiore, F. and Carniani, S. and Balmaverde, B. and Brusa, M. and Cicone, C. and Feruglio, C. and Gallazzi, A. and Mainieri, V. and Maiolino, R. and Nagao, T. and Nardini, E. and Sani, E. and Tozzi, P. and Zibetti, S.},
year={2019},
month=feb, 
pages={A146}
 }

@article{Wolfire_2022,
   title={Photodissociation and X-Ray-Dominated Regions},
   volume={60},
   ISSN={1545-4282},
   url={http://dx.doi.org/10.1146/annurev-astro-052920-010254},
   DOI={10.1146/annurev-astro-052920-010254},
   number={1},
   journal={Annual Review of Astronomy and Astrophysics},
   publisher={Annual Reviews},
   author={Wolfire, Mark G. and Vallini, Livia and Chevance, Mélanie},
   year={2022},
   month=aug, pages={247–318} }

@ARTICLE{Dopita_1995,
       author = {{Dopita}, Michael A. and {Sutherland}, Ralph S.},
        title = "{Spectral Signatures of Fast Shocks. II. Optical Diagnostic Diagrams}",
      journal = {\apj},
         year = 1995,
        month = dec,
       volume = {455},
        pages = {468},
          doi = {10.1086/176596},
       adsurl = {https://ui.adsabs.harvard.edu/abs/1995ApJ...455..468D}
}

@ARTICLE{McKee_1989,
       author = {{McKee}, Christopher F.},
        title = "{Photoionization-regulated Star Formation and the Structure of Molecular Clouds}",
      journal = {\apj},
         year = 1989,
        month = oct,
       volume = {345},
        pages = {782},
          doi = {10.1086/167950},
       adsurl = {https://ui.adsabs.harvard.edu/abs/1989ApJ...345..782M}
}

@ARTICLE{Padovani_2017,
       author = {{Padovani}, P. and {Alexander}, D.~M. and {Assef}, R.~J. and {De Marco}, B. and {Giommi}, P. and {Hickox}, R.~C. and {Richards}, G.~T. and {Smol{\v{c}}i{\'c}}, V. and {Hatziminaoglou}, E. and {Mainieri}, V. and {Salvato}, M.},
        title = "{Active galactic nuclei: what's in a name?}",
      journal = {\aapr},
         year = 2017,
        month = aug,
       volume = {25},
       number = {1},
          eid = {2},
        pages = {2},
          doi = {10.1007/s00159-017-0102-9},
archivePrefix = {arXiv},
       eprint = {1707.07134},
 primaryClass = {astro-ph.GA},
       adsurl = {https://ui.adsabs.harvard.edu/abs/2017A&ARv..25....2P}
}

@ARTICLE{Padovani_2018,
       author = {{Padovani}, Marco and {Ivlev}, Alexei V. and {Galli}, Daniele and {Caselli}, Paola},
        title = "{Cosmic-ray ionisation in circumstellar discs}",
      journal = {\aap},
         year = 2018,
        month = jun,
       volume = {614},
          eid = {A111},
        pages = {A111},
          doi = {10.1051/0004-6361/201732202},
archivePrefix = {arXiv},
       eprint = {1803.09348},
 primaryClass = {astro-ph.HE},
       adsurl = {https://ui.adsabs.harvard.edu/abs/2018A&A...614A.111P}
}

@ARTICLE{Spitzer_1968,
       author = {{Spitzer}, Lyman, Jr. and {Tomasko}, Martin G.},
        title = "{Heating of H i Regions by Energetic Particles}",
      journal = {\apj},
         year = 1968,
        month = jun,
       volume = {152},
        pages = {971},
          doi = {10.1086/149610},
       adsurl = {https://ui.adsabs.harvard.edu/abs/1968ApJ...152..971S}
}

@ARTICLE{Tommasin2008,
       author = {{Tommasin}, Silvia and {Spinoglio}, Luigi and {Malkan}, Matthew A. and {Smith}, Howard and {Gonz{\'a}lez-Alfonso}, Eduardo and {Charmandaris}, Vassilis},
        title = "{Spitzer IRS High-Resolution Spectroscopy of the 12 {\ensuremath{\mu}}m Seyfert Galaxies. I. First Results}",
      journal = {\apj},
         year = 2008,
        month = apr,
       volume = {676},
       number = {2},
        pages = {836-856},
          doi = {10.1086/527290},
archivePrefix = {arXiv},
       eprint = {0710.4448},
 primaryClass = {astro-ph},
       adsurl = {https://ui.adsabs.harvard.edu/abs/2008ApJ...676..836T}
}

@ARTICLE{Tommasin2010,
       author = {{Tommasin}, Silvia and {Spinoglio}, Luigi and {Malkan}, Matthew A. and {Fazio}, Giovanni},
        title = "{Spitzer-IRS High-Resolution Spectroscopy of the 12 {\ensuremath{\mu}}m Seyfert Galaxies. II. Results for the Complete Data Set}",
      journal = {\apj},
         year = 2010,
        month = feb,
       volume = {709},
       number = {2},
        pages = {1257-1283},
          doi = {10.1088/0004-637X/709/2/1257},
archivePrefix = {arXiv},
       eprint = {0911.3348},
 primaryClass = {astro-ph.CO},
       adsurl = {https://ui.adsabs.harvard.edu/abs/2010ApJ...709.1257T}
}

@ARTICLE{Weilbacher_2020,
       author = {{Weilbacher}, Peter M. and {Palsa}, Ralf and {Streicher}, Ole and {Bacon}, Roland and {Urrutia}, Tanya and {Wisotzki}, Lutz and {Conseil}, Simon and {Husemann}, Bernd and {Jarno}, Aur{\'e}lien and {Kelz}, Andreas and {P{\'e}contal-Rousset}, Arlette and {Richard}, Johan and {Roth}, Martin M. and {Selman}, Fernando and {Vernet}, Jo{\"e}l},
        title = "{The data processing pipeline for the MUSE instrument}",
      journal = {\aap},
         year = 2020,
        month = sep,
       volume = {641},
          eid = {A28},
        pages = {A28},
          doi = {10.1051/0004-6361/202037855},
archivePrefix = {arXiv},
       eprint = {2006.08638},
 primaryClass = {astro-ph.IM},
       adsurl = {https://ui.adsabs.harvard.edu/abs/2020A&A...641A..28W}
}

@ARTICLE{Tremonti_2004,
       author = {{Tremonti}, Christy A. and {Heckman}, Timothy M. and {Kauffmann}, Guinevere and {Brinchmann}, Jarle and {Charlot}, St{\'e}phane and {White}, Simon D.~M. and {Seibert}, Mark and {Peng}, Eric W. and {Schlegel}, David J. and {Uomoto}, Alan and {Fukugita}, Masataka and {Brinkmann}, Jon},
        title = "{The Origin of the Mass-Metallicity Relation: Insights from 53,000 Star-forming Galaxies in the Sloan Digital Sky Survey}",
      journal = {\apj},
         year = 2004,
        month = oct,
       volume = {613},
       number = {2},
        pages = {898-913},
          doi = {10.1086/423264},
archivePrefix = {arXiv},
       eprint = {astro-ph/0405537},
 primaryClass = {astro-ph},
       adsurl = {https://ui.adsabs.harvard.edu/abs/2004ApJ...613..898T}
}

@ARTICLE{Feltre_2016,
       author = {{Feltre}, A. and {Charlot}, S. and {Gutkin}, J.},
        title = "{Nuclear activity versus star formation: emission-line diagnostics at ultraviolet and optical wavelengths}",
      journal = {\mnras},
         year = 2016,
        month = mar,
       volume = {456},
       number = {3},
        pages = {3354-3374},
          doi = {10.1093/mnras/stv2794},
archivePrefix = {arXiv},
       eprint = {1511.08217},
 primaryClass = {astro-ph.GA},
       adsurl = {https://ui.adsabs.harvard.edu/abs/2016MNRAS.456.3354F}
}

@ARTICLE{Gonz_2013,
       author = {{Gonz{\'a}lez-Alfonso}, E. and {Fischer}, J. and {Bruderer}, S. and {M{\"u}ller}, H.~S.~P. and {Graci{\'a}-Carpio}, J. and {Sturm}, E. and {Lutz}, D. and {Poglitsch}, A. and {Feuchtgruber}, H. and {Veilleux}, S. and {Contursi}, A. and {Sternberg}, A. and {Hailey-Dunsheath}, S. and {Verma}, A. and {Christopher}, N. and {Davies}, R. and {Genzel}, R. and {Tacconi}, L.},
        title = "{Excited OH$^{+}$, H$_{2}$O$^{+}$, and H$_{3}$O$^{+}$ in NGC 4418 and Arp 220}",
      journal = {\aap},
         year = 2013,
        month = feb,
       volume = {550},
          eid = {A25},
        pages = {A25},
          doi = {10.1051/0004-6361/201220466},
archivePrefix = {arXiv},
       eprint = {1211.5064},
 primaryClass = {astro-ph.GA},
       adsurl = {https://ui.adsabs.harvard.edu/abs/2013A&A...550A..25G}
}

@software{newville_2015_11813,
  author       = {Newville, Matthew and
                  Stensitzki, Till and
                  Allen, Daniel B. and
                  Ingargiola,  Antonino},
  title        = {{LMFIT: Non-Linear Least-Square Minimization and 
                   Curve-Fitting for Python}},
  month        = oct,
  year         = 2015,
  publisher    = {Zenodo},
  version      = {0.8.0},
  doi          = {10.5281/zenodo.11813},
  url          = {https://doi.org/10.5281/zenodo.11813}
}

@INPROCEEDINGS{Weilbacher_2014,
       author = {{Weilbacher}, P.~M. and {Streicher}, O. and {Urrutia}, T. and {P{\'e}contal-Rousset}, A. and {Jarno}, A. and {Bacon}, R.},
        title = "{The MUSE Data Reduction Pipeline: Status after Preliminary Acceptance Europe}",
    booktitle = {Astronomical Data Analysis Software and Systems XXIII},
         year = 2014,
       editor = {{Manset}, N. and {Forshay}, P.},
       series = {Astronomical Society of the Pacific Conference Series},
       volume = {485},
        month = may,
        pages = {451},
          doi = {10.48550/arXiv.1507.00034},
archivePrefix = {arXiv},
       eprint = {1507.00034},
 primaryClass = {astro-ph.IM},
       adsurl = {https://ui.adsabs.harvard.edu/abs/2014ASPC..485..451W}
}

@article{Perez_Diaz_2021,
   title={Chemical abundances in the nuclear region of nearby galaxies from the Palomar Survey},
   volume={505},
   ISSN={1365-2966},
   url={http://dx.doi.org/10.1093/mnras/stab1522},
   DOI={10.1093/mnras/stab1522},
   number={3},
   journal={Monthly Notices of the Royal Astronomical Society},
   publisher={Oxford University Press (OUP)},
   author={Pérez-Díaz, B and Masegosa, J and Márquez, I and Pérez-Montero, E},
   year={2021},
   month=may, pages={4289–4309} }

@ARTICLE{Perez_Diaz_2022,
       author = {{P{\'e}rez-D{\'\i}az}, Borja and {P{\'e}rez-Montero}, Enrique and {Fern{\'a}ndez-Ontiveros}, Juan A. and {V{\'\i}lchez}, Jos{\'e} M.},
        title = "{Measuring chemical abundances in AGN from infrared nebular lines: HII-CHI-MISTRY-IR for AGN}",
      journal = {\aap},
         year = 2022,
        month = oct,
       volume = {666},
          eid = {A115},
        pages = {A115},
          doi = {10.1051/0004-6361/202243602},
archivePrefix = {arXiv},
       eprint = {2207.08718},
 primaryClass = {astro-ph.GA},
       adsurl = {https://ui.adsabs.harvard.edu/abs/2022A&A...666A.115P}
}

@ARTICLE{Perez_Diaz_2024,
       author = {{P{\'e}rez-D{\'\i}az}, Borja and {P{\'e}rez-Montero}, Enrique and {Fern{\'a}ndez-Ontiveros}, Juan A. and {V{\'\i}lchez}, Jos{\'e} M. and {Hern{\'a}n-Caballero}, Antonio and {Amor{\'\i}n}, Ricardo},
        title = "{Chemical abundances and deviations from the solar S/O ratio in the gas-phase interstellar medium of galaxies based on infrared emission lines}",
      journal = {\aap},
         year = 2024,
        month = may,
       volume = {685},
          eid = {A168},
        pages = {A168},
          doi = {10.1051/0004-6361/202348318},
archivePrefix = {arXiv},
       eprint = {2403.02903},
 primaryClass = {astro-ph.GA},
       adsurl = {https://ui.adsabs.harvard.edu/abs/2024A&A...685A.168P}
}

@ARTICLE{Perez_2014,
       author = {{P{\'e}rez-Montero}, E.},
        title = "{Deriving model-based T$_{e}$-consistent chemical abundances in ionized gaseous nebulae}",
      journal = {\mnras},
         year = 2014,
        month = jul,
       volume = {441},
       number = {3},
        pages = {2663-2675},
          doi = {10.1093/mnras/stu753},
archivePrefix = {arXiv},
       eprint = {1404.3936},
 primaryClass = {astro-ph.GA},
       adsurl = {https://ui.adsabs.harvard.edu/abs/2014MNRAS.441.2663P}
}

@ARTICLE{Perez_2019,
       author = {{P{\'e}rez-Montero}, E. and {Dors}, O.~L. and {V{\'\i}lchez}, J.~M. and {Garc{\'\i}a-Benito}, R. and {Cardaci}, M.~V. and {H{\"a}gele}, G.~F.},
        title = "{A bayesian-like approach to derive chemical abundances in type-2 active galactic nuclei based on photoionization models}",
      journal = {\mnras},
         year = 2019,
        month = oct,
       volume = {489},
       number = {2},
        pages = {2652-2668},
          doi = {10.1093/mnras/stz2278},
archivePrefix = {arXiv},
       eprint = {1908.04827},
 primaryClass = {astro-ph.GA},
       adsurl = {https://ui.adsabs.harvard.edu/abs/2019MNRAS.489.2652P}
}

@article{Dors_2022,
   title={Chemical abundances in Seyfert galaxies – IX. Helium abundance estimates},
   volume={514},
   ISSN={1365-2966},
   url={http://dx.doi.org/10.1093/mnras/stac1722},
   DOI={10.1093/mnras/stac1722},
   number={4},
   journal={Monthly Notices of the Royal Astronomical Society},
   publisher={Oxford University Press (OUP)},
   author={Dors, O L and Valerdi, M and Freitas-Lemes, P and Krabbe, A C and Riffel, R A and Amôres, E B and Riffel, R and Armah, M and Monteiro, A F and Oliveira, C B},
   year={2022},
   month=jun, pages={5506–5527} }

@article{Schawinski_2007,
    author = {Schawinski, Kevin and Thomas, Daniel and Sarzi, Marc and Maraston, Claudia and Kaviraj, Sugata and Joo, Seok-Joo and Yi, Sukyoung K. and Silk, Joseph},
    title = "{Observational evidence for AGN feedback in early-type galaxies}",
    journal = {Monthly Notices of the Royal Astronomical Society},
    volume = {382},
    number = {4},
    pages = {1415-1431},
    year = {2007},
    month = {11},
    issn = {0035-8711},
    doi = {10.1111/j.1365-2966.2007.12487.x},
    url = {https://doi.org/10.1111/j.1365-2966.2007.12487.x},
    eprint = {https://academic.oup.com/mnras/article-pdf/382/4/1415/3940026/mnras0382-1415.pdf},
}

@ARTICLE{Zhu_2023,
       author = {{Zhu}, Peixin and {Kewley}, Lisa J. and {Sutherland}, Ralph S.},
        title = "{A New Photoionization Model of the Narrow-line Region in Active Galactic Nuclei}",
      journal = {\apj},
         year = 2023,
        month = sep,
       volume = {954},
       number = {2},
          eid = {175},
        pages = {175},
          doi = {10.3847/1538-4357/acd757},
archivePrefix = {arXiv},
       eprint = {2305.12670},
 primaryClass = {astro-ph.GA},
       adsurl = {https://ui.adsabs.harvard.edu/abs/2023ApJ...954..175Z}
}

@BOOK{Rybicki_Lightman,
       author = {{Rybicki}, George B. and {Lightman}, Alan P.},
        title = "{Radiative processes in astrophysics}",
         year = 1979,
      publisher = {{John Wiley \& Sons}},
       adsurl = {https://ui.adsabs.harvard.edu/abs/1979rpa..book.....R}
}

@ARTICLE{Padovani_2009,
       author = {{Padovani}, M. and {Galli}, D. and {Glassgold}, A.~E.},
        title = "{Cosmic-ray ionization of molecular clouds}",
      journal = {\aap},
         year = 2009,
        month = jul,
       volume = {501},
       number = {2},
        pages = {619-631},
          doi = {10.1051/0004-6361/200911794},
archivePrefix = {arXiv},
       eprint = {0904.4149},
 primaryClass = {astro-ph.SR},
       adsurl = {https://ui.adsabs.harvard.edu/abs/2009A&A...501..619P}
}

@ARTICLE{Ferland_2013,
       author = {{Ferland}, G.~J. and {Porter}, R.~L. and {van Hoof}, P.~A.~M. and {Williams}, R.~J.~R. and {Abel}, N.~P. and {Lykins}, M.~L. and {Shaw}, G. and {Henney}, W.~J. and {Stancil}, P.~C.},
        title = "{The 2013 Release of Cloudy}",
      journal = {\rmxaa},
         year = 2013,
        month = apr,
       volume = {49},
        pages = {137-163},
          doi = {10.48550/arXiv.1302.4485},
archivePrefix = {arXiv},
       eprint = {1302.4485},
 primaryClass = {astro-ph.GA},
       adsurl = {https://ui.adsabs.harvard.edu/abs/2013RMxAA..49..137F}
}

@ARTICLE{Gonz_Alf_2018,
       author = {{Gonz{\'a}lez-Alfonso}, E. and {Fischer}, J. and {Bruderer}, S. and {Ashby}, M.~L.~N. and {Smith}, H.~A. and {Veilleux}, S. and {M{\"u}ller}, H.~S.~P. and {Stewart}, K.~P. and {Sturm}, E.},
        title = "{Outflowing OH$^{+}$ in Markarian 231: The Ionization Rate of the Molecular Gas}",
      journal = {\apj},
         year = 2018,
        month = apr,
       volume = {857},
       number = {1},
          eid = {66},
        pages = {66},
          doi = {10.3847/1538-4357/aab6b8},
archivePrefix = {arXiv},
       eprint = {1803.04690},
 primaryClass = {astro-ph.GA},
       adsurl = {https://ui.adsabs.harvard.edu/abs/2018ApJ...857...66G}
}

@ARTICLE{Bruzual_2003,
       author = {{Bruzual}, G. and {Charlot}, S.},
        title = "{Stellar population synthesis at the resolution of 2003}",
      journal = {\mnras},
         year = 2003,
        month = oct,
       volume = {344},
       number = {4},
        pages = {1000-1028},
          doi = {10.1046/j.1365-8711.2003.06897.x},
archivePrefix = {arXiv},
       eprint = {astro-ph/0309134},
 primaryClass = {astro-ph},
       adsurl = {https://ui.adsabs.harvard.edu/abs/2003MNRAS.344.1000B}
}

@article{Brinchmann_2004,
    author = {Brinchmann, J. and Charlot, S. and White, S. D. M. and Tremonti, C. and Kauffmann, G. and Heckman, T. and Brinkmann, J.},
    title = "{The physical properties of star-forming galaxies in the low-redshift Universe}",
    journal = {Monthly Notices of the Royal Astronomical Society},
    volume = {351},
    number = {4},
    pages = {1151-1179},
    year = {2004},
    month = {07},
    issn = {0035-8711},
    doi = {10.1111/j.1365-2966.2004.07881.x},
    url = {https://doi.org/10.1111/j.1365-2966.2004.07881.x},
    eprint = {https://academic.oup.com/mnras/article-pdf/351/4/1151/36455094/351-4-1151.pdf},
}

@article{Brinchmann_2013,
    author = {Brinchmann, Jarle and Charlot, Stéphane and Kauffmann, Guinevere and Heckman, Tim and White, Simon D. M. and Tremonti, Christy},
    title = "{Estimating gas masses and dust-to-gas ratios from optical spectroscopy}",
    journal = {Monthly Notices of the Royal Astronomical Society},
    volume = {432},
    number = {3},
    pages = {2112-2140},
    year = {2013},
    month = {05},
    issn = {0035-8711},
    doi = {10.1093/mnras/stt551},
    url = {https://doi.org/10.1093/mnras/stt551},
    eprint = {https://academic.oup.com/mnras/article-pdf/432/3/2112/16841480/stt551.pdf},
}

@ARTICLE{Dors_alone_2021,
       author = {{Dors}, Oli L.},
        title = "{Chemical abundances in Seyfert galaxies - VI. Empirical abundance calibration}",
      journal = {\mnras},
         year = 2021,
        month = oct,
       volume = {507},
       number = {1},
        pages = {466-474},
          doi = {10.1093/mnras/stab2166},
       adsurl = {https://ui.adsabs.harvard.edu/abs/2021MNRAS.507..466D}
}

@ARTICLE{Oliveira_2024,
       author = {{Oliveira}, C.~B. and {Krabbe}, A.~C. and {Dors}, O.~L. and {Zinchenko}, I.~A. and {Hernandez-Jimenez}, J.~A. and {Cardaci}, M.~V. and {H{\"a}gele}, G.~F. and {Ilha}, G.~S.},
        title = "{Chemical abundances of LINER galaxies - nitrogen abundance estimations}",
      journal = {\mnras},
         year = 2024,
        month = jun,
       volume = {531},
       number = {1},
        pages = {199-212},
          doi = {10.1093/mnras/stae1172},
archivePrefix = {arXiv},
       eprint = {2404.16631},
 primaryClass = {astro-ph.GA},
       adsurl = {https://ui.adsabs.harvard.edu/abs/2024MNRAS.531..199O}
}

@ARTICLE{Abazajian_2009,
       author = {{Abazajian}, Kevork N. and {Adelman-McCarthy}, Jennifer K. and {Ag{\"u}eros}, Marcel A. and {Allam}, Sahar S. and {Allende Prieto}, Carlos and {An}, Deokkeun and {Anderson}, Kurt S.~J. and {Anderson}, Scott F. and {Annis}, James and {Bahcall}, Neta A. and {Bailer-Jones}, C.~A.~L. and {Barentine}, J.~C. and {Bassett}, Bruce A. and {Becker}, Andrew C. and {Beers}, Timothy C. and {Bell}, Eric F. and {Belokurov}, Vasily and {Berlind}, Andreas A. and {Berman}, Eileen F. and {Bernardi}, Mariangela and {Bickerton}, Steven J. and {Bizyaev}, Dmitry and {Blakeslee}, John P. and {Blanton}, Michael R. and {Bochanski}, John J. and {Boroski}, William N. and {Brewington}, Howard J. and {Brinchmann}, Jarle and {Brinkmann}, J. and {Brunner}, Robert J. and {Budav{\'a}ri}, Tam{\'a}s and {Carey}, Larry N. and {Carliles}, Samuel and {Carr}, Michael A. and {Castander}, Francisco J. and {Cinabro}, David and {Connolly}, A.~J. and {Csabai}, Istv{\'a}n and {Cunha}, Carlos E. and {Czarapata}, Paul C. and {Davenport}, James R.~A. and {de Haas}, Ernst and {Dilday}, Ben and {Doi}, Mamoru and {Eisenstein}, Daniel J. and {Evans}, Michael L. and {Evans}, N.~W. and {Fan}, Xiaohui and {Friedman}, Scott D. and {Frieman}, Joshua A. and {Fukugita}, Masataka and {G{\"a}nsicke}, Boris T. and {Gates}, Evalyn and {Gillespie}, Bruce and {Gilmore}, G. and {Gonzalez}, Belinda and {Gonzalez}, Carlos F. and {Grebel}, Eva K. and {Gunn}, James E. and {Gy{\"o}ry}, Zsuzsanna and {Hall}, Patrick B. and {Harding}, Paul and {Harris}, Frederick H. and {Harvanek}, Michael and {Hawley}, Suzanne L. and {Hayes}, Jeffrey J.~E. and {Heckman}, Timothy M. and {Hendry}, John S. and {Hennessy}, Gregory S. and {Hindsley}, Robert B. and {Hoblitt}, J. and {Hogan}, Craig J. and {Hogg}, David W. and {Holtzman}, Jon A. and {Hyde}, Joseph B. and {Ichikawa}, Shin-ichi and {Ichikawa}, Takashi and {Im}, Myungshin and {Ivezi{\'c}}, {\v{Z}}eljko and {Jester}, Sebastian and {Jiang}, Linhua and {Johnson}, Jennifer A. and {Jorgensen}, Anders M. and {Juri{\'c}}, Mario and {Kent}, Stephen M. and {Kessler}, R. and {Kleinman}, S.~J. and {Knapp}, G.~R. and {Konishi}, Kohki and {Kron}, Richard G. and {Krzesinski}, Jurek and {Kuropatkin}, Nikolay and {Lampeitl}, Hubert and {Lebedeva}, Svetlana and {Lee}, Myung Gyoon and {Lee}, Young Sun and {French Leger}, R. and {L{\'e}pine}, S{\'e}bastien and {Li}, Nolan and {Lima}, Marcos and {Lin}, Huan and {Long}, Daniel C. and {Loomis}, Craig P. and {Loveday}, Jon and {Lupton}, Robert H. and {Magnier}, Eugene and {Malanushenko}, Olena and {Malanushenko}, Viktor and {Mandelbaum}, Rachel and {Margon}, Bruce and {Marriner}, John P. and {Mart{\'\i}nez-Delgado}, David and {Matsubara}, Takahiko and {McGehee}, Peregrine M. and {McKay}, Timothy A. and {Meiksin}, Avery and {Morrison}, Heather L. and {Mullally}, Fergal and {Munn}, Jeffrey A. and {Murphy}, Tara and {Nash}, Thomas and {Nebot}, Ada and {Neilsen}, Eric H., Jr. and {Newberg}, Heidi Jo and {Newman}, Peter R. and {Nichol}, Robert C. and {Nicinski}, Tom and {Nieto-Santisteban}, Maria and {Nitta}, Atsuko and {Okamura}, Sadanori and {Oravetz}, Daniel J. and {Ostriker}, Jeremiah P. and {Owen}, Russell and {Padmanabhan}, Nikhil and {Pan}, Kaike and {Park}, Changbom and {Pauls}, George and {Peoples}, John, Jr. and {Percival}, Will J. and {Pier}, Jeffrey R. and {Pope}, Adrian C. and {Pourbaix}, Dimitri and {Price}, Paul A. and {Purger}, Norbert and {Quinn}, Thomas and {Raddick}, M. Jordan and {Re Fiorentin}, Paola and {Richards}, Gordon T. and {Richmond}, Michael W. and {Riess}, Adam G. and {Rix}, Hans-Walter and {Rockosi}, Constance M. and {Sako}, Masao and {Schlegel}, David J. and {Schneider}, Donald P. and {Scholz}, Ralf-Dieter and {Schreiber}, Matthias R. and {Schwope}, Axel D. and {Seljak}, Uro{\v{s}} and {Sesar}, Branimir and {Sheldon}, Erin and {Shimasaku}, Kazu and {Sibley}, Valena C. and {Simmons}, A.~E. and {Sivarani}, Thirupathi and {Allyn Smith}, J. and {Smith}, Martin C. and {Smol{\v{c}}i{\'c}}, Vernesa and {Snedden}, Stephanie A. and {Stebbins}, Albert and {Steinmetz}, Matthias and {Stoughton}, Chris and {Strauss}, Michael A. and {SubbaRao}, Mark and {Suto}, Yasushi and {Szalay}, Alexander S. and {Szapudi}, Istv{\'a}n and {Szkody}, Paula and {Tanaka}, Masayuki and {Tegmark}, Max and {Teodoro}, Luis F.~A. and {Thakar}, Aniruddha R. and {Tremonti}, Christy A. and {Tucker}, Douglas L. and {Uomoto}, Alan and {Vanden Berk}, Daniel E. and {Vandenberg}, Jan and {Vidrih}, S. and {Vogeley}, Michael S. and {Voges}, Wolfgang and {Vogt}, Nicole P. and {Wadadekar}, Yogesh and {Watters}, Shannon and {Weinberg}, David H. and {West}, Andrew A. and {White}, Simon D.~M. and {Wilhite}, Brian C. and {Wonders}, Alainna C. and {Yanny}, Brian and {Yocum}, D.~R. and {York}, Donald G. and {Zehavi}, Idit and {Zibetti}, Stefano and {Zucker}, Daniel B.},
        title = "{The Seventh Data Release of the Sloan Digital Sky Survey}",
      journal = {\apjs},
         year = 2009,
        month = jun,
       volume = {182},
       number = {2},
        pages = {543-558},
          doi = {10.1088/0067-0049/182/2/543},
archivePrefix = {arXiv},
       eprint = {0812.0649},
 primaryClass = {astro-ph},
       adsurl = {https://ui.adsabs.harvard.edu/abs/2009ApJS..182..543A}
}

@ARTICLE{Marconi_2024,
       author = {{Marconi}, A. and {Amiri}, A. and {Feltre}, A. and {Belfiore}, F. and {Cresci}, G. and {Curti}, M. and {Mannucci}, F. and {Bertola}, E. and {Brazzini}, M. and {Carniani}, S. and {Cataldi}, E. and {D'Amato}, Q. and {de Rosa}, G. and {Di Teodoro}, E. and {Ginolfi}, M. and {Kumari}, N. and {Marconcini}, C. and {Maiolino}, R. and {Magrini}, L. and {Marasco}, A. and {Mingozzi}, M. and {Moreschini}, B. and {Nagao}, T. and {Oliva}, E. and {Scialpi}, M. and {Tomicic}, N. and {Tozzi}, G. and {Ulivi}, L. and {Venturi}, G.},
        title = "{HOMERUN: A new approach to photoionization modeling: I. Reproducing observed emission lines with percent accuracy and obtaining accurate physical properties of the ionized gas}",
      journal = {\aap},
         year = 2024,
        month = sep,
       volume = {689},
          eid = {A78},
        pages = {A78},
          doi = {10.1051/0004-6361/202449240},
archivePrefix = {arXiv},
       eprint = {2401.13028},
 primaryClass = {astro-ph.GA},
       adsurl = {https://ui.adsabs.harvard.edu/abs/2024A&A...689A..78M}
}

@ARTICLE{Blanc_2015,
       author = {{Blanc}, Guillermo A. and {Kewley}, Lisa and {Vogt}, Fr{\'e}d{\'e}ric P.~A. and {Dopita}, Michael A.},
        title = "{IZI: Inferring the Gas Phase Metallicity (Z) and Ionization Parameter (q) of Ionized Nebulae Using Bayesian Statistics}",
      journal = {\apj},
         year = 2015,
        month = jan,
       volume = {798},
       number = {2},
          eid = {99},
        pages = {99},
          doi = {10.1088/0004-637X/798/2/99},
archivePrefix = {arXiv},
       eprint = {1410.8146},
 primaryClass = {astro-ph.GA},
       adsurl = {https://ui.adsabs.harvard.edu/abs/2015ApJ...798...99B}
}

@ARTICLE{Thomas_2018,
       author = {{Thomas}, Adam D. and {Kewley}, Lisa J. and {Dopita}, Michael A. and {Groves}, Brent A. and {Hopkins}, Andrew M. and {Sutherland}, Ralph S.},
        title = "{Mixing between Seyfert and H II Region Excitation in Local Active Galaxies}",
      journal = {\apjl},
         year = 2018,
        month = jul,
       volume = {861},
       number = {1},
          eid = {L2},
        pages = {L2},
          doi = {10.3847/2041-8213/aacce7},
archivePrefix = {arXiv},
       eprint = {1806.06364},
 primaryClass = {astro-ph.GA},
       adsurl = {https://ui.adsabs.harvard.edu/abs/2018ApJ...861L...2T}
}


\begin{appendix} 
\onecolumn
\section{Aperture Properties}\label{ap_prop}

\begin{table*}[htb]
    \centering
    \caption{Aperture positions and minimum S/N for emission lines utilized in both optical and MIR bands, projected distance $D$ to the nucleus, and aperture category as defined in Fig.~\ref{fig:chosen_apertures}.}
    \scriptsize
    \setlength{\tabcolsep}{4pt} 
    \renewcommand{\arraystretch}{1.05} 
    \begin{tabular}{|c|c|c|c|c|c|}
        \hline
        \textbf{Aperture}
            & \makecell{\textbf{Position} \\ ($\Delta \alpha$, $\Delta \delta$)}
            & \makecell{\textbf{Min S/N}\\ \textbf{Optical}}
            & \makecell{\textbf{Min S/N} \\ \textbf{MIR}}
            & \makecell{\textbf{D} \\ $[\rm pc]$}
            & \makecell{\textbf{Category} \\ (Fig.~\ref{subfig:sketch})} \\
        \hline
         N   & ($0.0,\ 0.0$) & 7.8 & 12.1 & 0   & JR  \\
         2   & ($-0.7,\ -1.3$) & 6.8 & 5.4  & 293 & IJR \\
         3   & ($-0.9,\ 1.2$) & 7.7 & 7.3  & 297 & JR  \\
         4   & ($0.9,\ 1.3$) & \textbf{2.7}\textsuperscript{a} & 12.7 & 313 & IJR \\
         5   & ($1.1,\ -1.2$) & 11.7 & 7.7  & 323 & JR  \\
         6   & ($-1.9,\ 0.0$) & 8.0 & 7.7  & 376 & IJR \\
         7   & ($1.9,\ 0.2$ ) & 8.2 & 12.0 & 379 & IJR \\
         8   & ($0.1,\ -2.6$) & 4.8 & 7.0  & 516 & IJR \\
         9   & ($-0.1,\ 2.6$) & 8.2 & 4.5  & 516 & IJR \\
         10  & ($-2.3,\ -1.6$) & 5.7 & 4.0  & 555 & NJR \\
         11  & ($-2.7,\ 1.4$) & 7.8 & 6.4  & 603 & JR  \\
         12  & ($2.7,\ -1.6$) & 17.4 & 8.2  & 622 & JR  \\
         13  & ($2.9,\ 1.4$) & 8.2 & 10.7 & 638 & NJR \\
         14  & ($1.7,\ -2.8$) & 9.9 & 7.0  & 649 & IJR \\
         15  & ($-1.7,\ 2.8$) & 8.1 & \textbf{2.0}\textsuperscript{b}  & 649 & IJR \\
         16  & ($3.5,\ -0.2$) & 28.6 & 6.8  & 695 & IJR \\
         17  & ($-3.6,\ -0.8$) & 8.7 & \textbf{2.0}\textsuperscript{c} & 731 & NJR \\
        \hline
    \end{tabular}
    \begin{minipage}{0.85\textwidth}
    \centering
    \scriptsize
    \textsuperscript{a}Faint detection of [\ion{O}{I}]$\lambda6300\,\text{\AA}$, visually verified. Excluding this line, the next lowest S/N is 5.2.\\
    \textsuperscript{b}Faint detection of [\ion{Ar}{V}]$\lambda13.1\,\mu\text{m}$, visually verified. Excluding this line, the next lowest S/N is 5.7.\\
    \textsuperscript{c}Faint detection of [\ion{Ar}{III}]$\lambda21.8\,\mu\text{m}$, visually verified. Excluding this line, the next lowest S/N is 7.0.\\

    \end{minipage}
    \label{tab:aperture_sn}
\end{table*}

\section{Intermediate SED}\label{app_sed}

Here we present the intermediate AGN spectral energy SED produced with \textsc{Cloudy} 23.01 and introduced in \cite{Ferland_2020} versus the SED used in KFDS25. The intermediate SED was adopted in our new model grid to better reproduce high-ionization MIR emission lines.


\eliza{
More precisely, in the grid of models used for the AGN cases in KFDS25, the SED of the central photoionizing continuum is assumed to consist of a large blue bump from the accretion disc, radio emission from synchrotron radiation in the jet, infrared emission from dust, a soft X-ray excess, and a power-law component at hard X-rays. This SED may be expressed mathematically by Equation (2) in KFDS25.
In that formula, $F_\nu$ is the flux density as a function of frequency $\nu$; $\alpha_{\text{uv}}$ and $\alpha_{\text{x}}$ are the UV and X-ray spectral indices, respectively; $T_{\text{bb}}$ is the characteristic temperature of the big blue bump; $T_{\text{IR}}$ is the temperature corresponding to the infrared cutoff of the big blue bump; and $\alpha_{\text{ox}}$ describes the optical-to-X-ray spectral index. In KFDS25 AGN models, we adopted the following values: $\alpha_{\text{uv}} = -0.5$, $\alpha_{\text{x}} = -1.0$, $\alpha_{\text{ox}} = -1.4$, $T_{\text{BB}} = 10^5\,\mathrm{K}$, and $T_{\text{IR}} = 1.6 \times 10^3\,\mathrm{K}$. In the frequency range between the SED peak and $10^{16}$eV, we also find a slope of approximately $-1$ for the blue curve in Fig. \ref{fig:SEDs}.}

\eliza{
For the intermediate SED, utilized in the present work, we adopt the similarly named in \cite{Ferland_2020} SED,  based on the analysis by \cite{Jin_2012}, 
energy-conserving SEDs for unobscured AGN, grouped according to their Eddington ratio $L/L_\mathrm{Edd}$. The intermediate SED corresponds to a mean value of $\log (L/L_\mathrm{Edd}) \simeq -0.55$ and the template available in \textsc{Cloudy} 23.01 displays a shallower slope of approximately $-0.4$ (i.e., $\alpha_{\text{uv}} \simeq -0.4$) in the frequency range immediately after the SED peak up to $10^{17}$eV.}



\begin{figure}[ht!!!!!!!!!!!!!!!!!!!!!!!]
    \centering
    \includegraphics[width=0.45\linewidth]{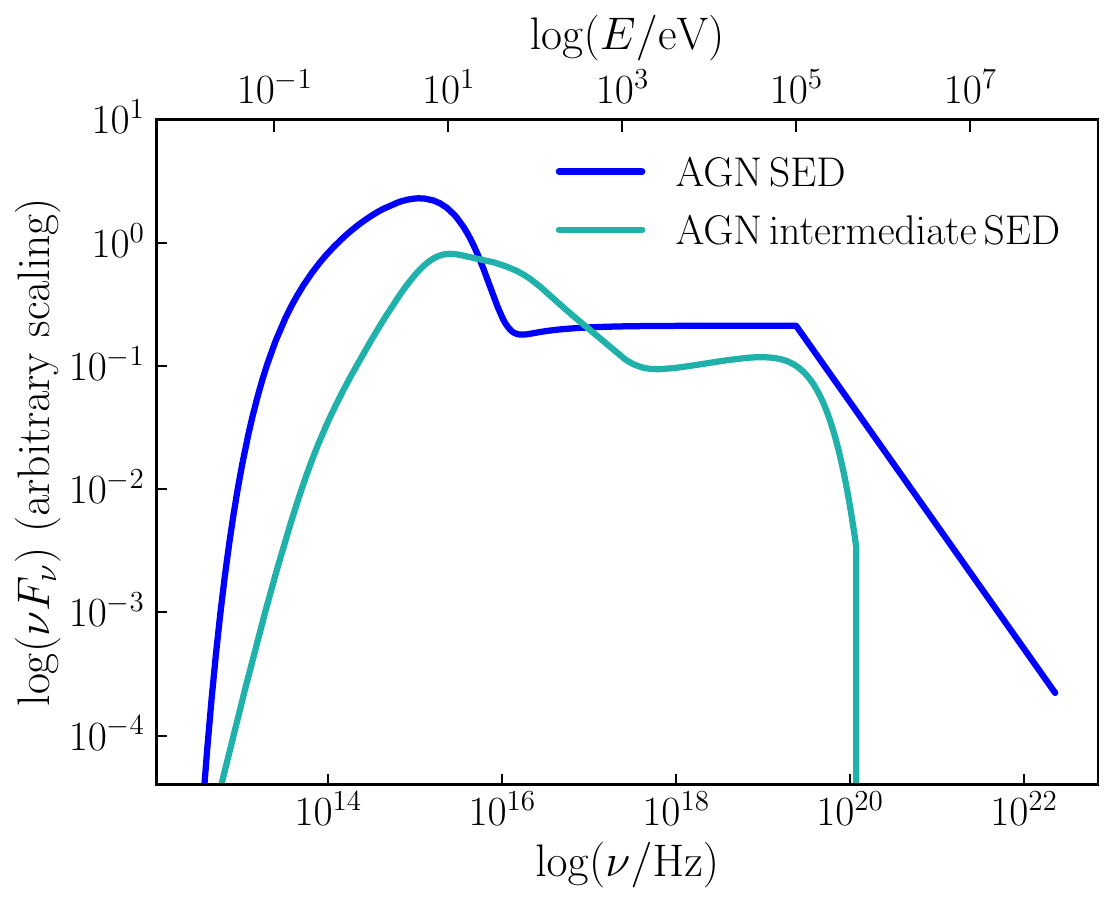}
  	\caption{\eliz{AGN SEDs simulated with \textsc{Cloudy}. The blue solid line represents the AGN SED used in KFDS25
    while the teal solid line represents the  AGN used in this paper. The y-axis is $\nu F_\nu$ in arbitrary scaling.}}
    \label{fig:SEDs}
    \vspace{-0.4cm}
\end{figure}

\clearpage

\section{BPT diagnostics with the intermediate SED}
\begin{figure*}[ht!!!!!!!!!!!!!!!]
    \centering
    \subfigure[]{\includegraphics[width=\textwidth]{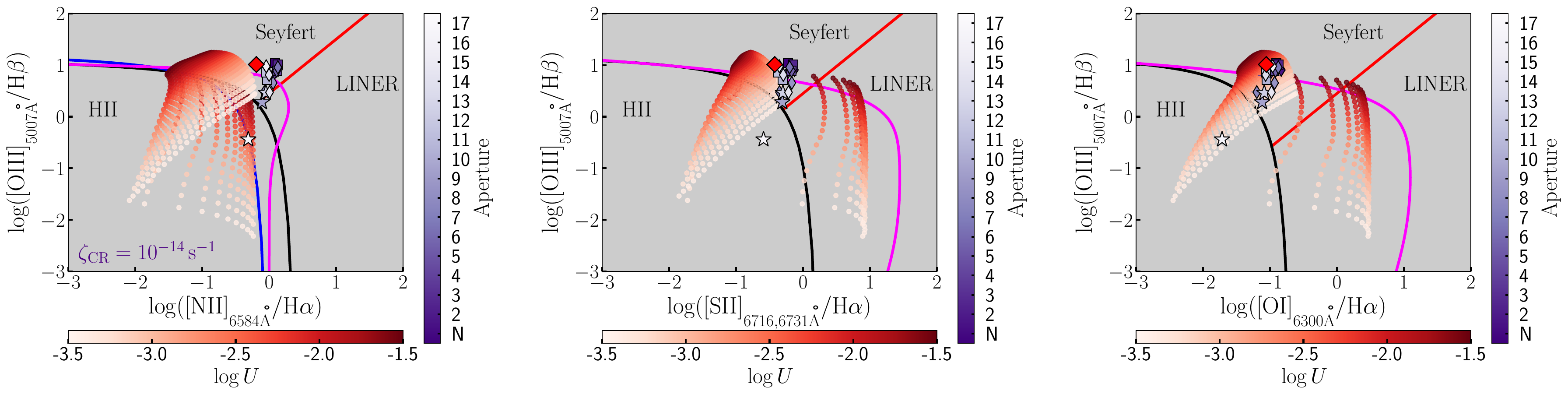}\label{subfig:U_5728_14}}
    \subfigure[]{\includegraphics[width=\textwidth]{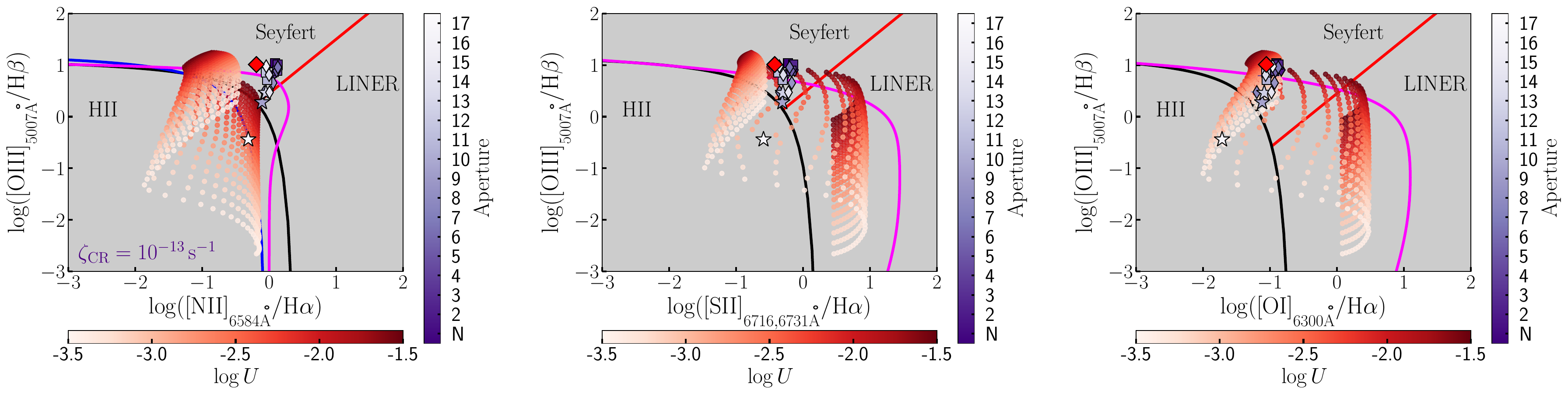}\label{subfig:U_5728_13}}
    \subfigure[]{\includegraphics[width=\textwidth]{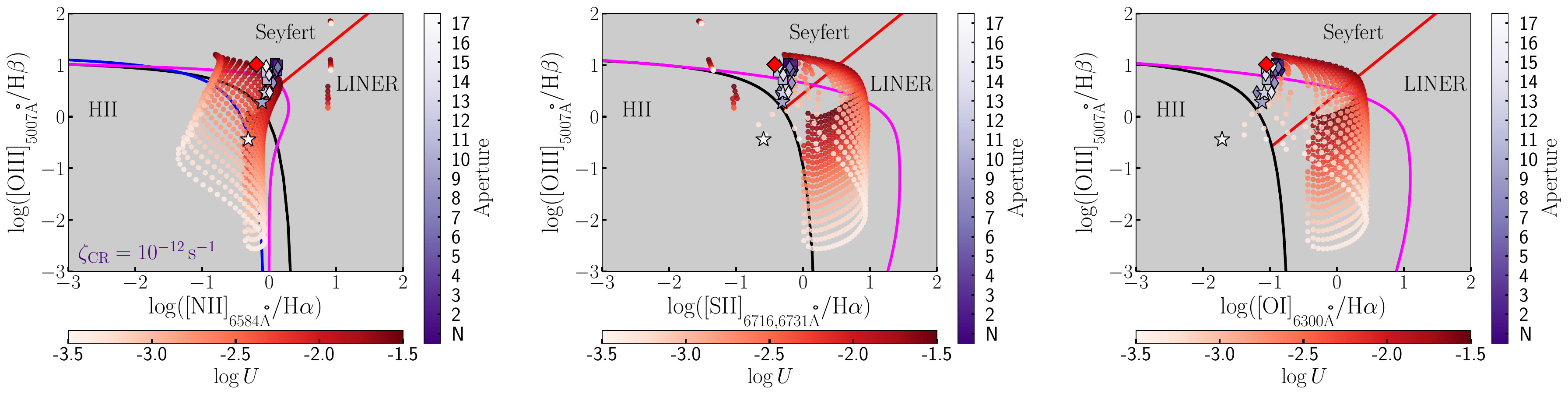}\label{subfig:U_5728_12}}
    \caption{BPT diagrams comparing AGN photoionization models with observations from selected apertures in NGC 5728 (Fig.~\ref{fig:chosen_apertures}). The BPT diagrams for [\ion{N}{ii}], [\ion{S}{ii}], and [\ion{O}{i}] are shown on the left, middle, and right, respectively. The different shades of purple, ranging from deep purple to pale lilac and/or white, represent the increasing distance from the nucleus, as also indicated by numbers, with "N" denoting the nuclear aperture. The different shapes-square, thin diamond, and star-represent the nucleus and/or jet impacted, intermediate, and distant areas, respectively. The different shades from white to deep red represent the range of ionization parameter values, $-3.5\leq \log U\leq -1.5$. The red diamonds represent the measured line ratios for the photoionization-dominated Seyfert 2 nucleus in NGC 1320. The Kewley, Kauffmann, Schawinski, and Koutsoumpou (SF\texorpdfstring{$\zeta$}{zeta}) lines correspond to the solid black, blue, red, and magenta lines, respectively. The panels from top to bottom correspond to $\zeta_\mathrm{CR}=10^{-14}\,\rm s^{-1},\,10^{-13}\,\rm s^{-1}$, and $10^{-12}\,\rm s^{-1}$, respectively.}\label{fig:5728_BPTS_U}
    \vspace{-0.2cm}
\end{figure*}

\clearpage

\section{Density estimates and extra MIR diagnostics}\label{appendix_gal}

Both [\ion{Ne}{v}] and [\ion{Ar}{v}] ratios serve as density diagnostics; in the sampled parameter space, they exhibit ratios of $0.5 \lesssim [\ion{Ne}{v}]/ [\ion{Ne}{v}]\lesssim 1.4$ and $1 \lesssim [\ion{Ar}{v}]/ [\ion{Ar}{v}]\lesssim 1.7$, respectively (see Fig. \ref{fig:pn}). These density values align well with the densities anticipated by the ratios obtained via our \textsc{Cloudy} simulations, as it can be seen in the first and second column of Fig. \ref{fig:5728_APPENDIX}, and with the initial hydrogen densities of our models (see Table \ref{tab:AGN_Model_Grid}).

\begin{figure*}[!ht]
    \centering
    \subfigure[]{\includegraphics[width=0.5\textwidth]{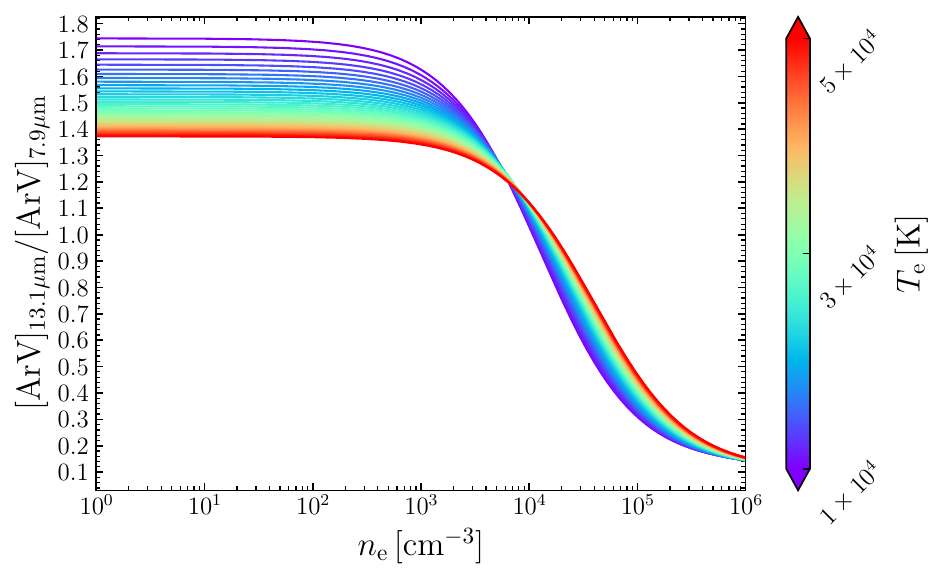}\label{subfig:Ar5_pn}}~
    \subfigure[]{\includegraphics[width=0.5\textwidth]{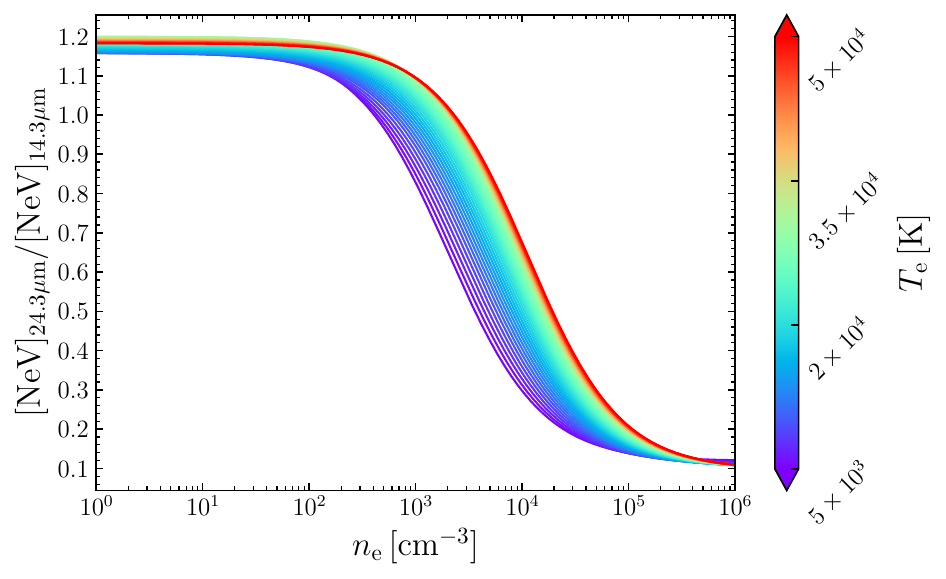}\label{subfig:Ne5_pn}}
    \caption{The 
    $[\ion{Ar}{V}]_{13.1\rm \mu m}/[\ion{Ar}{V}]_{7.9\rm \mu m}$ and $\rm [\ion{Ne}{V}]_{24.3\rm \mu m}/[\ion{Ne}{V}]_{14.3\rm \mu m}$ line ratios produced with \textsc{PyNeb} \citep{Luridiana_2014} as a function of electron density, $n_{\rm e}$, notated for different electron temperatures, $T_{\rm e}$.}
    \label{fig:pn}
    \vspace{-0.2cm}
\end{figure*}
\begin{figure*}[!!!!!!!!!!!!ht]
    \centering
    \includegraphics[width=0.95\textwidth]{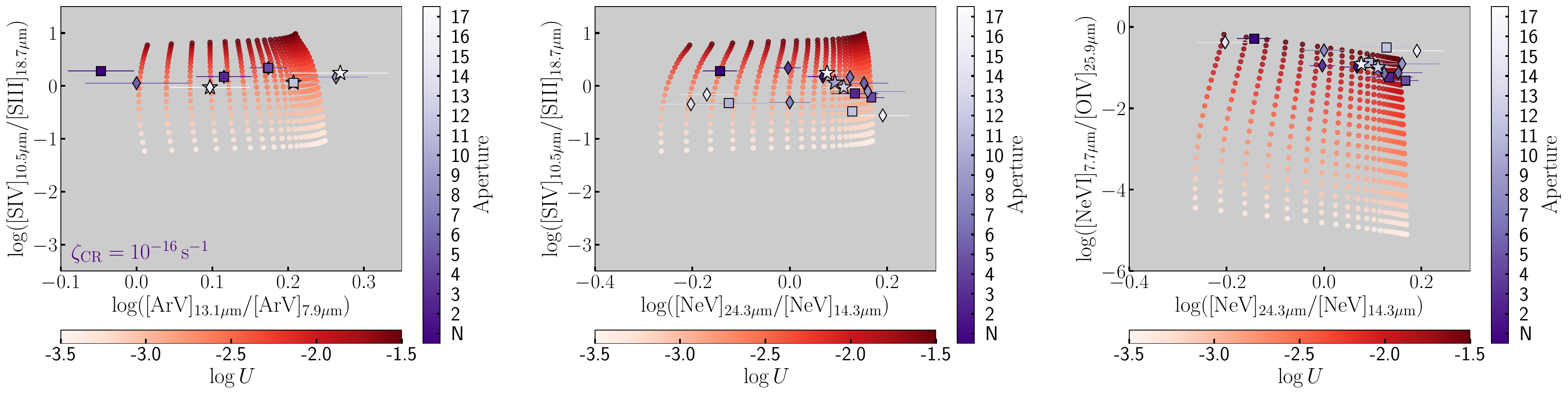}
    \includegraphics[width=0.95\textwidth]{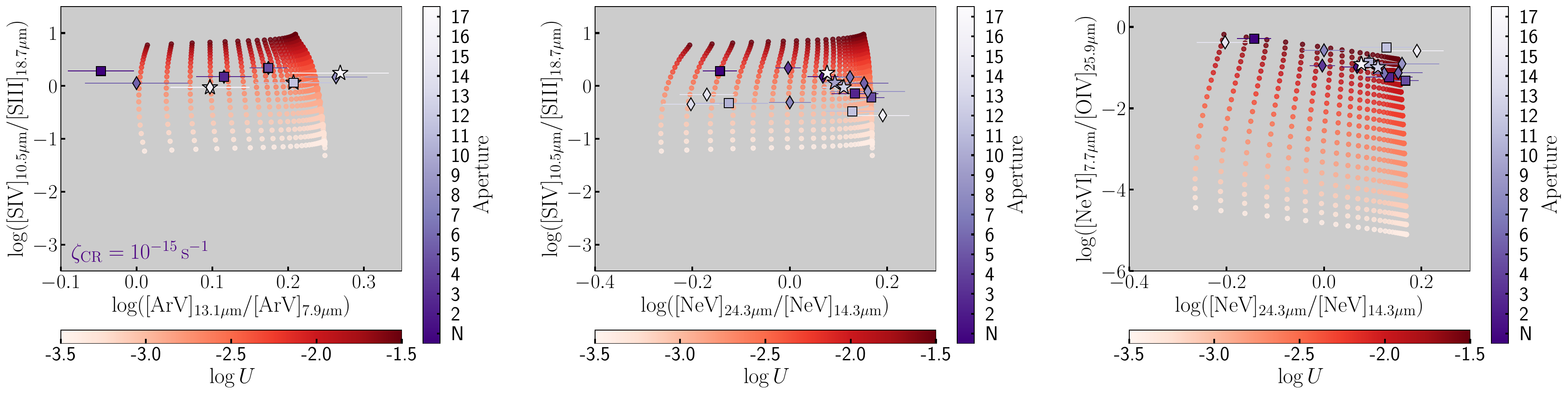}
    \includegraphics[width=0.95\textwidth]{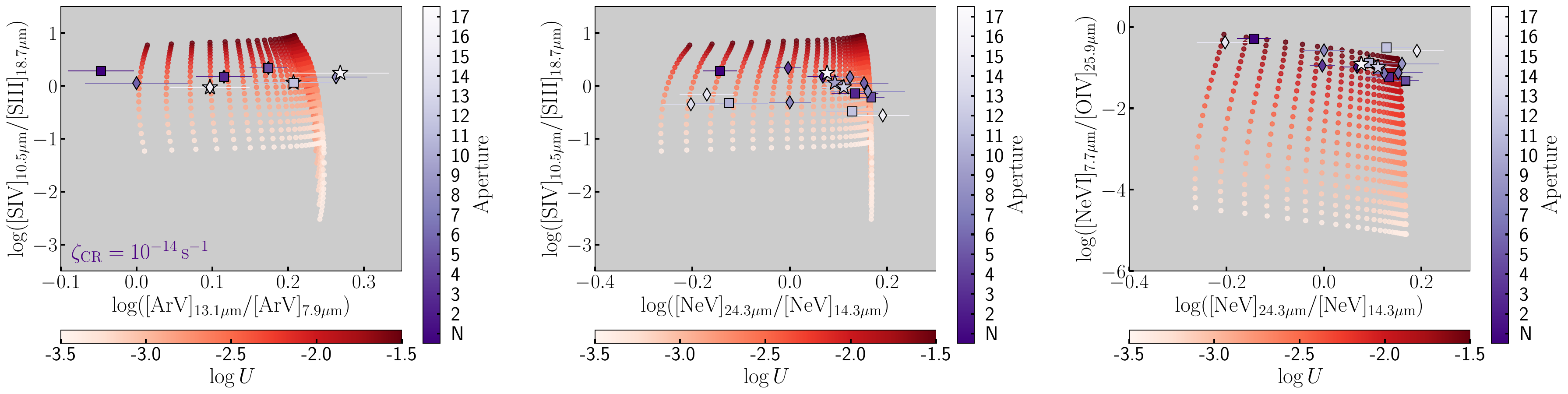}
    \includegraphics[width=0.95\textwidth]{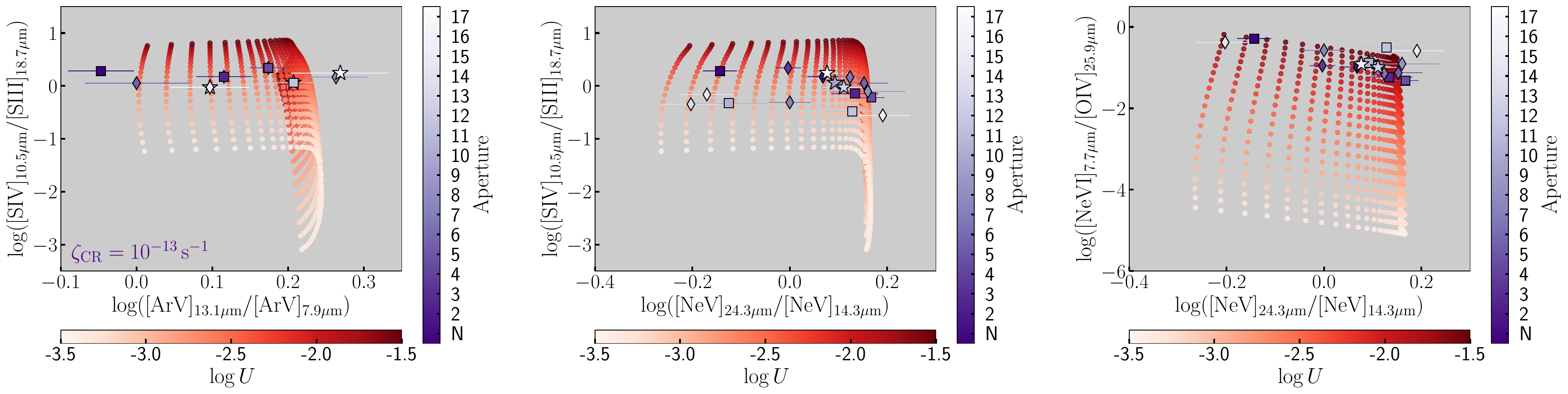}
    \includegraphics[width=0.95\textwidth]{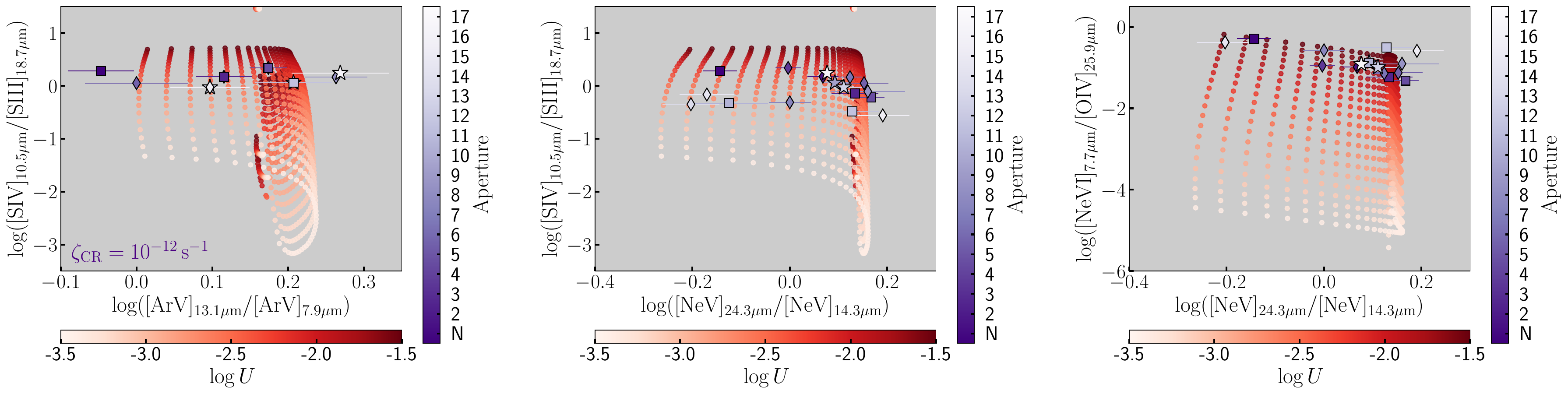}
   \caption{Diagrams with the AGN photoionization models compared with the observations from the selected apertures in NGC 5728 (Fig.\ref{fig:chosen_apertures}). 
  The different shades of purple, ranging from deep purple to pale lilac and/or white represent the increasing distance from the nucleus, as also denoted by numbers, with "N" corresponding to the nuclear aperture. The different shapes-square, thin diamond, and star, represent the nucleus and/or jet impacted, intermediate, and distant areas, respectively. The different shades from white to deep red represent the range of ionization parameter values, $-3.5\leq \log U\leq -1.5$. All the models have solar abundances. The panels from top to bottom correspond to  $\zeta_\mathrm{CR}=10^{-16}\,\rm s^{-1},\,10^{-15}\,\rm s^{-1},\,10^{-14}\,\rm s^{-1},\,10^{-13}\,\rm s^{-1}$, and $10^{-12}\,\rm s^{-1}$, respectively.}\label{fig:5728_APPENDIX}
\end{figure*}
In the first and second column of Fig. \ref{fig:5728_APPENDIX}, we present the [\ion{S}{iv}]$\lambda10.5\mu$m/[\ion{S}{iii}]$\lambda18.7\mu$m ratio plotted against the \ion{Ar}{v}]$\lambda13.1\mu$m/[\ion{Ar}{v}]$\lambda7.9\mu$m and the [\ion{Ne}{v}]$\lambda24.3\mu$m/[\ion{Ne}{v}]$\lambda14.3\mu$m ratios, respectively. The ratios \eliza{on the horizontal axes, [\ion{Ar}{v}]$\lambda13.1\mu$m/[\ion{Ar}{v}]$\lambda7.9\mu$m and [\ion{Ne}{v}]$\lambda24.3\mu$m/[\ion{Ne}{v}]$\lambda14.3\mu$m, are density tracers if the gas is assumed to be in thermodynamical equilibrium and}, appear relatively stable as the CR ionization rate increases from the top to the bottom row. \eliza{The vertical axis on both first and second column of Fig. \ref{fig:5728_APPENDIX}, the [\ion{S}{iv}]$\lambda10.5\mu$m/[\ion{S}{iii}]$\lambda18.7\mu$m ratio, is mildly affected by the increasing $\zeta_{\rm CR}$, due to the sensitivity of [\ion{S}{iii}] to CRs and to a lesser degree that of [\ion{S}{iv}] (see Fig. \ref{fig:mir_lines_struc}b, c) .} Finally, in the third column of Fig. \ref{fig:5728_APPENDIX}, the [\ion{Ne}{v}] ratio combined with the [\ion{Ne}{vi}]$\lambda7.7\mu$m /[\ion{O}{iv}]$\lambda25.9\mu$m  ratio, is unaffected by CRs suggesting that high-ionization lines are primarily influenced by photoionization.

\end{appendix}

\end{document}